\documentclass[12pt]{article}
\usepackage[utf8]{inputenc}

\usepackage{amsmath}
\usepackage{amssymb}
\usepackage{amsthm}
\usepackage{color}
\usepackage{bm}
\usepackage[colorlinks,linkcolor=blue,anchorcolor=green,citecolor=red]{hyperref}
\usepackage{appendix}
\usepackage{graphicx}
\usepackage{subfigure}
\usepackage{booktabs}
\usepackage{epstopdf}

\usepackage{subfigure}
\allowdisplaybreaks[4]
\usepackage[utf8]{inputenc}          
\usepackage[T1]{fontenc}  

\usepackage{url}
\usepackage{hyperref}

\title{Mathematical Model for Chemical Reactions in Electrolyte Applied to 
	Cytochrome $c$ Oxidase: an Electro-osmotic Approach}
\bigskip
\author{Shixin Xu\thanks{Duke Kunshan University, 8 Duke Ave, Kunshan, Jiangsu, China.}  
	\and Robert  Eisenberg \thanks{Department of Applied Mathematics, Illinois Institute of Technology, Chicago, IL, 60616, USA; Department of Physiology and Biophysics, Rush University, Chicago, IL, 60612, USA.}
	\and Zilong Song \thanks{ Math and Statistics Department, Utah State University,   Old Main Hill
		Logan, UT 84322.}
	\and  Huaxiong Huang\thanks{Corresponding author, Research Centre for Mathematics, Advanced Institute of Natural Sciences, Beijing Normal University (Zhuhai), China; BNU-
		HKBU United International College, Zhuhai, China; Department of Mathematics and Statistics York University, Toronto, ON, M3J 1P3, Canada (hhuang@uic.edu.cn). }}
\date{\today}
\usepackage{amsmath}
\usepackage{geometry}
\newcommand\bD{\boldsymbol{D}}
\newcommand\bE{\boldsymbol{E}}
\newcommand\bj{\boldsymbol{j}}

\newtheorem{rmk}{Remark}[section]

\usepackage[
backend=biber,
style=numeric,
sorting=ynt
]{biblatex}

\begin{document}
	
	\maketitle
	\begin{abstract}
A mathematical model for chemical reactions in electrolytes is developed using an Energy variational method consistent with classical thermodynamics. Electrostatics and chemical reactions are included in properly defined energetic and
dissipative functionals. The energy variation method is generalized to deal with open systems with
inputs of charge, mass, and energy. The open systems can transform input energy of one type into
output energy of another type. The generalized method is used to analyze the conversion
of electrical current into proton flow by cytochrome c oxidase, an important enzyme in mitochondria that
helps generate the 'energetic currency of life' ATP. The structure of the oxidase guides flows of current and mass that interact according to the energetic and dissipative functionals of the generalized theory. Kirchhoff's current law provides important coupling
when the enzyme is in its natural setting in the mitochondrial membrane. 
The natural function of the enzyme is the result.  Electron flows are converted into proton flows and gradients. 
	\end{abstract}
	\section{Introduction}

	History seems to have separated much of chemistry \cite{brush1976kind,garber1995maxwell} from the classical theory of fields \cite{simpson1998maxwell,zangwill2013modern,chree1908mathematical}.  Chemical reactions are found throughout the ionic solutions of biology and chemistry but they are usually described in a language apparently disjoint from that of classical field theory even though the reactants and products of chemical reactions are almost always charged and carry significant electrical current. The reactants, catalysts and enzymes of chemistry and biology depend on charge interactions for much of their function. 
	
	Field theory has much to offer chemistry, particularly in the study of charged systems
	as admirably reviewed in \cite{RN46052}.
	The Maxwell equations are as universal and precise as any theory and apply to chemical reactions in the plasmas of gases and ionic solutions, and liquids in general. Indeed, the Maxwell equations can be written without any adjustable parameters, implying the universal conservation of total current \cite{eisenberg2019kirchhoff,eisenberg2020electrodynamics}. A theory of the electromechanical response of charge to the electric
	field is needed in that case to make a complete description of the charged system, i.e. an electromechanical theory of polarization
	phenomena \cite{wang2021variational}.
	But some properties of the electromagnetic field (e.g., conservation
	of total current) are true entirely independent of the electromechanical response. The special systems that are almost completely described by conservation of current (without specification of charges) include the electronic circuits of our computers and many properties of the action potential of nerve and muscle fibers. Both systems are almost entirely specified by Kirchhoff's current law. The question then arises how do we fit chemical reactions into the framework of Kirchhoff's law and conservation of total current that is so remarkably simpler to implement than a full accounting of all the charges in a chemical system?

	Here we show one way to describe chemical reactions in ionic solutions with an extension of classical field theory that does not violate the traditions of either chemistry or electrodynamic field theory. In this work, we  take advantage of the  energy variation method \cite{xu2018osmosis,shen2020energy,shen2022energy}  that treats ionic solutions as complex fluids, with interactions, internal stored energy, flow, and dissipation like other complex fluids
	\cite{eisenberg2010energy,chen2020differential,liu2020molecular,li2020generalized,giga2017variational}. It begins with defining two functionals for the total energy and dissipation of the system,  and introducing the kinematic equations based on physical laws of conservation. The specific forms of the flux and stress functions in the kinematic equations are   obtained by taking the time derivative of the total energy functional and comparing with the  defined dissipation functional. More details of this method can be found in \cite{shen2020energy}.
	We use energy variation methods to link the electric field and reaction dynamics as Wang, et al, \cite{wang2020field,RN46052} have for reactions that do not involve charge or electrodynamics. 
	
	The generalization to charged systems allows us to study systems of some importance. We study active transporters of biological membranes. In a sense, we extend the electrical treatment of the membrane proteins called channels to deal with transporters. Conservation of current (in the form of Kirchhoff's law) provides the crucial coupling between the properties of disjoint sodium and potassium channel proteins that act independently in the atomic and molecular sense because they are well screened. Channels are many Debye lengths apart, shielded from each other by the ions, water dipoles (and quadrupoles), that also form the ionic atmosphere of proteins and lipid bilayer. The atomic scale function of these proteins is crucial to their biological function. Their coupling is just as important to their function, but the coupling of these channel proteins is not chemical. The coupling is provided by the cable equation \cite{eisenberg1975electrophysiology}
	as biophysicists have called the telegrapher equation version of the Maxwell equations (and Kirchhoff's law) used by Kelvin to design the Atlantic cable
	\cite{thompson1855theory}
	well before Maxwell wrote his equations.
	
	Active transport is one of the most important processes in life. It maintains the concentration gradients and membrane potentials that allow biological cells to function. Indeed, without active transport animal cells swell, burst, and die. Active transport powers the generation of ATP in both animals (in mitochondria, by oxidative phosphorylation) and plants (in chloroplasts, by photosynthesis). ATP is the currency of chemical energy in all plants and animals, storing in a small organic compound the energy from photosynthesis or oxidation. When hydrolyzed to ADP, the chemical energy is available for the myriad of dissipative processes essential to life. Nearly all of them use ATP as their energy source. Life is complex with many facets. The energy source of life is not.
	
	The coupling of flows in transporters in the inner membrane of mitochondria allows one substance to move uphill (against its gradient of electrochemical potential) using the energy from the downhill movement of another substance. Coupling is inherently about the relationship of flows, yet most analysis and simulations of active transporters do not explicitly include a variable for flow. Most do not use the electrodynamic equations for flow, whether conservation of total current or Kirchhoff’s law or a nonlinear version of Ohm’s law. Indeed, most analysis and simulations use methods that are derived assuming zero flow and do not include changes in potential or free energy associated with flow. The problems with this approach become apparent if one tries to analyze electron flow through a resistor or semiconductor diode or rectifier in that tradition. 
	
	We analyze an active transporter by studying the currents through it, as a specific example of our general field theory  for ionic flows with chemical reactions. We combine Kirchhoff’s law and chemical reaction energetics with diffusion, migration, and convection of ions and water to make an 'electro-osmotic' model. The name is chosen to emphasize the important role of electricity in this system, implying the need to deal with electrical flows by methods used to deal with electrical flows in other systems, like electrical and electronic circuits. This approach seems sensible for the systems like oxidative phosphorylation and photosynthesis where electron flows are involved. The analysis of electron flows has been well established in physics and engineering for more than a century. The analysis of ionic flows has been well established in membrane biophysics for some seventy years. We combine them here with chemical reactions hoping to construct a useful electro-osmotic theory of cytochrome $c$ oxidase. 
	
	We do not have to deal with the myriad of charges involved in the transport of current or the incalculable number of interactions of those charges. Analysis of current flow is all we use in this conservative coupling approach, following the practice of circuit analysis. We do not have to assume equilibrium or zero flow. We do not have to deal explicitly with the charges involved in the transport. Analysis of current flow is all we use. In particular, we do not assume equilibrium or near equilibrium flows, as has been done in previous analysis. Note that near equilibrium analysis (using the  Green-Kubo formalism, for example) is inappropriate for devices that function with large flows. These devices are not near equilibrium. The electronic devices of our digital technology function far from equilibrium and they are not analyzed by assuming nearly zero flow. Indeed, they usually use power supplies to maintain spatially nonuniform potentials, often described by inhomogeneous Dirichlet boundary conditions.  Traditionally, electronic devices are analyzed by studying small changes around a nonequilibrium operating point, which we hasten to add is maintained by large---not small--- flows from the power supplies. But even that linearization is not necessary nowadays because the full flow nonequilibrium problems can be solved conveniently with readily available software.
	
	We are certainly not the first to exploit the simplification provided by the analysis of current instead of charge. Circuit designers have used this approach 'forever' \cite{bush1929operational,RN45733}. 
	Charges are hardly mentioned when circuits are designed as a glance at textbooks shows \cite{RN45934,RN25358,RN21598,RN45662,RN28108,RN26593,RN45935}, perhaps most eloquently in symbolic circuit design \cite{RN45999}. Analysis of current---not charge---characterizes the study of ion channels since they were discovered as conductances by Hodgkin and Huxley some seventy years ago \cite{RN267,RN309,RN28957,RN12551,RN28654,RN291,}.
	Analysis of current is not a prominent feature of the study of active transporters, however, even in the work of Hodgkin and collaborators that started the physiological analysis of active transport in cell membranes (not whole epithelia or biochemical preparations), at much the same time as their work defining channels \cite{RN45995,RN9363,RN10470}.

	Most models of active transport include conformational changes of the protein that require a model to compute the spatial distribution of mass in the protein, as it changes during active transport \cite{RN29619,RN29611,}.
	The conformational changes usually provide \textbf{alternating} access to an occluded state that is not connected (or accessible) to either side of the protein. The occluded state blocks the conduction path (and incidentally often traps ions in the 'middle' of the transporter) and thus prevents backflux. Alternating access mechanisms create flux across the transporter protein without allowing backflux that would seriously degrade the efficiency of the transporter. Transporters were first thought to use quite different mechanisms from channels \cite{RN434,RN155}. 
	However, recent work \cite{RN23} shows otherwise.  Alternating access is apparently created by correlated motions of gates that account for activation and inactivation in classical voltage activated sodium channels \cite{RN29619,RN29607}. The physical basis of the gates is not discussed in the classical literature. Later in the paper, we speculate that the switches that provide alternating access might be like the switches of bipolar transistors. . 
	
	In this work, we use an electro-osmotic approach to describe cytochrome $c$ oxidase (or Complex IV). We choose cytochrome $c$ oxidase because experimental work and simulations of the highest quality have shown that “cytochrome $c$ oxidase is a remarkable energy transducer [i.e., coupled transporter of electrons and protons] that seems to work almost purely by Coulombic principles without the need for significant protein conformational changes” \cite{RN45750}. Cytochrome $c$ oxidase depends on an "occluded state" containing the reaction center(s) to prevent backflow as do other active transporters, but it uses some type of the 'water-gate' mechanism proposed in \cite{RN4995,RN45750}. The alternating access in cytochrome $c$ oxidase occurs without conformation change (of the spatial distribution of mass), in marked contrast to the usual alternating access models of transporters. Perhaps the gate in the oxidase is like the switch in a semiconductor (diode) rectifier. The switches might be rectifiers produced by spatial distributions of permanent charge, of opposite signs, as rectification (and switching) is produced in PN diodes, and bipolar transistors. Diode rectifiers depend on changes in shape (i.e., conformation) of the electric field, not changes in the distribution (i.e., conformation) of mass. This idea is outlined in the Discussion Section following \cite{RN24,RN15949,RN6246}.

	The rest of the paper is organized as follows. In section 2, we derive the general three dimensional field equations for an ionic system with reaction and use it to create a general framework of an electro-osmotic model. In section 3, we propose a specific, simplified  model for cytochrome $c$ oxidase.  In section 4, we carry out computational studies of our cytochrome $c$ oxidase model and explore the effects of various conditions on the transport of protons across the mitochondria membrane. In section 5, we conclude our paper with a discussion of our cytochrome $c$ oxidase model and future directions.

	\section{Derivation of Electro-osmotic Model }
	
	We mainly focus on a mathematical model of  elementary reactions
	\begin{equation}\label{reacton_de}
		\alpha_1 C_1^{z_1}+\alpha_2 C_2^{z_2}+\alpha_3 C_3^{z_3}\overset{k_f}{\underset{k_r}{\rightleftharpoons}} \alpha_4 C_4^{z_4},
	\end{equation}
	where $k_f$ and $k_r$ are two constants for forward and reverse directions,   $[C_i]$  is  the concentration  of $i^{th}$ species, respectively. Here  $\alpha_i$  is the stoichiometric coefficient, $z_i$ is the valence of $i^{th}$ species and together they satisfy 
	\begin{equation}\label{zrelation}
		\sum_{i=1}^3 \alpha_iz_i = \alpha_4z_4.
	\end{equation} 
	In particular, we have in mind a case where an active transporter ('pump') uses the energy supplied by a chemical reaction to pump molecules. Later, we will focus on the reaction for cytochrome $c$ oxidase, i.e., for Complex IV of the respiratory chain
	
	\begin{equation}
		2 H^{+}+\frac 1 2 O_2+2e^{-}\overset{k_f}{\underset{k_r}{\rightleftharpoons}}  H_2O. 
	\end{equation}
	
	According to the conservation laws, we have the following conservation of chemical elements (like sodium, potassium and chloride). Note that this conservation is in addition to the conservation of mass, because nuclear reactions that change one element into another are prohibited in our treatment, as in laboratories and most of life. 
	\begin{subequations}\label{conservation of element}
		\begin{align}
			&\frac{d}{dt} (\alpha_4 [C_1] +\alpha_1 [C_4])=0,\\
			&\frac{d}{dt} (\alpha_4 [C_2] +\alpha_2 [C_4])=0,\\
			&\frac{d}{dt} (\alpha_4 [C_3] +\alpha_3 [C_4])=0.
		\end{align}
	\end{subequations}

	In order to derive a thermal dynamical consistent model, the Energy Variation Method \cite{shen2020energy} is used.
	Based on the laws of conservation of elements and Maxwell equations, we have the following kinematic  system  
	\begin{equation}\label{assumption_de}
		\left\{
		\begin{array}{l}
			\frac{d [C_1]}{d t} =-\nabla\cdot \boldsymbol{j}_1 -\nabla\cdot \boldsymbol{j}_p-\alpha_1 \mathcal{R},\\
			\frac{d [C_2]}{d t} = -\nabla\cdot \boldsymbol{j}_2 -\alpha_2\mathcal{R},\\
			\frac{d [C_3]}{d t} =-\nabla\cdot \boldsymbol{j}_3 - \alpha_3 \mathcal{R},\\
			\frac{d [C_4]}{d t} =-\nabla\cdot \boldsymbol{j}_4 + \alpha_4 \mathcal{R},\\
			\nabla\cdot(\bD) = \sum_{i=1}^4z_i[C_i]F,\\
			\nabla\times \bE = \boldsymbol{0},
		\end{array}
		\right.
	\end{equation}
	where $\boldsymbol{j}_l, l=1,2,3,4$ are the passive fluxes and $\boldsymbol{j}_p$ is the pump flux, $\mathcal{R}$ is reaction rate function. All these variables are unknown and will be derived by using the Energy Variational method.  
	
	$\bj_{ex}$ is the flux of electrons supplied from an external source. In mitochondiral membranes this will include special pathways linking one enzyme and one complex to another by the movement of lipid soluble or water soluble electron donors and water soluble electron donors or acceptors like the quinones. 
	
	$\bD$ is Maxwell's electrical displacement field and $\bD = \varepsilon_0\varepsilon_r \bE$ with electric field $\bE$, dielectric constant $\varepsilon_0$ and relative dielectric constant $\varepsilon_r$. 
	The equation $ \nabla\times \bE = \boldsymbol{0}$
	implies that there exists a $\phi$ such that $\bE = -\nabla \phi$.  
	We consider a system with structure and boundary conditions defined on that structure. 
	
	The structures are given to us by structural biologists. The structures are decorated with molecules (proteins and lipids for the most part) that use particular atomic arrangements to channel physical forces into physiological function.  We describe a constant flux for one species, say $C_3$, that serve as the input of the system. In cytochrome $c$ oxidase the input is electrons carried on the heme groups of cytochrome oxidase. 
	
	\begin{equation}\label{bd_flux}
		\left\{
		\begin{array}{ll}
			\boldsymbol{j}_i\cdot\boldsymbol{n} = j_{i,extra}, i=1\cdots, 4, & \mbox{on~} \partial\Omega,\\
			\boldsymbol{D}\cdot\boldsymbol{n} = 0, & \mbox{on~} \partial\Omega.
		\end{array}\right.
	\end{equation}
	\begin{rmk}
		By multiplying $z_i e$ on both sides of the first three equations and $-e$ on both sides of the fourth equation, we have 
		\begin{eqnarray}
			\frac{d}{dt}(\nabla \cdot\bD) &=&\sum_{i=1}^4z_iF\frac{d[C_i]}{dt} \\
			&=& -\sum_{i=1}^4\nabla\cdot(z_iF\bj_i)-z_iF\nabla\cdot\bj_p- (z_1\alpha_1+z_2\alpha_2+z_3\alpha_3-z_4\alpha_4)F\mathcal{R}\\
			&=& -\sum_{i=1}^4 \nabla\cdot(z_iF\bj_i)-z_iF\nabla\cdot\bj_p, 
		\end{eqnarray}
		which is consistent with the electrostatic Maxwell equations. Treatment of transient problems, involving displacement currents is needed to deal with some important experimental work \cite{RN30714,RN30642,RN6359,RN30008,RN30276,RN45724}. 
	\end{rmk}
	
	The total energetic functional  is defined as the summation of entropies of mixing, \cite{RN46053}
	internal energy and electrical static energy.  
	\begin{eqnarray}\label{totalenergy}
		E_{tot} &=& E_{ent}+E_{int}+E_{ele}\nonumber\\
		&= &\sum_{i=1}^4 \int_{\Omega}RT\left\{ [C_i]\left(\ln{\left(\frac{[C_i]}{c_0}\right)}-1 \right)\right\} dx+\int_{\Omega} \sum_{i=1}^4[C_i]U_idx 
		+
		\int_{\Omega} \frac{\boldsymbol{D}\cdot\boldsymbol{E}}{2}dx.
	\end{eqnarray}
	
	Then the chemical potentials can be calculated from the variation of total energy 
	\begin{equation}
		\tilde\mu_l=\frac{\delta E_{tot}}{\delta [C_i]} = RT\ln\frac{[C_i]}{c_0} +U_i  +z_l\phi e, l=1,\cdots, 4.
	\end{equation}

	It is assumed in the present work that  dissipation of the system energy is due to passive diffusion, chemical reaction and the  pump.    Accordingly, the total dissipation functional $\Delta$ is defined as follows 
	\begin{eqnarray}
		\Delta = \int_{\Omega}\left\{\sum_{j=1}^4|\bj_i|^2+  RT\mathcal{R}\ln \left(\frac{\mathcal{R}}{k_{r}\left(\frac{[C_4]}{c_0}\right)^{\alpha_4}}+1\right)\right\} dx  -\int_{\Omega} f_p  dx, 
	\end{eqnarray}
	where $f_p = f_p(\mathcal{R}, \mu,x)\ge 0 $ is the term from the pump.

	Open systems  in which some fluxes flow in or out, entering or leaving the system altogether, have distinctive energy dissipation laws that differ from those of closed systems. The natural mitochondrion is an UNclamped system, in which the electrical potential assumes whatever value satisfies the field equations. The sum of all currents across the membrane of the natural mitochondrion is zero (including the capacitive displacement current) as it is in small biological cells. Many experiments are done in voltage clamped systems. In these the sum of the currents does not equal zero just as the sum of currents in the classical Hodgkin Huxley experiments was not zero. Of course, the ratio of fluxes will be different in the clamped and unclamped cases, as we document at length later in this paper.
	
	In the natural Unclamped mitochondrion,  we have the following generalized energy dissipation law  
	\begin{equation}
		\frac{dE_{tot}}{dt}  = J_{E,\partial\Omega}-\Delta.
	\end{equation}
	Here $J_{E,\partial\Omega}$ is the rate of boundary energy absorption or release that measures the energy of flows that enter or leave the system altogether through the boundary. 
	Recall that  the chemical potential of a species is the energy that can be absorbed or released due to a change of the number of particles of the given species and $J_i\cdot n $ is the total number of $i^{th}$ particles passing through the boundary, per area per unit time. We define $J_{E,\partial\Omega}$ as follows
	\begin{equation}\label{J_Edefinition}
		J_{E,\partial\Omega}=\int_{\partial\Omega} \sum_{i=1}^4 \tilde{\mu}_i \bj_i\cdot\boldsymbol{n} dS.
	\end{equation}
	
	In general, different types of boundary conditions can be written in the following general format
	\begin{equation}
		\bj_i\cdot\boldsymbol{n} = g_i(f([C_i])-f([C_i]_{extra})),  
	\end{equation}  where $g_i$ is the conductance of $i^{th}$ species on the boundary, $[C_i]_{extra}$ is the fix reservoir's concentration of $i^{th}$ species and $f$ is some specific function.   Then the rate of boundary energy absorption or release is 
	\begin{equation}
		J_{E,\partial\Omega}= \int_{\partial\Omega} \sum_{i=1}^4g_i\tilde{\mu}_i(f([C_i])-f([C_i]_{extra})).
	\end{equation}

	In this case energy can change both because of the flux across the boundary and also because of the change in dissipation.
	and 
	\begin{equation}
		\frac{dE}{dt}-J_{E,\partial\Omega} = -\Delta.    
	\end{equation} 
	
	\begin{rmk} Boundary Conditions, Structure, Evolution, and Engineers
		
		These boundary conditions serve as the link
		between general field equations and structures
		that serve as devices. Structures are chosen
		and devices designed (by evolution or engineers) so these boundary conditions
		are satisfied. The boundary conditions are chosen so devices have almost the same properties no matter where they are placed in a network. The structures and boundary conditions on those structures are not automatic properties of nature. The structures are decorated with (i.e., include) specific substructures (like power transistors) that exploit arrangements of atoms (like doping charges) to create properties that are useful. The properties are summarized by boundary conditions located on the structures provided by evolution and engineers. These boundary conditions help make the idea of a component useful. They help ensure that
		a component in one part of a system does what it
		does in another part of the system and so can be
		described by a 'transfer function' independent of 
		its location in the system. 
		
		It is clear that channels
		and transporters in biological systems behave as components. 
		Indeed, most of classical physiology and biophysics is devoted
		to identifying such components, on a wide variety of length scales from 
		atoms to organisms, and studying how they interact in the hierarchy
		of structures that make animals and plants \cite{RN7109,RN45615,RN28910,RN26033,RN45909,RN45950,RN134,RN295}.
	\end{rmk}
	
	\begin{rmk} 
		\begin{enumerate}
			\item A closed system allows no flux across the boundaries. It has the following no-flux boundary conditions
			\begin{equation}\label{bd_noflux}
				\left\{
				\begin{array}{ll}
					\boldsymbol{j}_i\cdot\boldsymbol{n} = 0~ i =1,2,3,4, & \mbox{on~} \partial\Omega,\\
					\boldsymbol{D}\cdot\boldsymbol{n} = 0, & \mbox{on~} \partial\Omega.
				\end{array}\right.
			\end{equation} 
			In a closed system,   $J_{E,\partial\Omega} = 0$ and the energy dissipation law is 
			$$\frac{dE_{tot}}{dt} = -\Delta.$$
			In a closed system, the energy changes into dissipation. That is the only way energy can change in a closed system.
			
			\item An open system has flow across the boundaries. An open system might have constant inflow/outflow 
			$\bj_i\cdot\boldsymbol{n} = J_{i,extra} $. 
			In that case,
			$g_i = \frac{J_{i,extra}}{f([C_i])-f([C_i]_{extra})}$, and
			$$J_{E,\partial\Omega} =\int_{\partial\Omega}\sum_{i}^4 \tilde{\mu}_i J_{i,extra} dS .$$
			\item For the  Dirichlet boundary condition $[C_i]=[C_i]_{extra}$ on $\partial\Omega$, the flux $\bj_i\cdots\boldsymbol{n}$ is unknown and part of the solution. In this case, $J_{E,\partial\Omega}$ is an unknown flux needed to ensure that the Dirichlet condition $[C_i]=[C_i]_{extra}$ is obeyed on $\partial\Omega$. 
			
			It is very important to understand this requirement. In reality, i.e., in experiments and their models, supplying the unknown flux requires specialized instrumentation, for example, a patch clamp amplifier in a voltage clamp setup. Almost always, that flux is supplied at one location in space. In that way a classical voltage clamp can be established. However, if one wishes to "clamp" a field, one must control the potential at many locations. Each location requires a different flux and thus a different amplifier and different electrodes to supply that flux. Without such complicated apparatus, it is almost impossible to maintain a constant field in space  \cite{han1993superconducting}.
			Indeed, it is nearly impossible to maintain any pre-specified field because it is practically impossible to apply different fluxes at different locations. If one assumes a constant field in a theory, without such apparatus in an experiment, one is in effect introducing flux into the calculation and model that is not present in the experimental setup. One is introducing an artifactual flux likely to produce artifactual conclusions that are not relevant to the original experiment. \cite{eisenberg2010computing,eisenberg1998ionic}.
		\end{enumerate}
	\end{rmk}.

	By taking the time derivative of total energy function \eqref{totalenergy}, we have 
	\begin{eqnarray}
		\frac{dE_{tot}}{dt} &=&\int_{\Omega}\sum_{i=1}^4\left\{\mu_i \frac {d[C_i]}{dt}\right\}dx +\int_{\Omega} \bE\cdot\frac{d\bD}{dt}dx\nonumber\\
		&=&\int_{\Omega}\sum_{i=1}^4\left\{\mu_i \frac {d[C_i]}{dt}\right\}dx -\int_{\Omega}\nabla \phi\cdot\frac{d\bD}{dt}dx\nonumber\\
		&=&  \int_{\Omega}\sum_{i=1}^4\left\{\mu_i \frac {d[C_i]}{dt}\right\}dx+\int_{\Omega} \phi\nabla\cdot\left(\frac{d\bD}{dt}\right)dx\nonumber\\
		&=&  \int_{\Omega}\sum_{i=1}^4\left\{\mu_i \frac {d[C_i]}{dt}\right\}dx +\int_{\Omega} \phi F\sum_{i=1}^4\left\{z_i \frac {d[C_i]}{dt}\right\} dx\nonumber\\
		&=&  \int_{\Omega} \sum_{i=1}^4\left\{\tilde{\mu}_i \frac {d[C_i]}{dt}\right\}dx\nonumber\\
		&=&-\int_{\Omega}\sum_{i=1}^4\left\{\tilde{\mu}_i\nabla\cdot \boldsymbol{j}_i\right\}dx -\int_{\Omega}\tilde{\mu}_1 \nabla\cdot \bj_p dx
		- \int_{\Omega}\mathcal{R}(\sum_{i=1}^3\alpha_i \tilde{\mu}_i-\alpha_4\tilde{\mu}_4 )dx\nonumber \nonumber\\
		&=&\int_{\Omega}\sum_{i=1}^4\left\{\nabla\tilde{\mu}_i\cdot \boldsymbol{j}_i\right\}dx+\int_{\Omega}\nabla\tilde{\mu}_1 \cdot \bj_p dx -\int_{\Omega}\mathcal{R}(\sum_{i=1}^3\alpha_i \mu_i  -\alpha_4 \mu_4)dx+\int_{\partial\Omega} \sum_{i=1}^4 \mu_i \bj_i\cdot\boldsymbol{n} dS,
	\end{eqnarray}
	where    $\mu_i = RT\ln\frac{[C_i]}{c_0} +U_i$  and  Eq. \eqref{zrelation} is used.

	By comparing with the dissipation functional, we have 
	\begin{subequations} 
		\begin{align}
			&\bj_i = -\frac{D_i}{RT}[C_i]\nabla\tilde\mu_i,~ i=1,2,3, \label{fluxi} \\
			&RT\ln\left(\frac{\mathcal{R}}{k_{r}\left(\frac{[C_4]}{c_0}\right)^{\alpha_4}}+1\right) =   \sum_{i=1}^3\alpha_i \mu _i  -\alpha_4 \mu_4\label{reactionrateele}.
		\end{align}
	\end{subequations}
	And the corresponding energy influx rate is 
	\begin{equation}
		J_E = \sum_{i=1}^4\int_{\partial\Omega} \tilde{\mu}_i j_{i,extra}. 
	\end{equation}
	For the pump flux, if we assume flux is only along the $z$ direction, then,
	\begin{equation}
		\bj_p = (0,0,\frac{f_p}{\partial_z\mu_1}). \label{pumpmu1}
	\end{equation}

	At equilibrium,  we have 
	\begin{equation}
		\left\{\begin{array}{l}
			\bj_i = \nabla [C_i]_{eq}+\frac{z_iF}{RT}[C_i]_{eq}\nabla\phi_{eq} =0,\nonumber\\
			\sum_{i=1}^3\alpha_i \mu_i([C_i]_{eq} ) -\alpha_4\ \mu_4([C_4]_{eq} ) = 0,
		\end{array}
		\right.
	\end{equation}
	The last equation means 
	\begin{eqnarray}
		0 =  RT \ln \left(\frac{\Pi^3_{i=1} (\frac{[C_i]_{eq}}{c_0})^{\alpha_i}}{(\frac{[C_4]_{eq}}{c_0})^{\alpha_4}}\right)   +(\sum_{i=1}^3\alpha_i U_i  -\alpha_4 U_4) ,\label{equilibriumcondition}
	\end{eqnarray}
	
	According to the definition of equilibrium constant $k_{eq}$, 
	\begin{equation}
		k_{eq} = \frac{\Pi^3_{i=1} (\frac{[C_i]_{eq}}{c_0})^{\alpha_i}}{(\frac{[C_4]_{eq}}{c_0})^{\alpha_4}}
	\end{equation}
	
	Eq. \eqref{equilibriumcondition} yields
	
	\begin{equation}\label{keqele}
		\ln{k_{eq}} = e^{-\frac{\Delta U}{RT}},
	\end{equation}
	with $\Delta U = \sum_{i=1}^3\alpha_i U_i  -\alpha_4 U_4$ .

	Then combining Eqs. \eqref{reactionrateele} and \eqref{keqele} yields
	\begin{eqnarray}
		\ln\left(\frac{\mathcal{R}}{k_{r}\left(\frac{[C_4]}{c_0}\right)^{\alpha_4}}+1\right) &=& \ln \left(\frac{\Pi_{i=1} \left(\frac{[C_i]}{c_0}\right)^{\alpha_i}}{\left(\frac{[C_4]}{c_0}\right)^{\alpha_4}}\right), 
	\end{eqnarray}
	which implies 
	\begin{equation}
		\mathcal{R} = k_f\left(\frac{[C_1]}{c_0}\right)^{\alpha_1}\left(\frac{[C_2]}{c_0}\right)^{\alpha_2}\left(\frac{[C_3]}{c_0}\right)^{\alpha_3}-k_{r}\left(\frac{[C_4]}{c_0}\right)^{\alpha_4},\nonumber
	\end{equation}
	where $k_f = \frac{k_r}{k_{eq}}$ \cite{wang2020field}.
	
	\begin{rmk}
		Here $k_{eq}$ is dimensionless. $k_r$ and $k_f$ are with unit $s^{-1}$ \cite{ozcan2022equilibrium}. 
	\end{rmk}
	

	Then the whole system is as follows 
	
	\begin{equation}\label{model_ele}
		\left\{
		\begin{array}{l}
			\frac{d [C_1]}{d t} =\nabla\cdot (D_1 \nabla [C_1]+D_1\frac{z_1F}{RT}[C_1]\nabla\phi) -\partial_z j_p-\alpha_1 \mathcal{R},\\
			\frac{d [C_2]}{d t} =\nabla\cdot (D_2 \nabla [C_2]+D_2\frac{z_2F}{RT}[C_2]\nabla\phi)-\alpha_2 \mathcal{R},\\
			\frac{d [C_3]}{d t} =\nabla\cdot (D_3 \nabla [C_3]+D_3\frac{z_3F}{RT}[C_3]\nabla\phi)-\alpha_3\mathcal{R},\\
			\frac{d [C_4]}{d t} =\nabla\cdot (D_4 \nabla [C_4]+D_4\frac{z_4F}{RT}[C_4]\nabla\phi)+\alpha_4 \mathcal{R},\\
			-\nabla\cdot(\varepsilon_0\varepsilon_r \nabla \phi) = \sum_{i=1}^4z_iF[C_i],
		\end{array}
		\right.
	\end{equation}
	with
	\begin{eqnarray}
		\mathcal{R} = k_f\left(\frac{[C_1]}{c_0}\right)^{\alpha_1}\left(\frac{[C_2]}{c_0}\right)^{\alpha_2}\left(\frac{[C_3]}{c_0}\right)^{\alpha_3}-k_{r}\left(\frac{[C_4]}{c_0}\right)^{\alpha_4}.
	\end{eqnarray}
	
	
	
	
	and boundary conditions 
	\begin{equation}\label{bd_flux}
		\left\{
		\begin{array}{ll}
			\boldsymbol{j}_i\cdot\boldsymbol{n} = j_{extra}, i=1\cdots 4, & \mbox{on~} \partial\Omega,\\
			\boldsymbol{D}\cdot\boldsymbol{n} = 0, & \mbox{on~} \partial\Omega.
		\end{array}\right.
	\end{equation}

	\begin{rmk}
		If we assume that one of the reactants is an electron,  for instance  $C_3$,  
		and  supplied by an thin  electrode along z direction, the density of electron    $[C_3]=\rho_e = \rho(z)\delta(x_0, y_0)$.  Then the model is changed to
		\begin{equation}\label{model_eletron}
			\left\{
			\begin{array}{l}
				\frac{d [C_1]}{d t} =\nabla\cdot (D_1 \nabla [C_1]+D_1\frac{z_1e}{RT}[C_1]\nabla\phi) -\partial_{z}(j_p) -\alpha_1 \mathcal{R}\delta(x_0, y_0),\\
				\frac{d [C_2]}{d t} =\nabla\cdot (D_2 \nabla [C_2]+D_2\frac{z_2e}{RT}[C_2]\nabla\phi)-\alpha_2 \mathcal{R}\delta(x_0, y_0),\\
				\frac{d [C_4]}{d t} =\nabla\cdot (D_4 \nabla [C_4]+D_4\frac{z_4e}{RT}[C_4]\nabla\phi)+\alpha_4 \mathcal{R}\delta(x_0, y_0),\\
				\frac{d\rho(z)}{dt} = -\partial_z \bj_e -\alpha_3\mathcal{R}\delta(x_0, y_0),\\
				-\nabla\cdot(\varepsilon_0\varepsilon_r \nabla \phi) = \sum_{i=1,2,4}z_ie[C_i] -F\rho(z)\delta(x_0, y_0).
			\end{array}
			\right.
		\end{equation}

	\end{rmk}

	\begin{rmk} When the reaction and ions are in an electrolyte,  the fluid effect needs to  be taken into consideration. In this case, the energy functional is changed to be
		\begin{eqnarray}
			E_{tot} &=&E_{kin}+ E_{ent}+E_{int}+E_{ele}\nonumber\\
			&= &\int_{\Omega}\frac{\rho |\mathbf{u}|^2}2 dx +\sum_{i=1}^4 \int_{\Omega}RT\left\{ [C_i]\left(\ln{\left(\frac{[C_i]}{c_0}\right)}-1 \right)\right\} dx+\int_{\Omega} \sum_{i=1}^4[C_i]U_idx    
		\end{eqnarray}
		and the dissipation functional is changed to 
		\begin{eqnarray}
			\Delta = \int_{\Omega}2\eta|\boldsymbol{D}_{\eta}|^2 dx + \int_{\Omega}\left\{\sum_{j=1}^4|\bj_i|^2  +RT\mathcal{R}\ln \left(\frac{\mathcal{R}}{k_{r}\left(\frac{[C_4]}{c_0}\right)^{\alpha_4}}+1\right)\right\} dx  -\int_{\Omega} f_p  dx,
		\end{eqnarray}
		where $\boldsymbol{D}_{\eta} = \frac{\nabla\boldsymbol{u} + (\nabla\boldsymbol{u})^T}{2}$ and  $\boldsymbol{u}$ is the velocity. 
		We can use the Energy variation method to get the diffusion-reaction-convection model as follows 
		\begin{equation}\label{model_ele_full}
			\left\{
			\begin{array}{l}
				\frac{\partial [C_1]}{\partial t} +\nabla\cdot([C_1]\boldsymbol{u}) =\nabla\cdot (D_1 \nabla [C_1]+D_1\frac{z_1F}{RT}[C_1]\nabla\phi) -\partial_z j_p-\alpha_1 \mathcal{R},\\
				\frac{\partial [C_2]}{\partial t} +\nabla\cdot([C_2]\boldsymbol{u}) =\nabla\cdot (D_2 \nabla [C_2]+D_2\frac{z_2F}{RT}[C_2]\nabla\phi)-\alpha_2 \mathcal{R},\\
				\frac{\partial [C_3]}{\partial t} +\nabla\cdot([C_3]\boldsymbol{u}) =\nabla\cdot (D_3 \nabla [C_3]+D_3\frac{z_3F}{RT}[C_3]\nabla\phi)-\alpha_3\mathcal{R},\\
				\frac{\partial [C_4]}{\partial t} +\nabla\cdot([C_4]\boldsymbol{u})=\nabla\cdot (D_4 \nabla [C_4]+D_4\frac{z_4F}{RT}[C_4]\nabla\phi)+\alpha_4 \mathcal{R},\\
				-\nabla\cdot(\varepsilon_0\varepsilon_r \nabla \phi) = \sum_{i=1}^4z_iF[C_i], \\
				\rho (\frac{\partial\boldsymbol{u}  }{\partial t} +(\boldsymbol{u}\nabla)\cdot \boldsymbol{u} +\nabla p= \nabla(\eta (\nabla\boldsymbol{u}+ (\nabla\boldsymbol{u})^T)) -(\sum_{i=1}^4 z_iF[C_i])\nabla\phi\\
				\nabla\cdot \boldsymbol{u} = 0. 
			\end{array}
			\right.
		\end{equation}
	\end{rmk}
	
	Note we are not here considering transient problems in which charge is stored in polarization fields. These will be studied separately so we can deal with the important experiments reported in \cite{RN30714,RN30642,RN6359,RN30008,RN30276,RN45724}.  Transient problems are obviously important if reactions are studied on the atomic scale of distance and time (angstroms and femtoseconds) because the polarization currents are large. Dealing with those currents requires use of a universal form of the Maxwell equations combined with an appropriate model of the stress strain relation of charge in a viscoelastic structure, commonly called polarization. Speaking loosely, the transient problems can be dealt with in circuits by a generalization of Kirchhoff’s law
	\cite{RN45998,eisenberg2019kirchhoff} to describe the actual transient currents that flow through an ideal resistor \cite{eisenberg2018current}. 
	
	It is important to realize that currents (and fluxes) cannot be computed by methods that assume the
	currents and fluxes are zero. Electrostatics cannot compute currents because currents and fluxes involve time and electrostatics does not \cite{sugitani2008theoretical}. Electrostatics does not include Maxwell's Ampere law that is the universal coupler of current to  electric and magentic fields. In the context of cytochrome $c$ oxidase these issues come to the fore. Models without electron or proton current as variables do not describe the 'transfer function' of the transporter being studied. Models cannot calculate Ohm's law (for system with large and small currents and electrical potentials) if the models assume currents are zero.
	
	In fact, using  a formulation of electrodynamics that explicitly
	involves current is straightforward, as engineers have known for a very long time, going back to Heaviside \cite{belevitch1962summary,darlington1984history} and are worked out in practical detail in \cite{RN45998}.
	Kirchhoff's current law allows analysis of systems of great importance, without dealing with charges explicitly. That is why analysis of electronic circuits does not need to use distributions of charges but rather uses Kirchhoff's current law or its generalization, conservation of total current. 
	Kirchhoff's current law is an exact corollary of the Maxwell equations themselves, if current includes the displacement current 
	\cite{RN45998,eisenberg2018current,eisenberg2019kirchhoff}.
	It might seem that another corollary of the Maxwell equations, the continuity equation, can be used instead of Kirchhoff's law for total current. And it is certainly true that the continuity equation of electrodynamics contains
	the same information as conservation of (total) current, all conjoined with
	the Maxwell equations. But that information is not useful when enormous numbers of charges are involved as in cytochrome $c$ oxidase or in other macroscopic scale systems like the electronic circuits of our computers. The information implicit in the flux of charges is only usable when written as total current that is conserved perfectly whenever the Maxwell equations are valid. This formulation using conservation of total current
	does not require explicit treatment of charges. The continuity equation does require the explicit treatment of charges, and their significant interactions, whether involving two charge interactions, three charge, .... or the interactions of an entire cluster expansion. The significant interactions of charge are difficult to understand or even enumerate and more difficult to compute \cite{RN46056,RN46055}. Kirchhoff's current law is easy to understand and trivial to compute.

	\section{An Electro-osmotic Model of cytochrome $c$ oxidase}
\begin{figure}[!ht]
			\centering
			\begin{subfigure}[] 
				{\includegraphics[width=3.in]{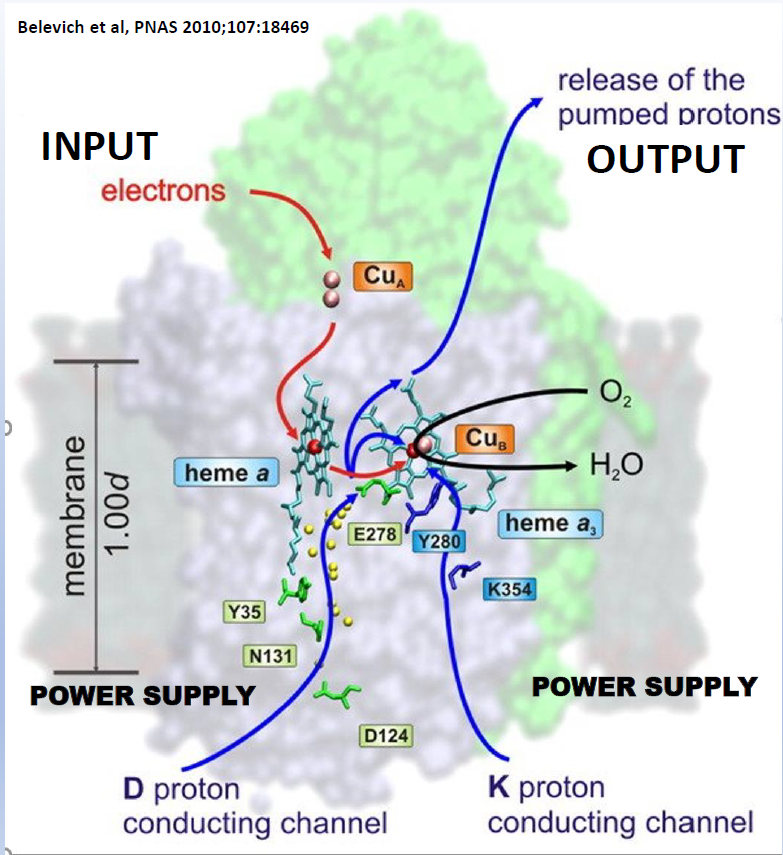}}
			\end{subfigure}
			
			\begin{subfigure}[]{
					\includegraphics[width=0.45\textwidth]{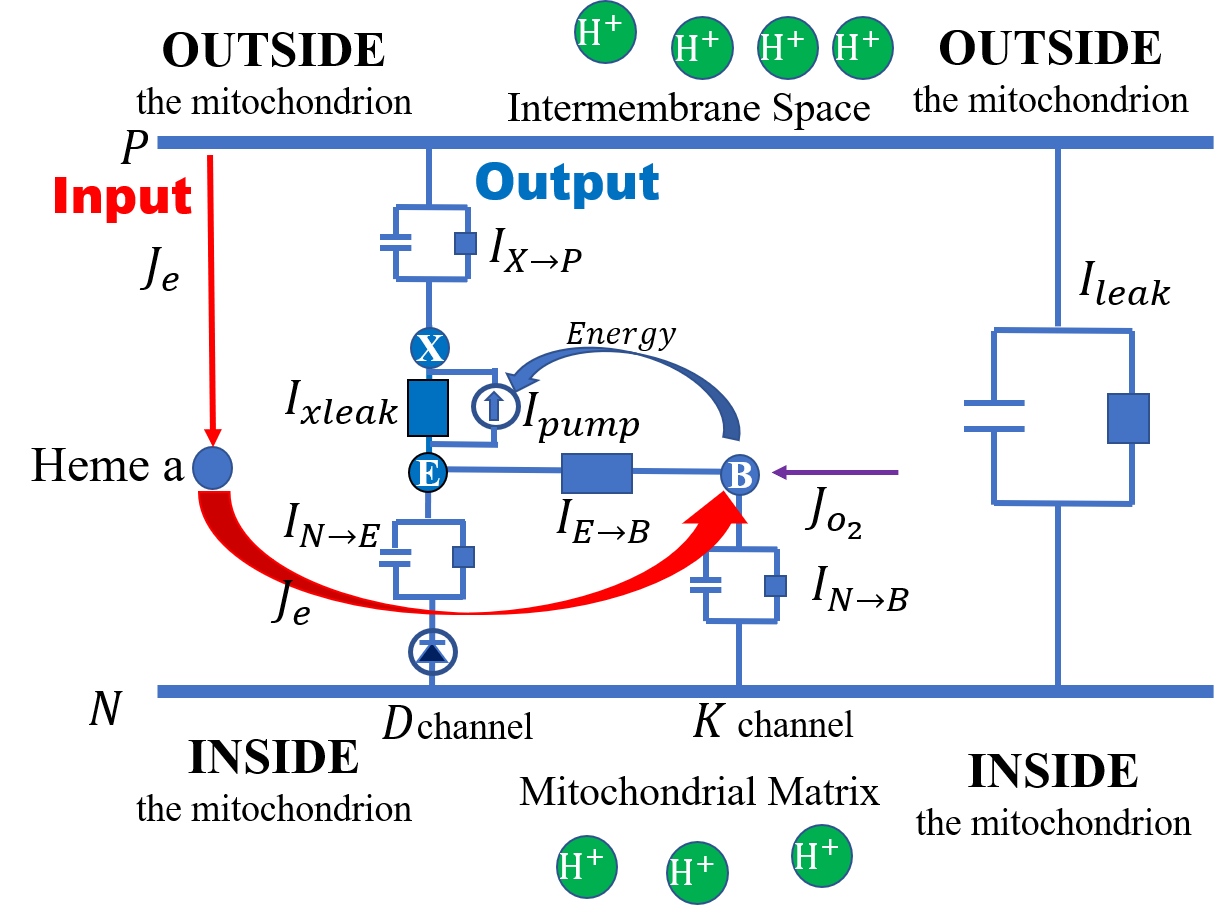}}
			\end{subfigure}
			\begin{subfigure}[]{
					\includegraphics[width=0.45\textwidth]{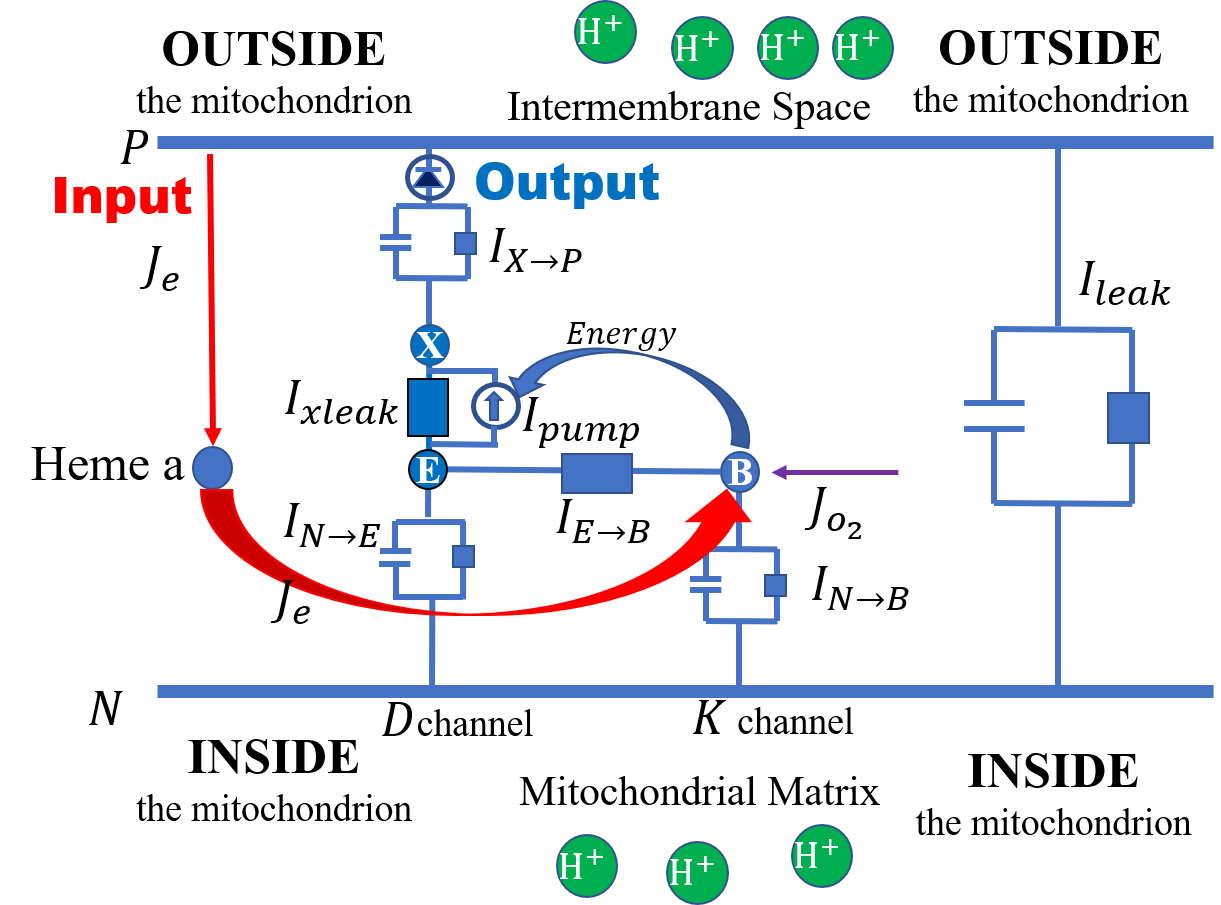}}
			\end{subfigure}
			\caption{Schematic. (a) Complex IV structure; (b)
				Circuit diagram of model \eqref{current_model} - \eqref{curentdef}; (c) Circuit diagram when the rectifier is located between the protein loading site PLS and P side. The drawing of  (a) is taken from  \cite{belevich2010initiation} according to the policies of the Proceedings of the National Academy of Sciences  \text{https://www.pnas.org/author-center/publication-charges\#author-rights-and-permissions} examined on June 18, 2022. We thank the authors for providing all of us such a helpful figure.
			}	\label{fig:Schematic}
		\end{figure}

	Here we propose a specific model of cytochrome $c$ oxidase (or Complex IV) as an example so our approach can be seen in action. 
	The schematic structure of cytochrome $c$ is shown in Fig.\ref{fig:Schematic} a, where both two channels from mitochondria matrix (inside), D and K,  are taken into consideration. Here,  $E$ denotes the end of $D$ channel. And the end of K channel is asumped to be the binuclear center (BNC) denoted by $B$ where the chemical reaction \eqref{HOreaction} occurs.  The protons accumulated in E are transported to the BNC and the proton loading site (PLS), denoted by X. A pump is located between E and PLS. The pump provides energy that comes from concentration gradients, namely gradients of chemical potential at BNC. Then finally, the proton is pushed out from PLS to the inter membrane space, outside the mitochondrion. 
	
	It is clear that
	this model is incomplete at best, and in some sense wrong, at worst. We depend on our experimental colleagues to help us correct and improve the model, for example, by including mechanisms we have over simplified. Enormous detail of the
	chemical reactions is described in the literature, with more intermediates being reported frequently. We do not include these intermediates.

		
		
		Let $\mathcal{\rho}_e = \rho_0\delta(x_0,y_0,z_0, t)$. 
		Integrating the diffusion-reaction equation 
		\begin{equation}
			\frac{d [C]_i}{dt} = -\nabla\cdot \bj_i -\alpha_i \mathcal{R},
		\end{equation}
		in the complex IV compartment yields 
		\begin{equation}
			\eta\frac {d\bar{C}_i}{dt} = J_i^{in} - J_i^{out}-\alpha_i \mathcal{R},
		\end{equation}
		where $\eta_{mat}$, $
		\eta_{ims}$ and $\eta$ are the volumes of mitochondrial compartment, inter membrane space and reaction compartment, respectively. 
		
		The chemical reaction in the cytochrome $c$ oxidase Complex IV is  
		\begin{equation}\label{HOreaction}
			2 H^{+}+\frac 1 2 O_2+2e^{-}\overset{k_f}{\underset{k_r}{\rightleftharpoons}}  H_2O. 
		\end{equation}
		
		For simplicity, we follow the Hodgkin Huxley tradition and  fix the proton concentrations  in the mitochondrial matrix (inside) and the inter membrane space outside the mitochondria so they do not vary with time or flow. More general treatments in which concentrations are changed with time by flow are possible as have been done in even more complex structures, Such analysis has been done in a bi-domain model of the lens of the eye and a tri-domain model of the optic nerve and glia \cite{RN28691,RN30228,RN30228,RN30601,RN30649}.

		Here for simplicity, we assume that the concentration of oxygen at the B site is a constant.  
		If the
		oxygen varies with time, an additional equation can be used to describe the dynamics of oxygen. The properties of that term will be
		determined either directly by experimentation  or by a
		higher resolution model as in \cite{yamashita2012insights}. We do not expect the extra term to introduce significant mathematical, numerical, or computational difficulties. 
		
		
		Many variables are needed to keep track of all the potentials and concentrations in the variouis regions of our model. The concentrations and potentials  at E242; at BNC; and at the proton loading site (PLS), are different inside and outside the mitochondria, as is the electron concentration (see Fig.\eqref{fig:Schematic} (a)). They are described by the variables 
		$[H]_E$, $[H]_B$, $[H]_x$,$\phi_E$, $\phi_B$, $\phi_X$, $\phi_N$, $\phi_P$  and $\rho_e$, respectively.

		\begin{subequations}\label{current_model}
			\begin{align}
				&\frac{d[H]_E} {dt} =\frac {S_v} F(I_{N2 E}-I_{E2 X}-I_{E2 B}),\label{HEequation} \\
				&\frac{d[H]_B} {dt}= \frac {S_v} F( I_{E2 B}+I_{N2 B})-2\mathcal{R},\label{HBquation} \\
				&\frac{d[H]_X} {dt}= \frac {S_v} F( I_{E2 X}-I_{X2 P} ),\label{HXquation} \\ 
				&\frac{d\rho_e}{dt} = \frac {-S_v} F I_e-2\mathcal{R}, \label{rhoequation}\\
				&C_E \frac{d(\phi_E-\phi_N)}{dt} = (I_{N2 E}-I_{E2 X}-I_{E2 B}), \\
				&C_B \frac{d(\phi_B-\phi_N)}{dt} = I_{E2 B} +I_{N2 B}+I_e,  \\
				&C_X  \frac{d(\phi_X-\phi_P)}{dt}  = ( I_{E2 X}-I_{X2 P}),\\
				&  C_m\frac{d(\phi_N-\phi_P)}{dt} +I_{leak} +I_{X2 p}+I_e = 0, 
			\end{align}
		\end{subequations}
		
		with currents
		\begin{subequations}\label{curentdef}
			\begin{align}
				& I_{N2 B}  = g_K(\phi_N-\phi_B-\frac{RT}{F}\ln\frac{[H]_B}{[H]_N})=\frac{g_K}{F}(\mu_N-\mu_B),\\
				&I_{E2 B} = g_B(\phi_E-\phi_B-\frac{RT}{F}\ln\frac{[H]_B}{[H]_E})=\frac{g_B}{F}(\mu_E-\mu_B), \\
				&I_{E2 X}  =I_{pump} +I_{xleak},\\
				&I_{leak} = g_m (\phi_N-\phi_P -E_{other} ),  \\
				&I_e = -FJ_e,\\
				&I_{xleak} =  -g_E(\mu_X-\mu_E),\\
				& I_{pump} = \left\{
				\begin{array}{cc}
					g_{pump}max(R_c,0) (\mu_X-\mu_E),       \mu_X-\mu_E< \delta_{th},  \\
					g_{pump}max(R_c,0)\delta_{th} \exp{\left(- \frac{(\mu_X-\mu_E)}{\varepsilon}\right)},       \mu_X-\mu_E\ge \delta_{th}, \end{array} 
				\right.\label{pumpcurrent}\\
				&\mathcal{R} = k_f[H^+]^{2}[O_2]^{1/2}\rho_e^2-k_{r}[H_2O].\label{reaction}
			\end{align}
		\end{subequations} 
		We follow the review of Wikström \cite{RN45750} and implement switching functions without invoking conformation changes of the distribution of mass. We treat cytochrome c oxidase as a Coulombic system and use rectifiers to implement the switching functions that provide alternating access of an occluded state. 
		Here we discuss  two cases. In one case, a rectifier between $N$ and $E$ blocks the proton flows. In the other case, a rectifier between $X$ and $P$  blocks the backward proton flows. 
		Then, the currents $I_{N2 E}$ and $I_{X2 P}$ are modelled in the follows two cases.
		\begin{itemize}
			
			\item Case 1: the rectifier is between $N$ and $E$ as shown in Fig. \ref{fig:Schematic}b
			\begin{subequations}\label{rect_n2e}
				\begin{align}
					& I_{N2 E}  =  max\left(g_D \left(\phi_N-\phi_E-\frac{RT}{F}\ln\frac{[H]_E}{[H]_D}\right), -SW_{0}\right)=max\left(\frac{g_D}{F}(\mu_N-\mu_E),-SW_{0}\right)\label{swith},\\
					&I_{X2 P}  = g_X(\phi_X-\phi_P-\frac{RT}{F}\ln\frac{[H]_P}{[H]_X})=\frac{g_X}{F}(\mu_X-\mu_P),
				\end{align}
			\end{subequations}
			\item Case 2: the rectifier is between $X$ and outside as shown in Fig. \ref{fig:Schematic}c
			\begin{subequations}\label{rect_x2p}
				\begin{align}
					& I_{N2 E}  =  g_D \left(\phi_N-\phi_E-\frac{RT}{F}\ln\frac{[H]_E}{[H]_D}\right) \label{swith2},\\
					&I_{X2 P}  =max\left( g_X(\phi_X-\phi_P-\frac{RT}{F}\ln\frac{[H]_P}{[H]_X}),-SW_{0}\right)=max\left(\frac{g_X}{F}(\mu_X-\mu_P),-SW_{0}\right),
				\end{align}
			\end{subequations}
		\end{itemize}
		where $SW_{0}$ is the threshold for the turn-off of the rectifier and $SW_{0}=0$ stands for perfect rectifier. We reiterate that pn junctions are used to rectify the movement of the pseudo-ions holes and electrons throughout our digital circuitry. Analogous distributions of permanent charge provided by acid and base side chains of proteins produce rectification of charge movement in ionic systems.

		We use Kirchhoff's law and the conductance formulation of Hodgkin and Huxley. Complex properties are hidden by a nonlinear time dependent version of Ohm's law and modelled by a conductance, as did Hodgkin and Huxley \cite{RN267,RN309,RN28957,RN12551,RN28654,RN291,}. 
		Alternating access (with its implied occluded state) is described by a switching function for the D channel using equation \eqref{swith}. This is a classical rectifier function and (when $SW_0 =0$), allows current only to flow from D to E: no backward flow is allowed. 
		
		Many properties of the model depend on the pump current between E242 and the Proton Loading Site PLS. We assume that in the ordinary situation the pump strength depends on the reaction rate and the chemical reaction difference between two sites.  However, in the less ordinary situation, when the difference is too large,  the pump may not be able to overcome the barrier. A turn-off threshold is assigned to the pump for that reason. We assume that when the difference in chemical potential  $\mu_x - \mu_E$ is greater than the threshold, the pump current decreases exponentially to zero as it turns off. More realistic, and complex properties of the pump will undoubtedly be needed to explain some functions of cytochrome c oxidase. They can easily be incorporated into our model, as these properties are measured and modelled.

		Our `electro-osmotic' model is a `master equation' approach building on the work of Hummer and 
		Kim, \cite{kim2012proton,kim2007kinetic,kim2009kinetic}
		but showing how to exploit conservation of current in the form of Kirchhoff's current law. This approach is used throughout electrical and electronic engineering to design semiconductor devices, as textbooks document ($op. cit.$) perhaps most eloquently in the modern automated circuit design literature built on Kirchhoff's law \cite{RN45999}. Currents are sufficient for such automated design. Charges are not needed except in switched-capacitor networks (p. 64 of \cite{RN45999}) . 
		
		We extend the classical use of Kirchhoff's law that forms the foundation of circuit design to include chemical reactions. We must include chemical reactions to drive currents of electrons, protons and other ions because that is how cytochrome $c$ oxidase functions. The essential function of cytochrome $c$ oxidase is to convert a flow of electrons  to a flow of protons from inside the mitochondrion to outside it. The electrons that are inputs to the cytochrome $c$ oxidase are presented to the enzyme attached to the heme group of cytochrome $c$ itself.
		
		The existing literature analyzes these systems without explicitly dealing with currents, making the task either impossible (by using a theory that assumes a zero value for the fluxes being studied) or very difficult (by involving a staggering number of charges). Using currents instead of charges avoids these difficulties and has the added advantage of automatically satisfying Maxwell's equations, if total current is used in Kirchhoff's law.
		
		This approach is incomplete because it does not deal with all the charge in the circuit formed by cytochrome $c$ oxidase But \textbf{those details of charge are not needed in the design of electronic circuits.} That simple fact can be verified by examining textbooks of circuit design (as already cited).
		
		In circuit analysis of this type, some questions about charges need not be asked: for example, the atomic mechanism of current flow (particularly electron flow) can be ignored. The function of the circuit is  independent of the details of the components of current in wires, for example, with only a few exceptions \cite{RN45998}. Thus, we de-emphasize the atomic details of the various pathways that provide electron flow to the main reaction centers. For us, these pathways are wires. The atomic and chemical details of electron flow in these wires are known in breathtaking detail and we are sorry that we do not seem to need to use these magnificent results, but, at this resolution, we do not. 
		
		A key biological result is that some of the coupling so important to understand the electro-osmotic properties of a mitochondrion depends
		on the macroscopic conservation of current, i.e., Kirchhoff's law applied to the entire mitochondrion. The application of Kirchhoff's law to mitochondrial transport, and active transport in general, is not common in the literature. But Kirchhoff's law has been used in another branch of biophysics for a long time, for some eighty years. Kirchhoff's law is the keystone of the analysis of ion channels. Kirchhoff's law is the keystone that supports the structure of the Hodgkin approach to the action potential by balancing the various components of current, summing them to zero
		in the appropriate (finite) geometries, like those of mitochondria, the way the keystone of an arch sums mechanical forces.

		Conservation of current provides the coupling in other biophysical applications, e.g., generation and
		propagation of the action potential, linking atomic scale properties of ion channels of one type 
		to properties (e.g., opening) of another type.
		In a classical action potential, the opening of sodium channels is coupled to the opening of other sodium channels, and to the closing of potassium channels by the electric field, not by anything else. There is no steric or chemical interaction between the channels. The coupling  is essential to the function of the nerve cell, but that coupling is described by a version of the Maxwell equations (called the cable or telegrapher's equation) not by equations of chemical kinetics. The ion channels of the action potential act independently in the chemical sense because they are so far apart, without opportunity for short range or chemical interactions. The ion channels are not independent, in the physiological or physical sense, however. Rather they are coupled by the electric field. The electrical field is that which satisfies the Maxwell equations, or their equivalent, Kirchhoff's current law.
		
		What we propose here is in the tradition of Hodgkin's treatment of ionic channels, but we
		include the chemical reactions that are the essence of oxidative phosphorylation and the life and function of mitochondria.
		We are more than aware that a detailed analysis of alternating access, the occluded states, and the switching function
		is needed to understand cytochrome $c$ oxidase. That analysis needs the currents flowing to be analyzed along with the atomic detail of the water-gate switch  \cite{RN4995,RN45750}, in our view. The switches act on currents and of course satisfy conservation of current.
		An analysis of charges cannot easily guarantee conservation of current, and classical chemical analysis precludes large currents because assuming equilibrium or near equilibrium conditions is  clearly inappropriate for a system like Reaction Center IV designed for the efficient handling of large flows of electrons and protons.
		Here we describe alternating access with the classical equation of a rectifier to highlight the possibility that occluded states
		and alternating access are the biochemical names for what engineers call rectification. 
		It is important to realize that rectification is an automatic unavoidable consequence of the distribution of doping in semiconductors,  for example in the classical PN diode. This rectification occurs with no change in the spatial distribution of mass (i.e., with no change in what is usually called conformation)
		and so it is compatible with the view cited above that cytochrome c oxidase functions without changes in the spatial distribution of mass, i.e., without what is classically called conformation change. In the rectification mechanism, the switching (rectification) occurs because of a radical change in the distribution of electrical potential, which in turn allows current flow in one direction and not another. The distribution of potential depends on the distribution of mobile electrons which have almost no mass. The conformation of the potential profile and thus the electric field creates a barrier for current flow in one direction but not in another, because of the effects of doping (permanent charge) and mobile charge combined in the Maxwell Gauss law, or the Poisson equation. This system is rather complex, although completely understood and used in literally billions of places in each of our computers, The system involves diffusive and electrical movement of electrons (and holes) driven by the gradients of chemical potential (e.g., concentration) and electrical potential. As the electrical potential changes sign, diffusive and electrical flow changes. As concentrations change, diffusive and electrical flow change in other ways. All interact through the changing fields of electrical and chemical potential. A few pages (not just a few words or sentences) are needed to explain how each kind of movement (diffusive, electrical, holes and electrons) contribute to rectification. See textbooks of semiconductor circuit design, e.g, \cite{colinge2005physics,pierret1996semiconductor}. It is also important to consult 
		research articles \cite{laux1999revisiting,haggag2000analytical} to understand the oversimplifications of the textbook discussions and to validate them. 
		More elaborate patterns of doping, starting with the PNPN designs (thyristors, Silicon Controlled Rectifiers = SCR) are used in  power transistors. Analogous spatial distributions of permanent charge (and acid and base side chains) might be used to implement switches in cytochrome c oxidase.
		
		We note rectification arising from the distribution of permanent charge in a protein was proposed by Mauro a very long time ago \cite{RN23400,RN23397}. Such rectifiers of ionic current have even more complex properties than semiconductor rectifiers because concentrations of current carriers in biological solutions can be changed independently of electrical potential, which is not often
		the case in analogous semiconductor systems.  Ionic rectifiers were built a long time ago
		using a biological protein as a template
		\cite{miedema2007biological} and are now used routinely in 
		the ionic channels of nanotechnology \cite{RN45776} 
		even in a hybrid  chip that can enable a scalable integrated ionic circuit platform for micro-total-analytical systems \cite{RN45777}.
		
		The switch of Reaction Complex IV is likely to involve both the distribution of permanent charge (mostly acid and base side chains),
		and chemical interactions as described in the water-gate model  \cite{RN4995,RN45750}, perhaps also involving spatial distribution of dielectric properties as well \cite{RN30417}. 
		It seems premature to attack this problem here, as important as it is for the function of cytochrome $c$ oxidase, and all alternating access transporters, for that matter. Here, we simply describe the rectification without further analysis of how it arises from the distribution of permanent charge and other properties of the transporter structure.
		
		Of course, other possibilities exist. 
		Alternating access might arise, for example, from bubbles in the conduction pathway, as we are studying in other work \cite{RN46054}.
		Two bubbles might act as coupled activation and inactivation gates, correlated to provide alternating access to an occluded state, for example. 
		
		\section{Computational Studies}
		In this section, we carry out several computational studies to explore the effects of various conditions on proton transport efficiency.  
		The initial values and default parameters are listed in Table 1-2. 
		
		\begin{table}[]
			\begin{center}
				\begin{tabular}{|c|c|c|}
					\hline
					Variable & Notations & Values (with Unit) \\
					\hline
					$E_{242}$ site $H^+$ concentration    & $[H]_{E}$ & 0.01196 $\mu M$ \\
					\hline
					BNC site $H^+$ concentration    & $[H]_{B}$ &0.01682 $\mu M$ \\
					\hline
					PLS site $H^+$ concentration    & $[H]_{X}$ & 0.01441 $\mu M$ \\
					\hline
					BNC site electric  density  & $\rho_{e}$& 0.01166  $\mu M$ \\
					\hline
					$E_{242}$ site electric potential   & $\phi_{E}$ & -5 $mV$ \\
					\hline
					BNC site electric potential   & $\phi_{B}$ & -14.1562 $mv$ \\
					\hline
					PLS site electric potential   & $\phi_{X}$&  200 $mv$ \\
					\hline
					N site electric potential   & $\phi_{N}$&  0 $mv$ \\
					\hline
					P site electric potential   & $\phi_{P}$&  160 $mv$ \\
					\hline
				\end{tabular}
				\caption{Default Initial Values}
			\end{center}
		\end{table}

		\begin{table}[]
			\begin{center}
				\begin{tabular}{|c|c|c|}
					\hline
					Variable & Notations & Values (with Unit) \\
					\hline
					$E_{242}$ site effective  capacitance      & $C_D$& 1E-1 $fA ms/mV /(\mu m)^2 $\\
					\hline
					BNC site effective  capacitance     & $C_B$& 1E-1 $fA ms/mV /(\mu m)^2 $\\
					\hline
					PLS site effective  capacitance      & $C_X$&1E-1 $fA ms/mV /(\mu m)^2 $ \\
					\hline
					Membrane  capacitance      & $C_X$&7.5E-2 $fA ms/mV /(\mu m)^2 $ \\
					\hline
					D channel conductance for $H^+$ & $g_D$&  3.75E-3$ pS/(mum)^2$ \\
					\hline
					K channel conductance for $H^+$     & $g_K$& 1E-3 $ pS/(\mu m)^2$\\
					\hline
					E2B channel conductance for $H^+$     & $g_B$&  5E-2 $ pS/(\mu m)^2$\\
					\hline
					E2X channel conductance for $H^+$     & $g_E$&  1E-3 $ pS/(\mu m)^2$\\
					\hline
					E2X Pump rate for $H^+$     & $g_P$&  369 $pS ms/(\mu m)^2 \mu M$\\
					\hline
					X2P channel conductance for $H^+$     & $g_X$&  9.8E-4 $ pS/(\mu m)^2$\\
					\hline
					Membrane conductance for leak     & $g_m$&  1E-5$ pS/(\mu m)^2$\\
					\hline
					Mito. matrix $H^+$ concentration    & $[H]_{mat}$ & 0.01 $\mu M$ \\
					\hline
					Mito. inner membrane space $H^+$ concentration    & $[H]_{ims}$ & 0.063 $\mu M$ \\
					\hline
					Nernst Potential due to other Ions & $E_{Other}$ & $-160mV$ \\
					\hline
					Reaction site $[O_2]$ concentration   & $[O_{2}]$ & 0.0028 $\mu M$  \\
					\hline
					Reaction site $[H_2O]$ concentration   & $[H_2O]$ & 0 $\mu M$ \\ 
					\hline
					Electron current &$I_e$ & -5.24 $fA$\\
					\hline
					Forward reaction rate coefficient &   $k_f$ & 1333
					\\
					\hline
					Backward reaction rate coefficient &   $k_r$ & 0.005 \\
					\hline
					surface volume ratio & 
					$S_v$ &1000\\
					\hline 
					Potential Threshold & $\delta_{th}$ & 210 mv\\
					\hline
					Decay rate & $\varepsilon$& 1 $(ms)^{-1}$ \\
					\hline
				\end{tabular}
				\caption{Parameters}
			\end{center}
		\end{table} 
		\subsection{Effect of electron current: Input to Output Relations}
		
\begin{figure}[!ht]
			\centering
			\begin{subfigure}[]{
					\includegraphics[width=2.5in]{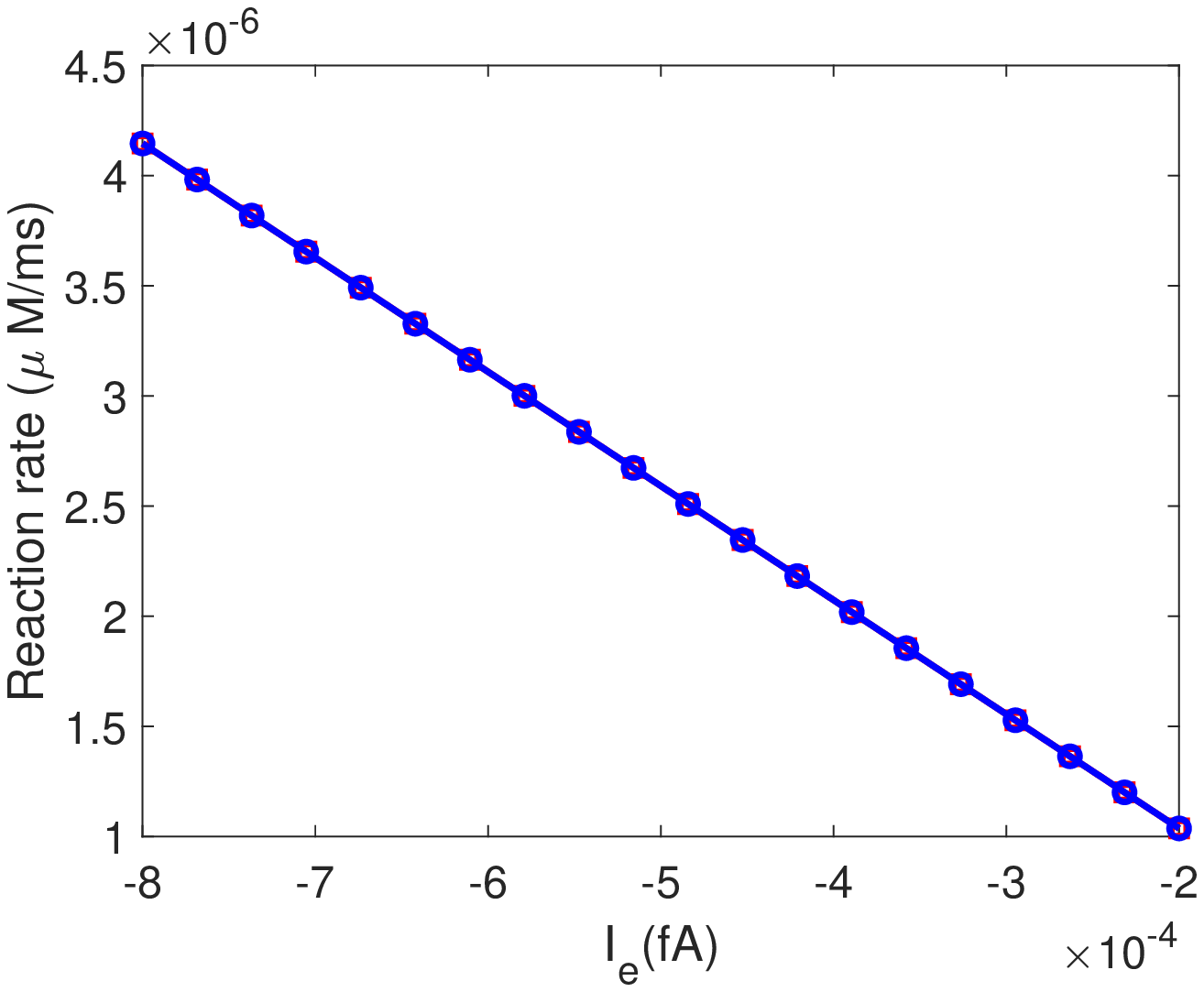}}
			\end{subfigure}
			\begin{subfigure}[]{
					\includegraphics[width=2.5in]{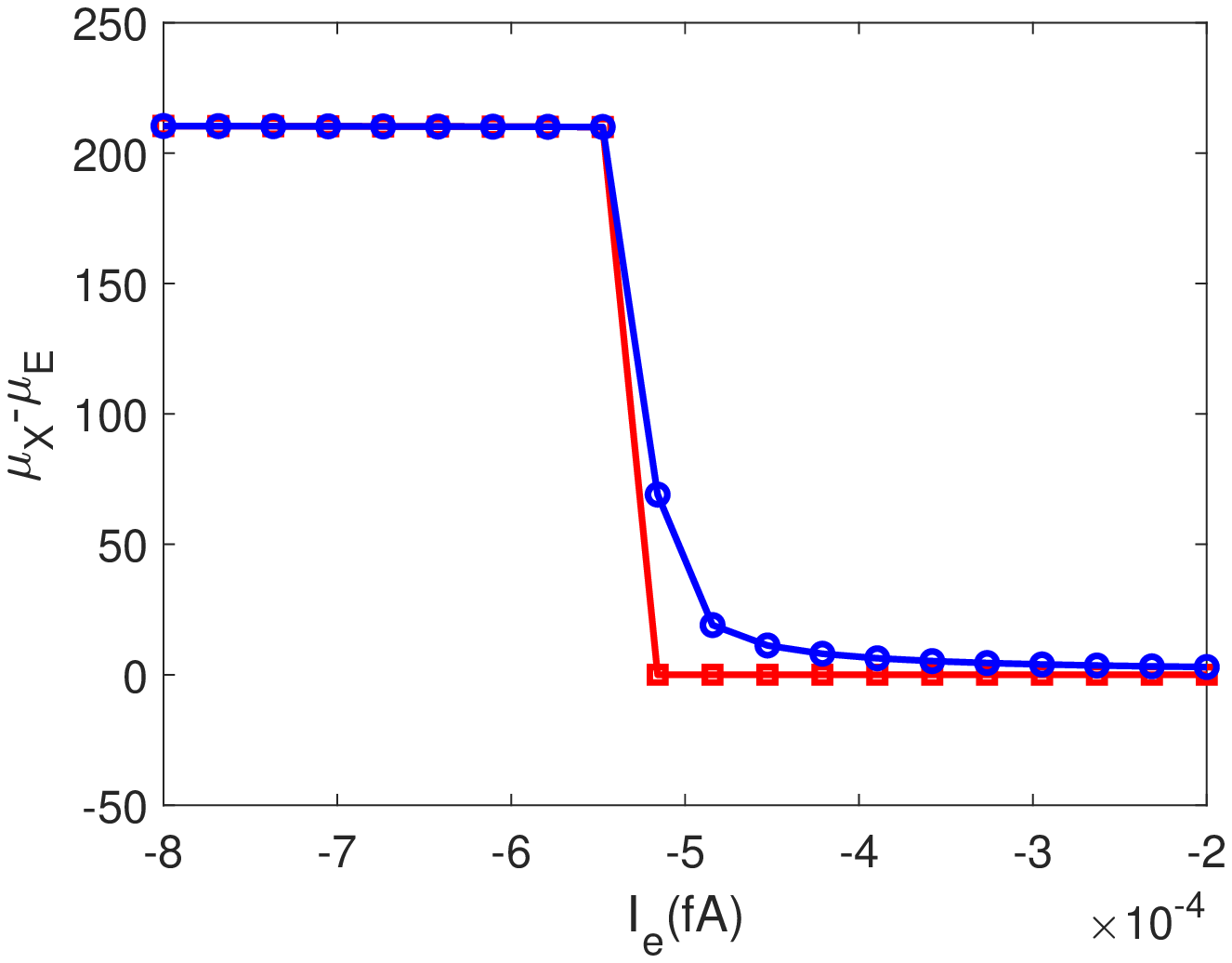}}
			\end{subfigure}
			\begin{subfigure}[]{
					\includegraphics[width=2.5in]{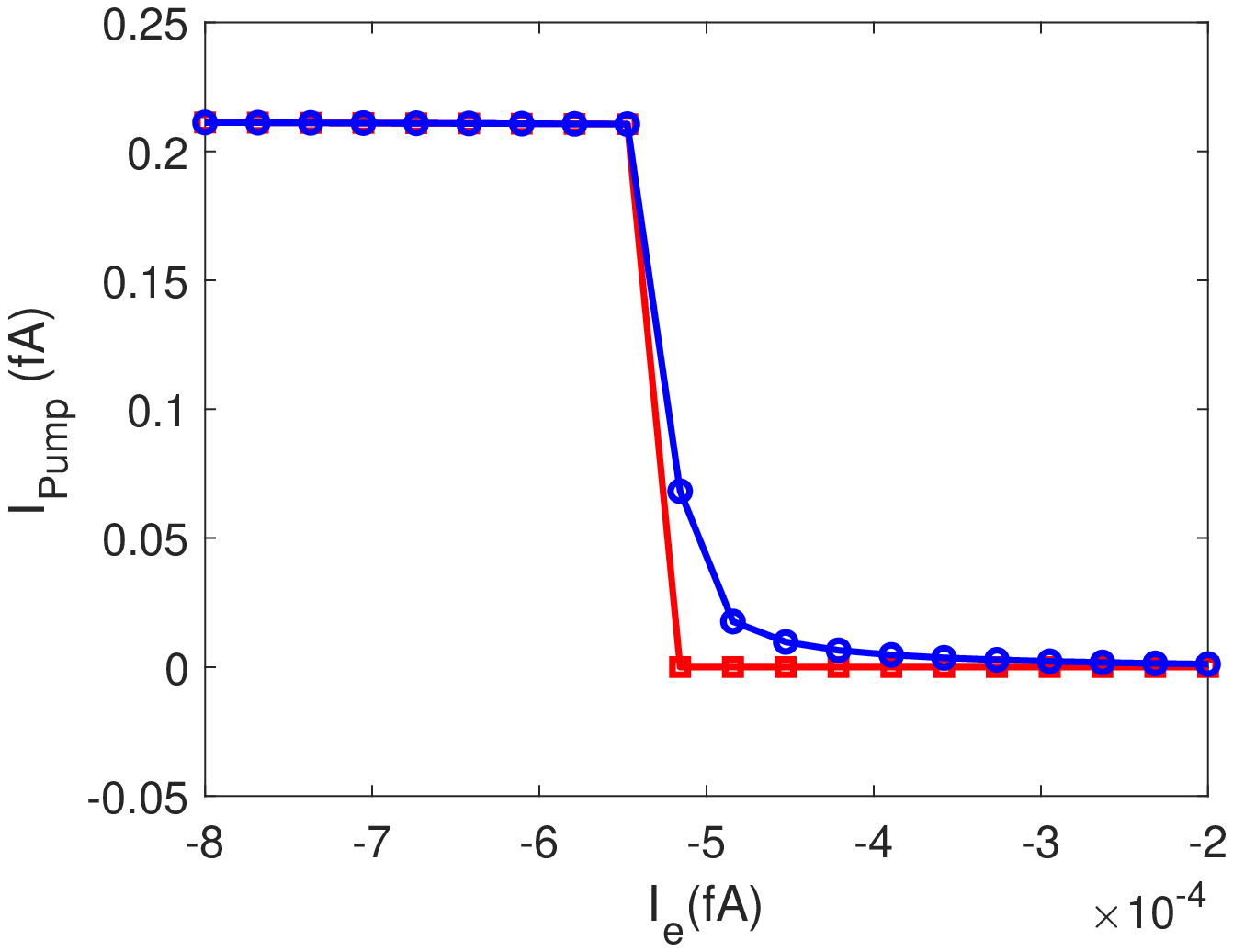}}
			\end{subfigure}
			\begin{subfigure}[]{
					\includegraphics[width=2.5in]{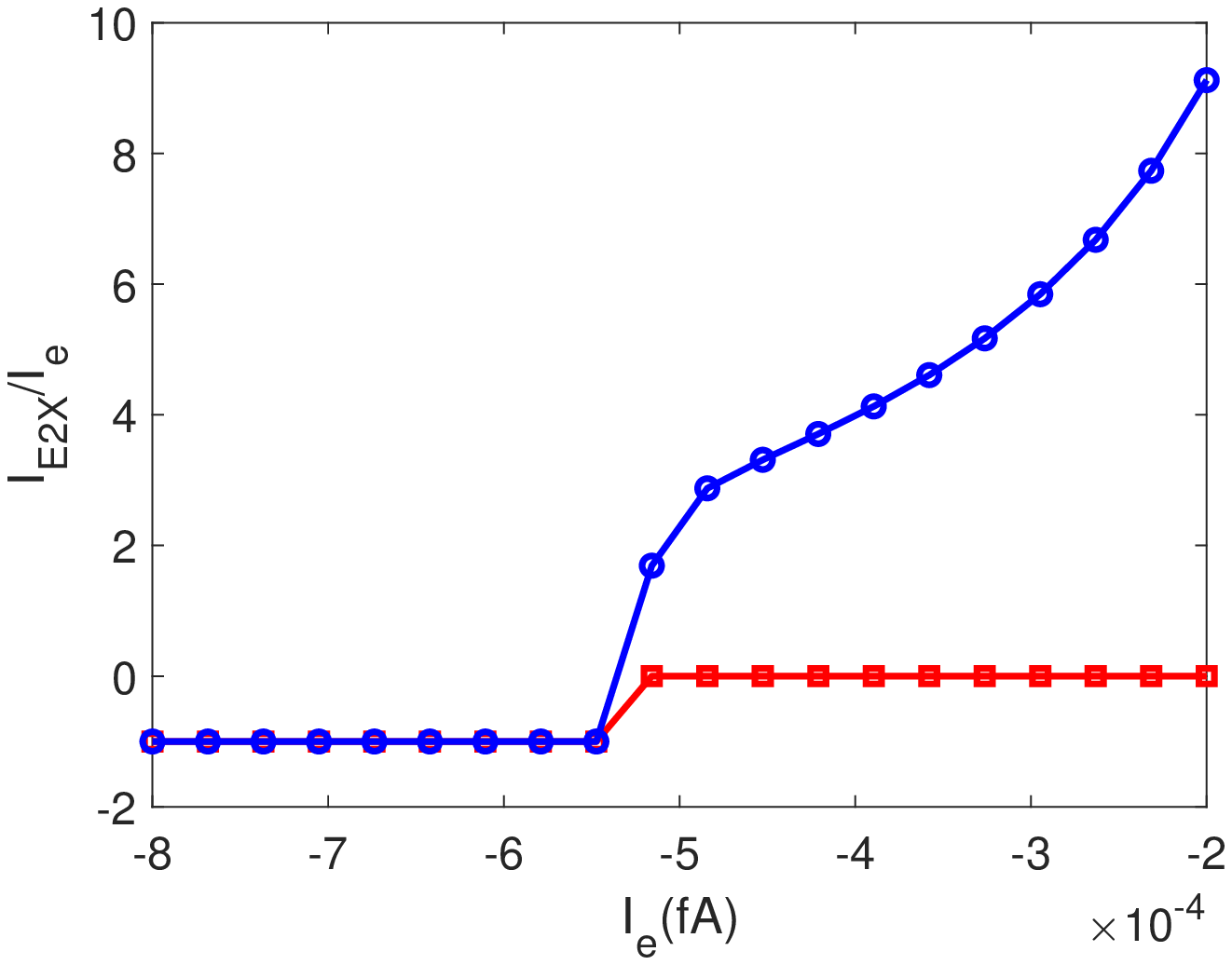}}
			\end{subfigure}
			\begin{subfigure}[]{
					\includegraphics[width=2.5in]{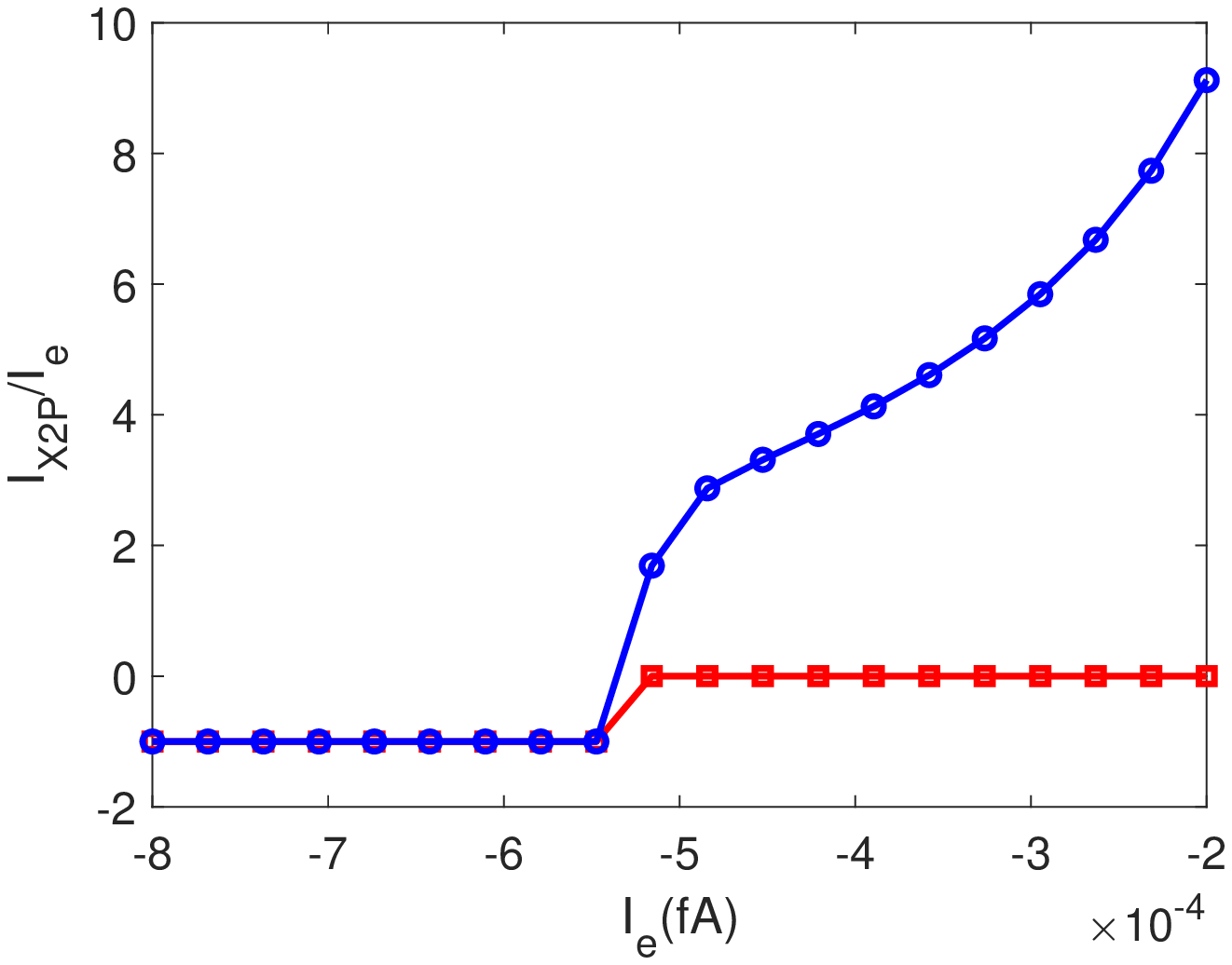}}
			\end{subfigure}
			\begin{subfigure}[]{
					\includegraphics[width=2.5in]{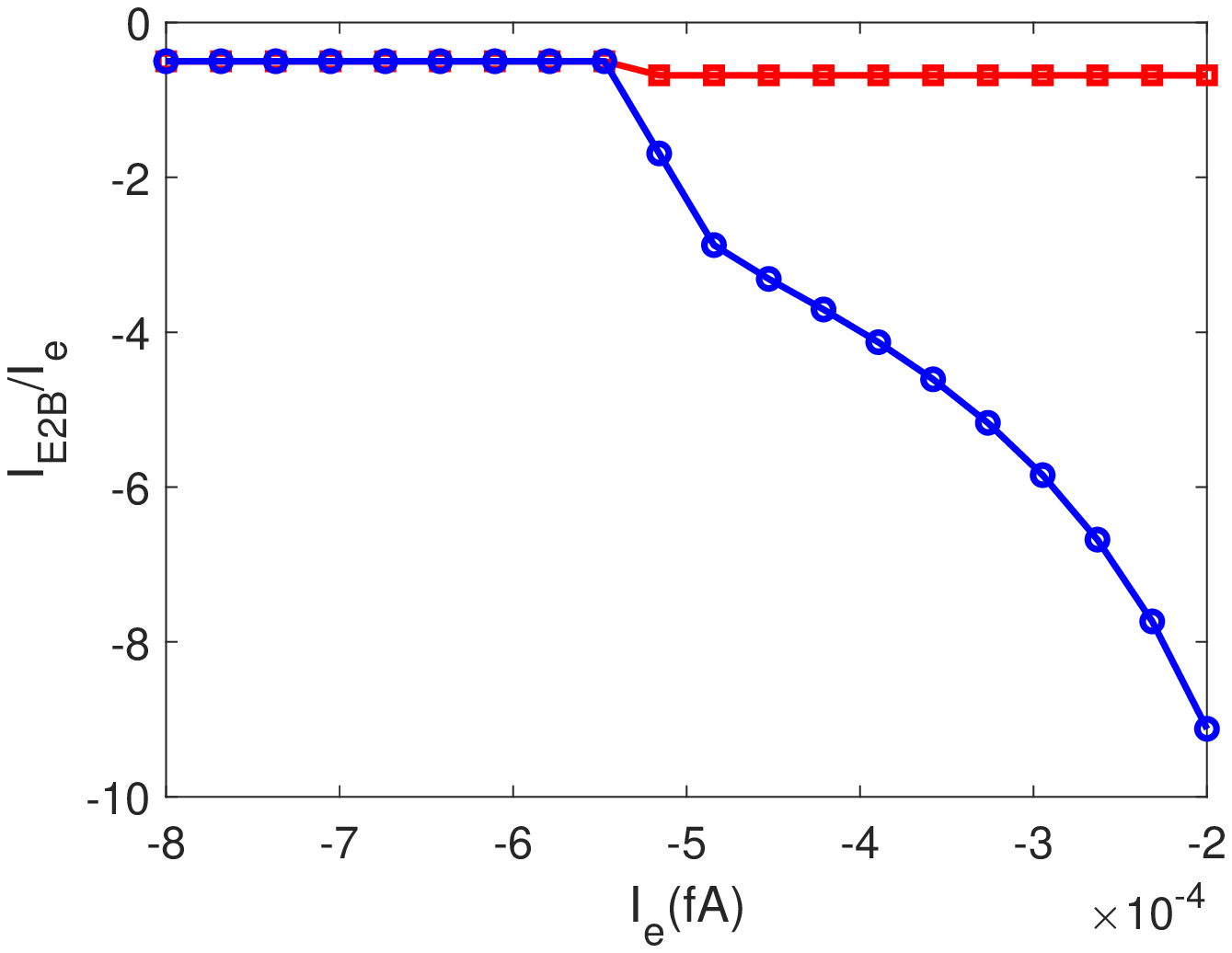}}
			\end{subfigure}
			\begin{subfigure}[]{
					\includegraphics[width=2.5in]{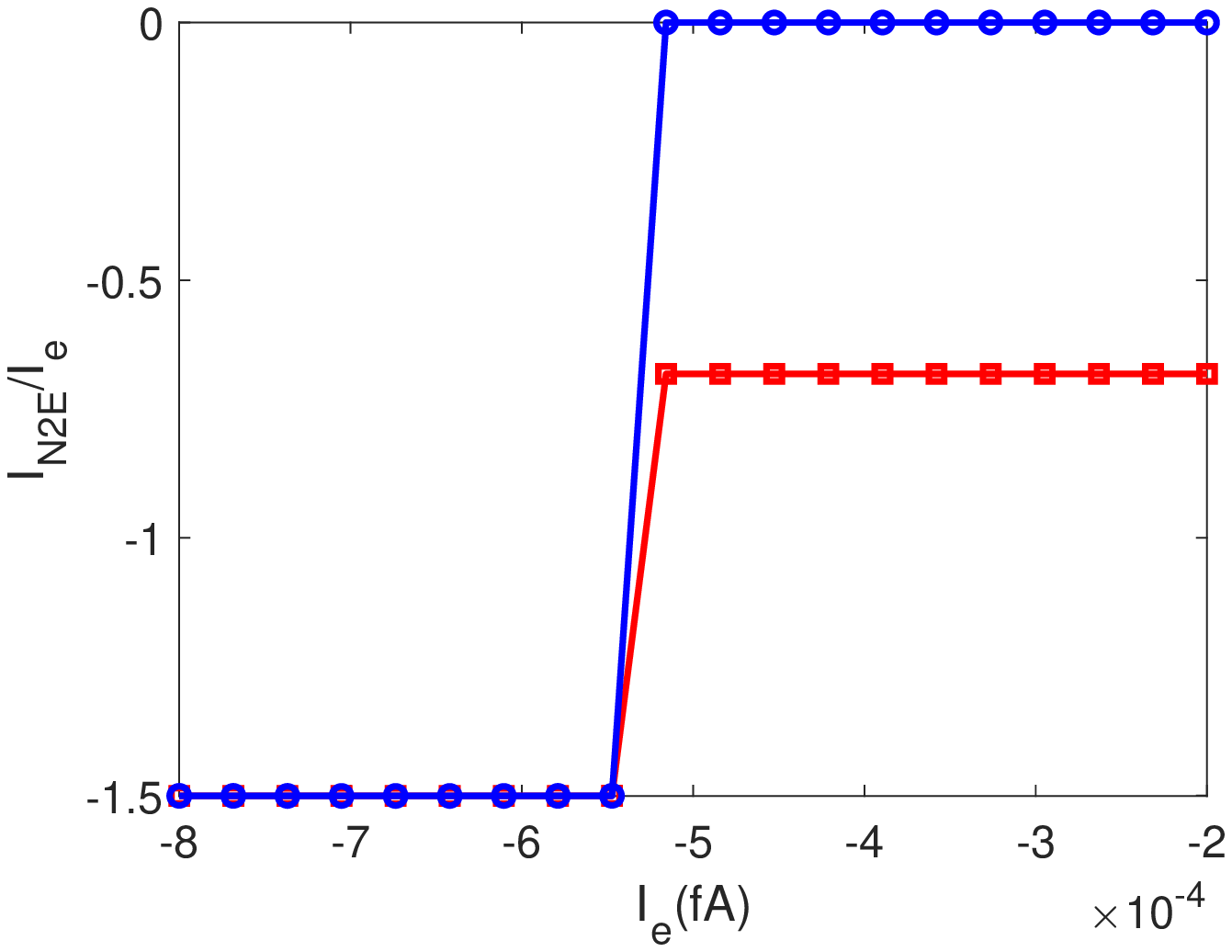}}
			\end{subfigure}
			\begin{subfigure}[]{
					\includegraphics[width=2.5in]{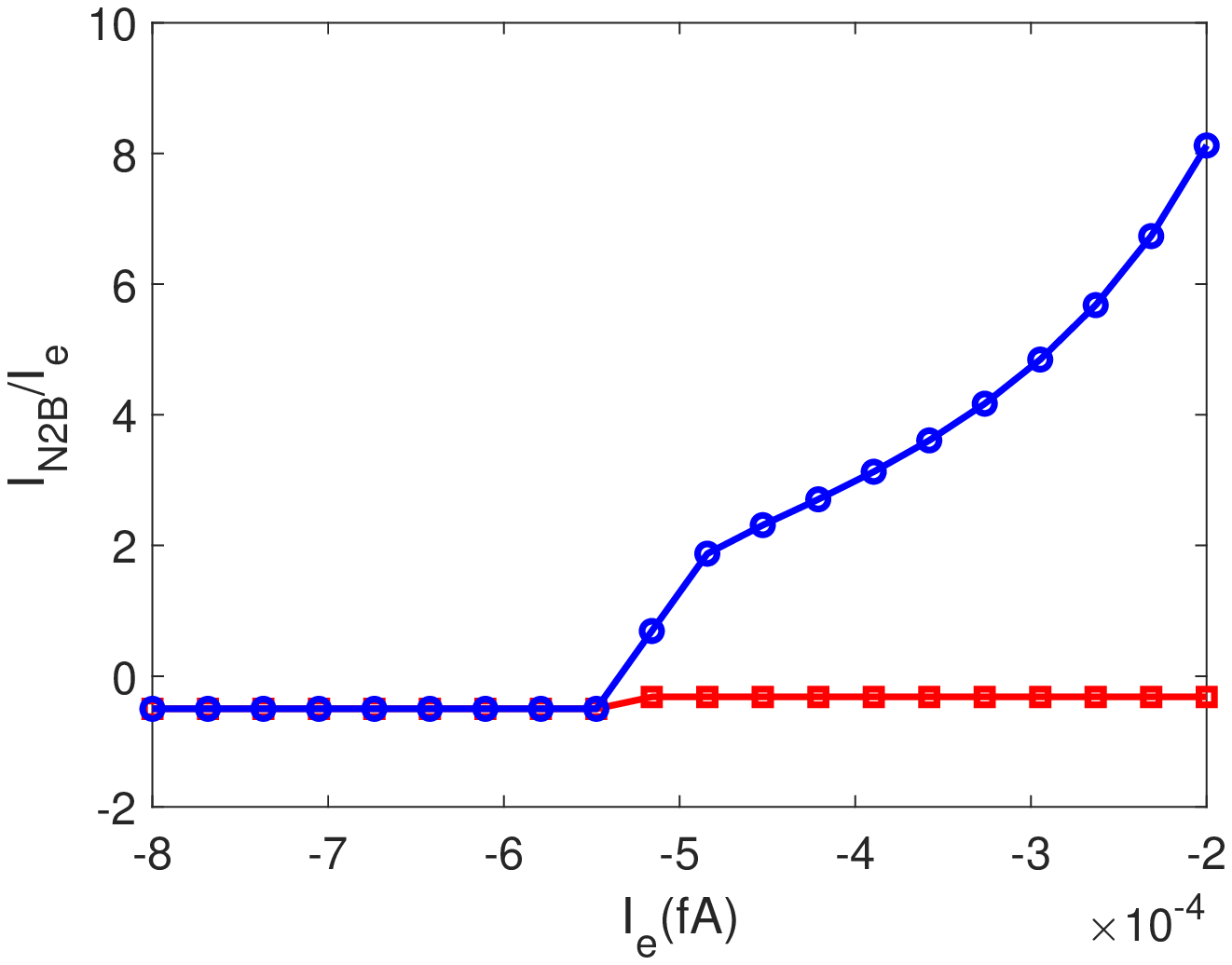}}
			\end{subfigure}
			\caption{Pump efficiency at equilibrium states with different electron current and different threshold. (a) Reaction rate;  (b) $\mu_X -\mu_E$; (c) $I_{Pump}$;(d) $I_{E2X}/I_e$; (e) $I_{X2P}/I_e$; (f) $I_{E2B}/I_e$; (g) $I_{N2E}/I_e$; (h) $I_{N2B}/I_e$. Red line: Switch between X and outside; Blue dash line: Switch between inside and E.}
			\label{fig:ratio_Ie_eq}
		\end{figure}

		We first check Figs.\ref{fig:ratio_Ie_eq}      the effect of electron current $I_e$ on the  efficiency of Complex IV. The case 1 (inside to E rectifier)  results are represented by blue circle and the case 2 (X to outside rectifier) results are represented by red square. 
		
		The ratios between currents and the supplied electron current are measures of the transfer function or 'gain' of cytochrome c oxidase. According to the previous study \cite{RN45750}, the  ratios are $\frac{I_{X2P}}{I_e} = -1$, $\frac{I_{E2X}}{I_e} = -1$ and $\frac{I_{E2P}}{I_{N2E}+I_{N2B}} = -0.5$ at the normal state.   These ratios mean that (nominally) each input electron will bring 2 protons from the  N side. One of the protons is used for the chemical reaction and the other one is pumped to the P side becoming an output in that way.   
		
		Fig. \ref{fig:ratio_Ie_eq}  (d)-(h) confirm that when the electron supply is sufficient ($|I_e|\ge 5.24 fA$) these ratios can be maintained.  
		However, if the input electron current decreases (in magnitude), the reaction rate decreases linearly (see Fig. \ref{fig:ratio_Ie_eq} (a)) since $\mathcal{R} = \frac{-S_v}{2F}I_e$ at equilibrium according to Eq. \eqref{rhoequation}. 
		The pump strength depends on reaction rate (Eq.\eqref{pumpcurrent}), so the pump current $I_{Pump}$ decreases hand in hand with reaction rate. 
		
\begin{figure}[!ht]
			\centering
			\begin{subfigure}[]{
					\includegraphics[width=3.in]{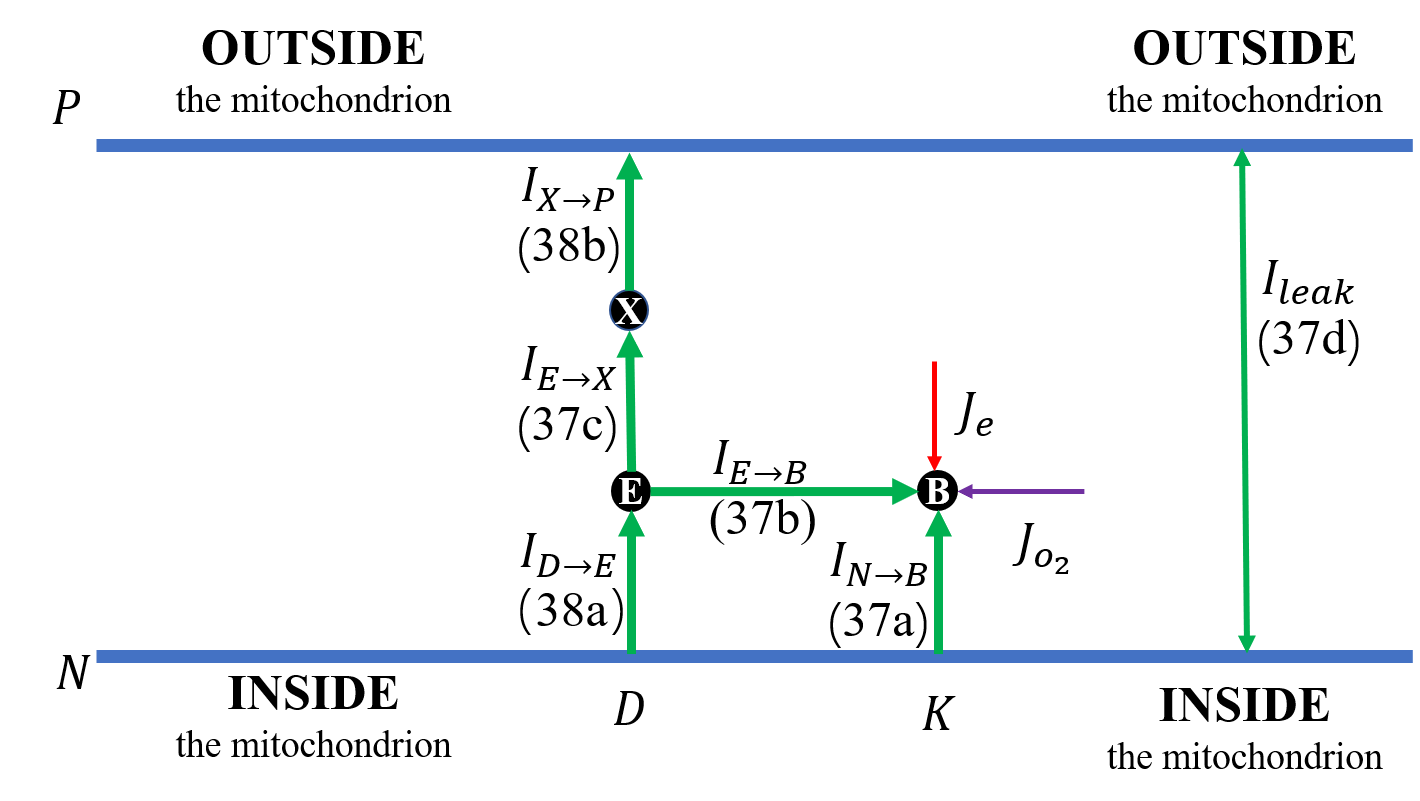}}
			\end{subfigure}
			\begin{subfigure}[]{
					\includegraphics[width=3.in]{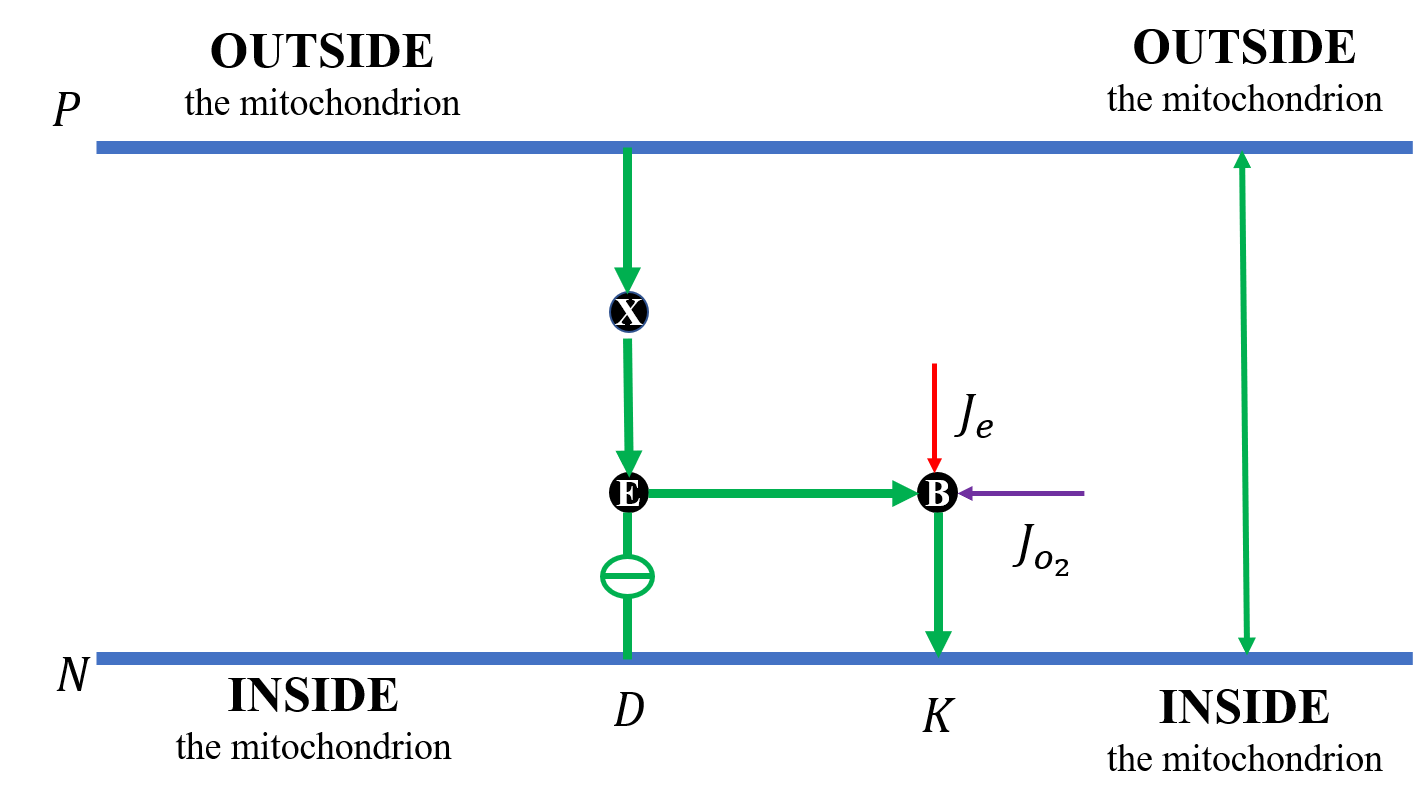}}
			\end{subfigure}
			\begin{subfigure}[]{
					\includegraphics[width=3.in]{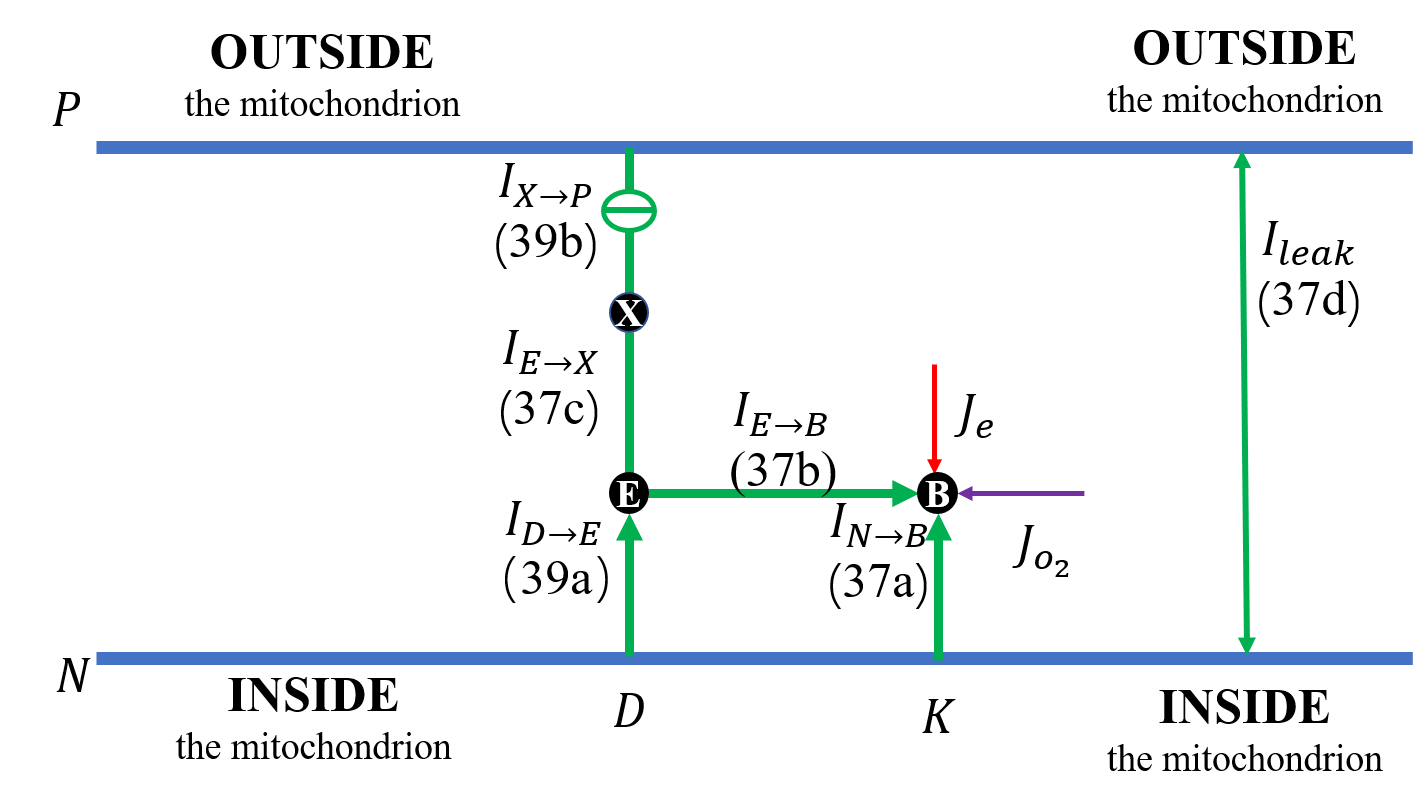}}
			\end{subfigure}
			\caption{ Schematic for the proton flow pattern.  Equation numbers that define arrows are shown, e.g., eq. (38b) for $I_{X 2 P}$.  (a) Normal state;  (b) Backward flow state with Perfect switch  between N and E242; (c) Flow state with perfect switch between the proton Loading Site PLS and outside. }
			\label{fig:protonflow}
		\end{figure}

		Beyond a threshold,  the total current between E242 and PLS $I_{E2x}$ becomes negative, $provided$ the rectifier is located $between$ the inside and E site (blue lines with circles). The the protons leak back form PLS to E242. This `leak back' can be seen in Fig. \ref{fig:ratio_Ie_eq}  (d), where the positive ratio means means that the $I_{E2X}$ is negative (because $I_e$ is negative, with our sign conventions). Proton back flow from outside to the proton loading site PLS occurs in this case. The accumulated protons in E242 increase the chemical potential $\mu_E$ to be  greater than $\mu_N$ and $\mu_E$, which leads to more current from E242 to the reaction cite B (see Fig. \ref{fig:ratio_Ie_eq}  (f)) and back flow from reaction cite to N side (see Fig.\ref{fig:ratio_Ie_eq}  (h)). The rectifier blocks the direct back flow from E242 to the N side. The urrent $I_{N2E}$ becomes zeros (see Fig. \ref{fig:ratio_Ie_eq}  (g)).  In this case, the proton flow pattern is shown in Fig. \ref{fig:protonflow} (b).  
		
		The location of the rectifier is important. Behavior is different when the rectifier is moved. When the rectifier is between the protein loading site PLS and outside (red lines with squares), the backward flow from outside to PLS is blocked to be zero.  Then the current $I_{E2X}$ is also zero at the equilibrium, according to Eq. \eqref{HXquation}, which means $\mu_E = \mu_X$ as shown in Fig.  \ref{fig:ratio_Ie_eq} (b). The protons are still transported from inside to E242 then to BNC. $\frac{I_{N2E}}{I_e}=-0.68$  through D channel and directly to  BNC with ratio $\frac{I_{N2B}}{I_e}=-0.32$ through the K channel. Fig.   \ref{fig:protonflow}(c) shows  the proton flow pattern in this situation. 
		
		We suspect that the rectifier between the protein loading site PLS and outside is  closer to the real biology setup, because it blocks backward flow. For that reason, we mainly present the results with the rectifier  between PLS and outside. Our approach can of course handle almost any location or properties of the rectifier/switch once they are specified by experiment or higher resolution models.

		\subsection{Effect of  Proton Concentration }
		In this section, we study the effect of proton concentrations in the intermembrane space (outside) by increasing from the default value  $0.06 \mu M$ to $0.15 \mu M$.  
		
\begin{figure}[!ht]
			\centering
			\begin{subfigure}[]{
					\includegraphics[width=2.5in]{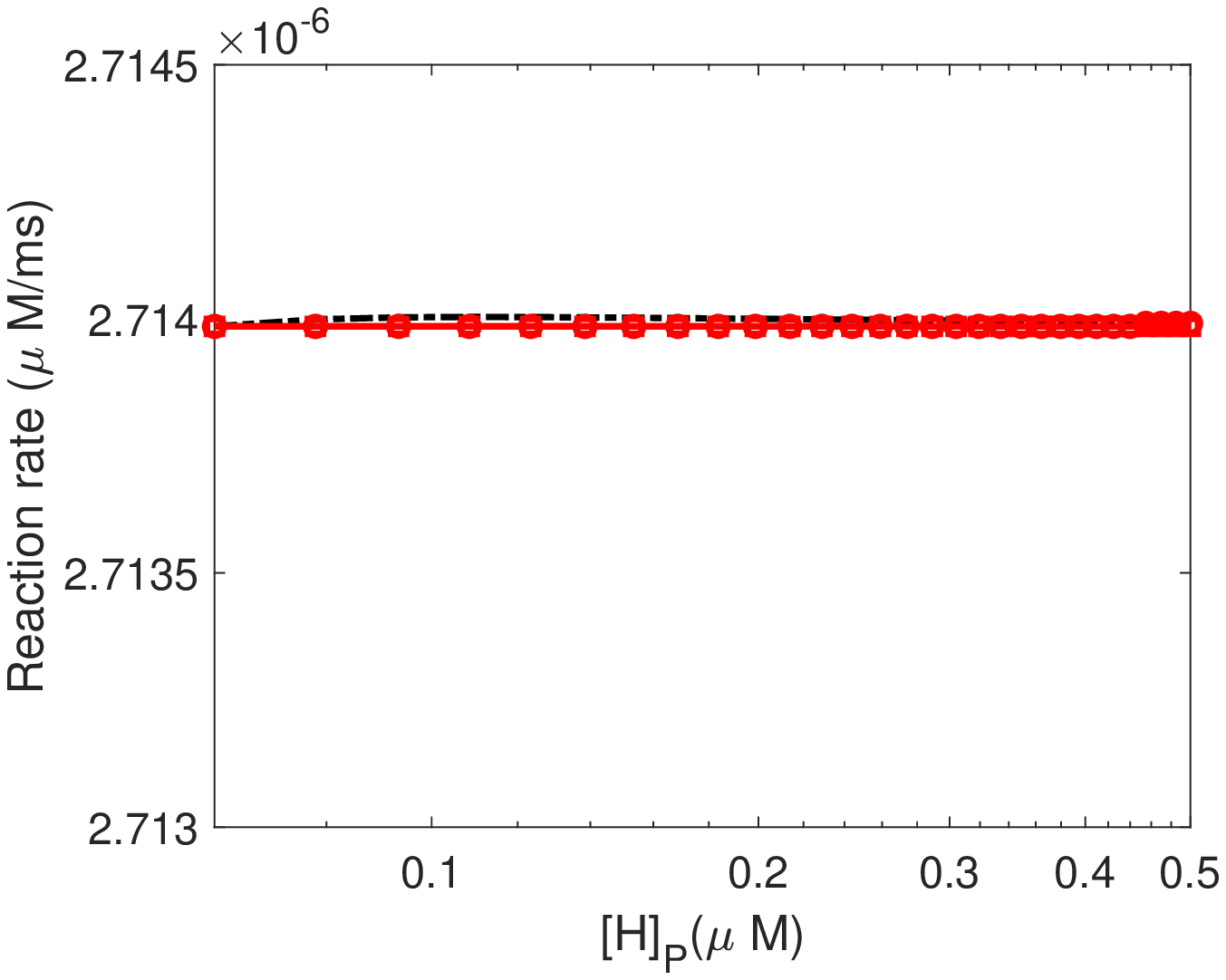}}
			\end{subfigure}
			\begin{subfigure}[]{
					\includegraphics[width=2.5in]{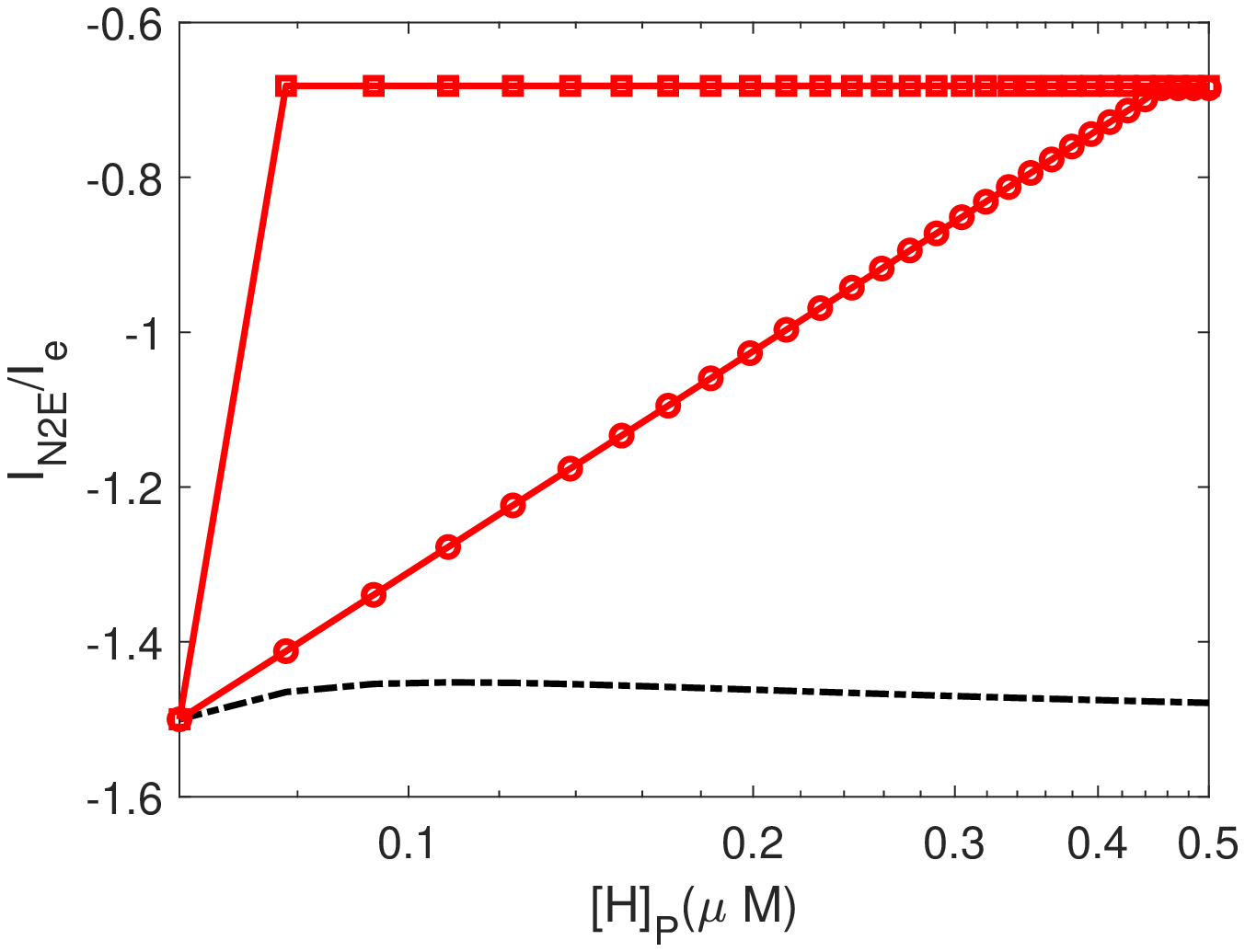}}
			\end{subfigure}
			\begin{subfigure}[]{
					\includegraphics[width=2.5in]{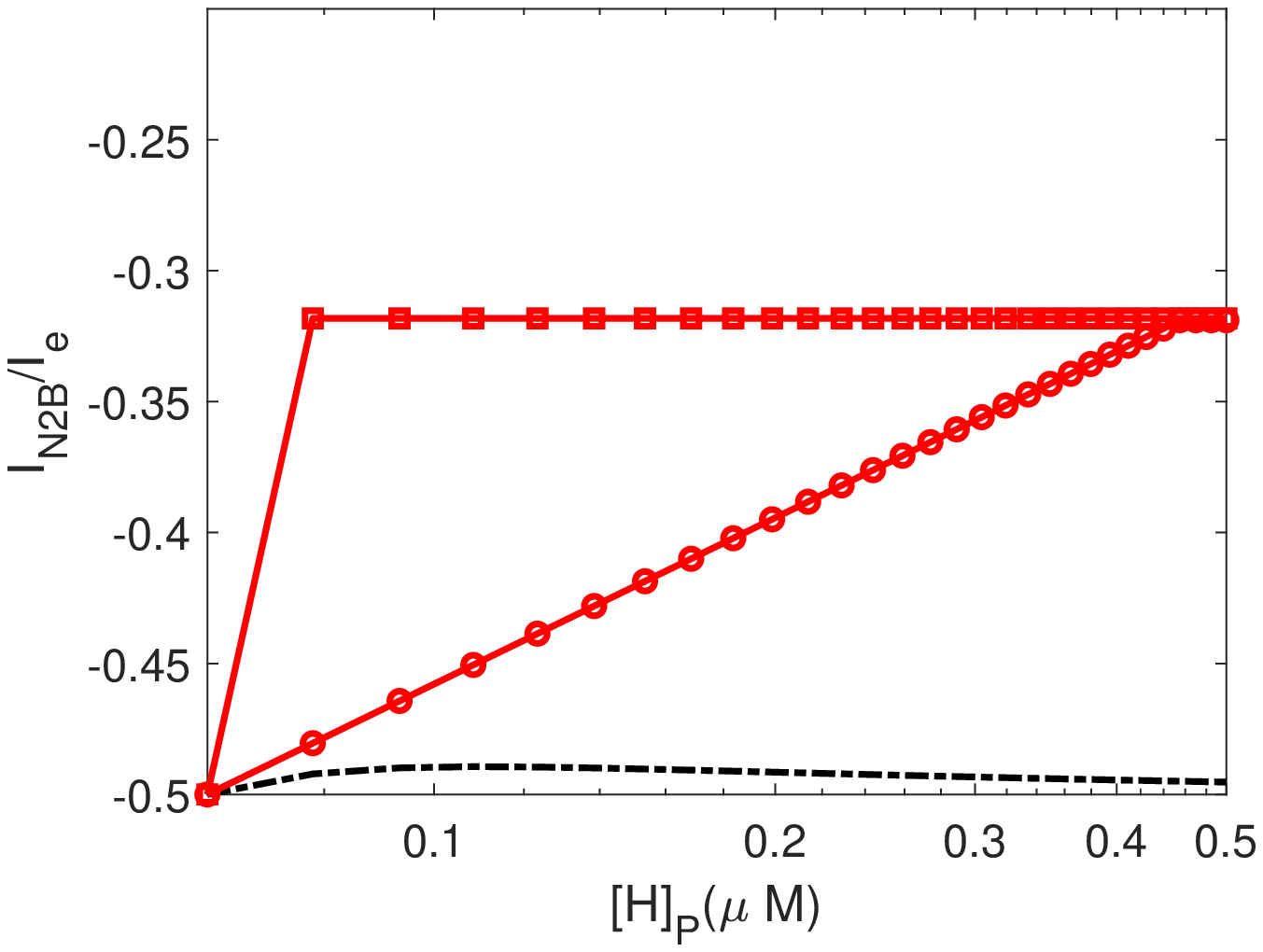}}
			\end{subfigure}
			\begin{subfigure}[]{
					\includegraphics[width=2.5in]{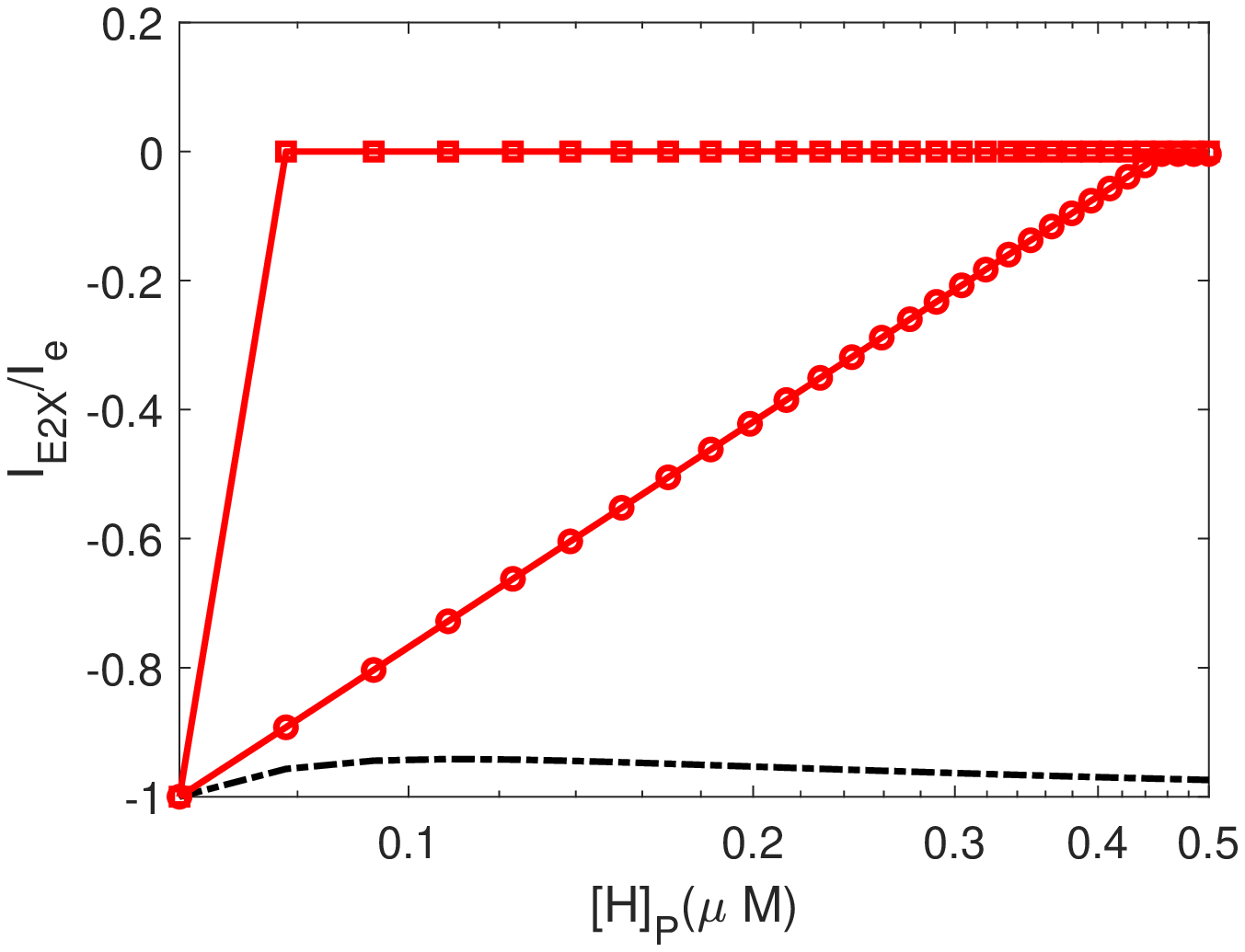}}
			\end{subfigure}
			\begin{subfigure}[]{
					\includegraphics[width=2.5in]{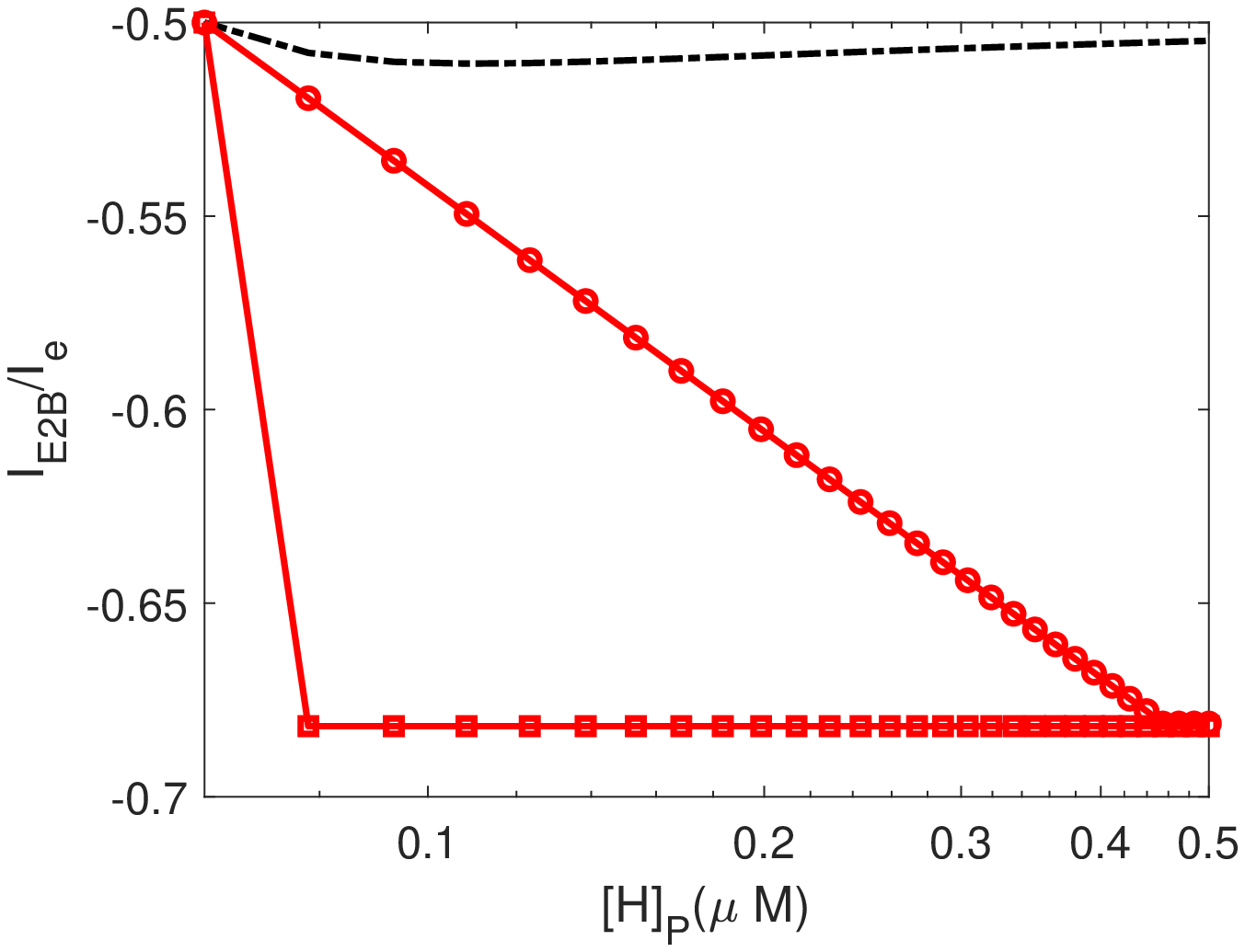}}
			\end{subfigure}
			\begin{subfigure}[]{
					\includegraphics[width=2.5in]{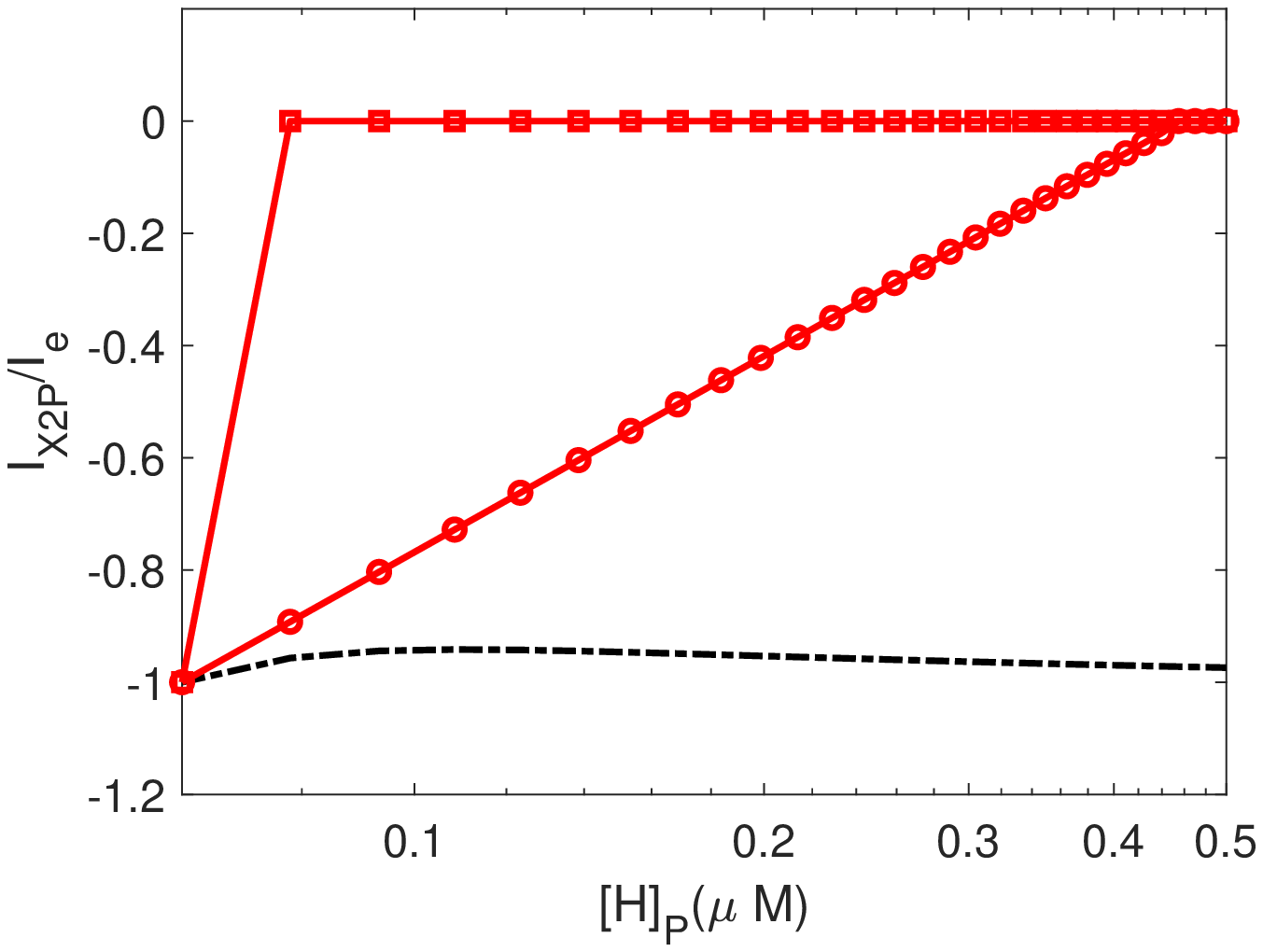}}
			\end{subfigure}
			\caption{Pump efficiency at equilibrium states with different proton concentration and different leak conductance. (a) Reaction rate;  (b)   $I_{N2E}/I_e$; (c) $I_{N2B}/I_e$; (d) $I_{E2X}/I_e$; (e) $I_{E2B}/I_e$; (f) $I_{X2P}/I_e$.  Black dash line: $g_m =0$; Red line with circle: $g_m=10^{-5}$;  Red line with square: $g_m = 10^{-3}$.    }
			\label{fig:Pump_p_eq}
		\end{figure}

\begin{figure}[!ht]
			\centering
			\begin{subfigure}[]{
					\includegraphics[width=3.in]{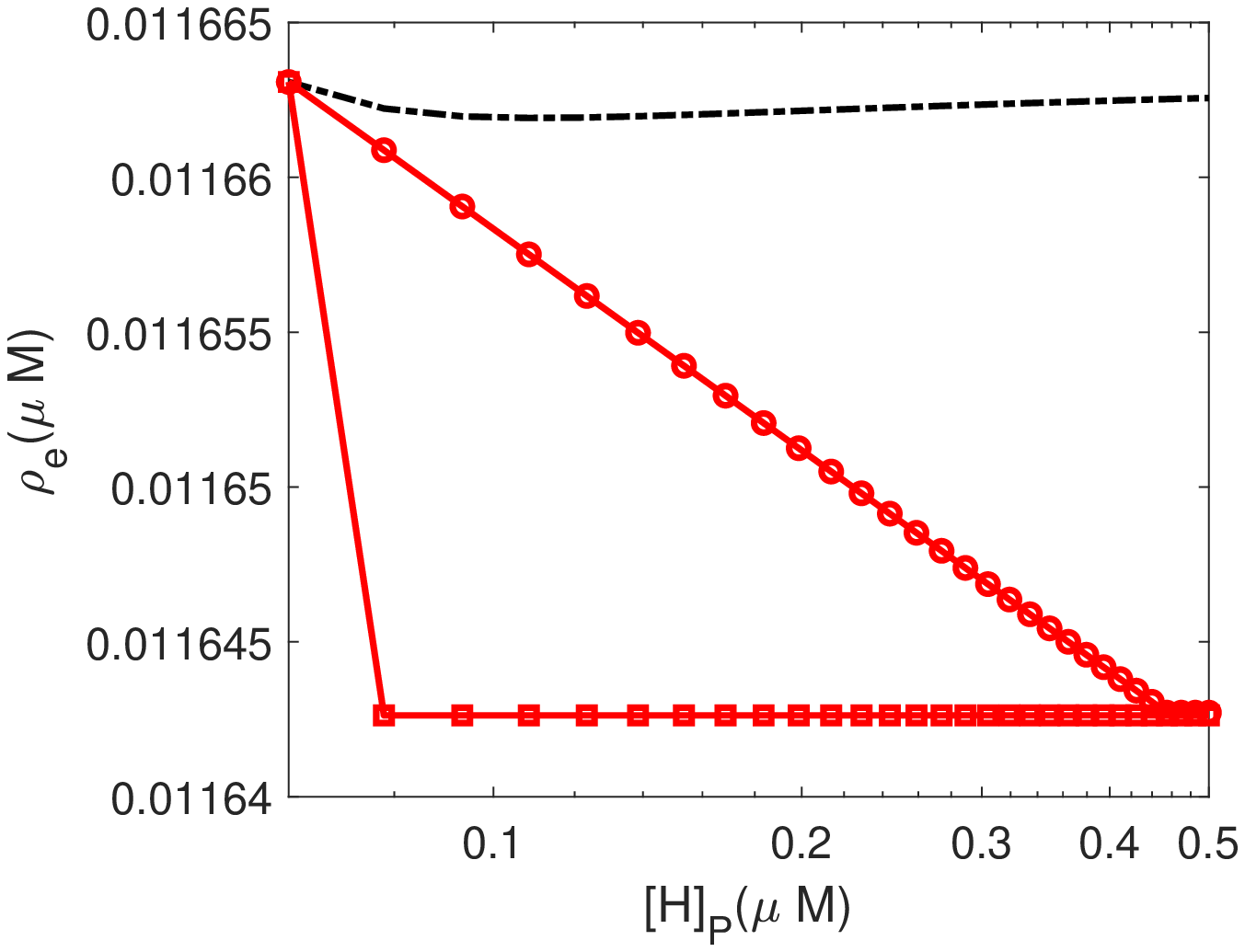}}
			\end{subfigure}
			\begin{subfigure}[]{
					\includegraphics[width=3.in]{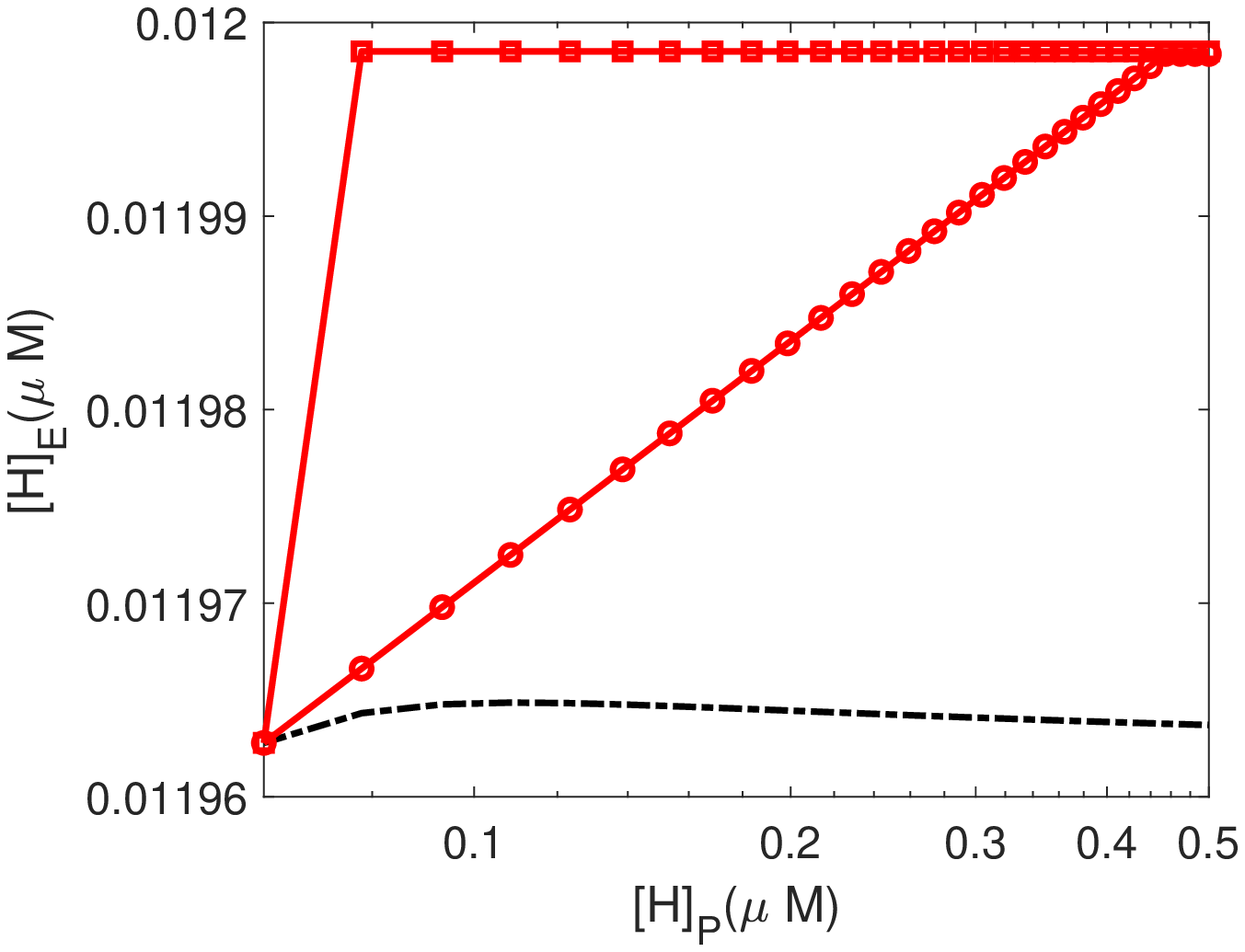}}
			\end{subfigure}
			\begin{subfigure}[]{
					\includegraphics[width=3.in]{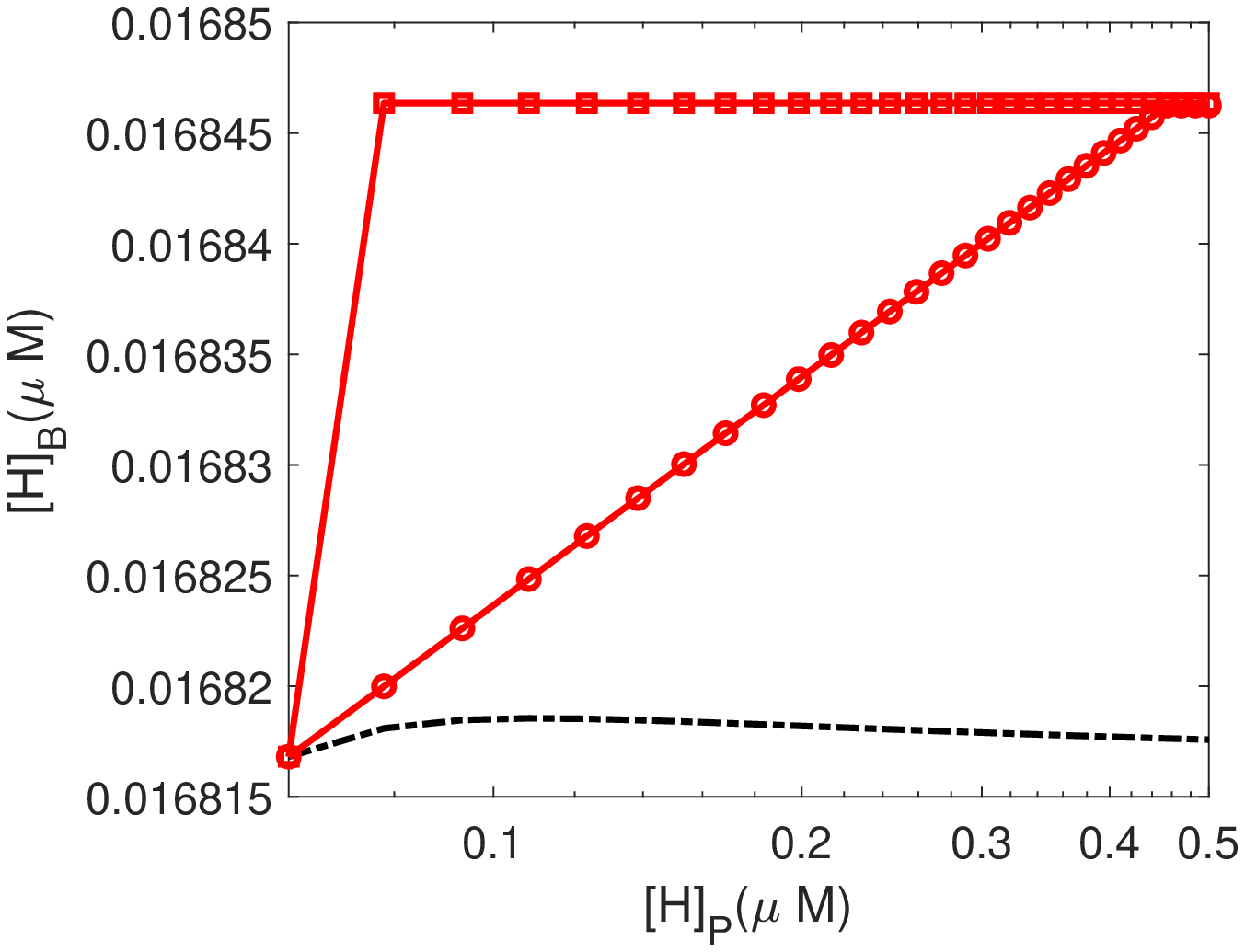}}
			\end{subfigure}
			\begin{subfigure}[]{
					\includegraphics[width=3.in]{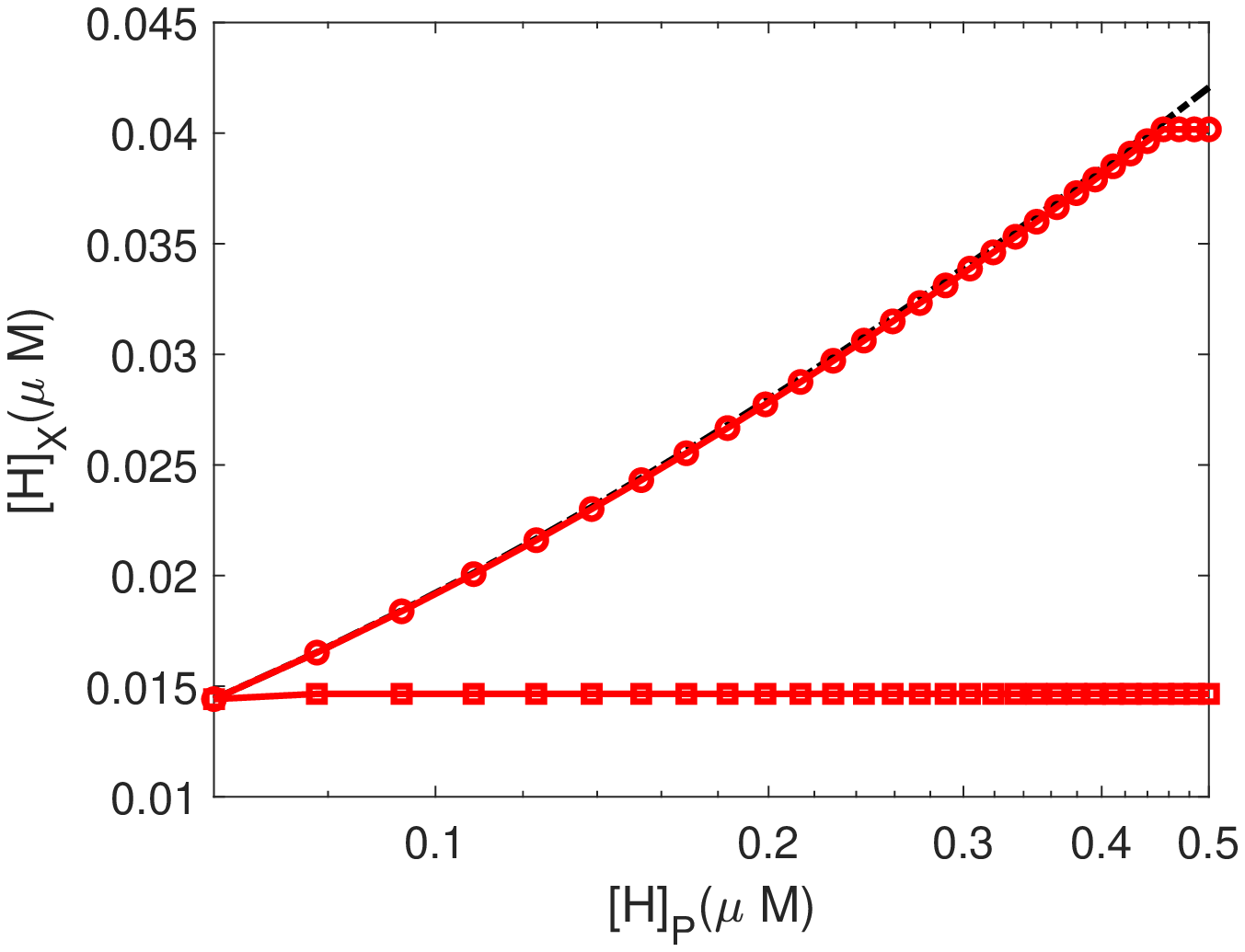}}
			\end{subfigure}
			\caption{Concentration at equilibrium states with different proton concentration and different leak conductance. (a) Electron concentration $\rho_e$;  (b) $[H]_E$; (c) $[H]_B$;(d) $[H]_X$.  Black dash line: $g_m =0$; Red line with circle: $g_m=10^{-5}$;  Red line with square: $g_m = 10^{-3}$.  }
			\label{fig:Concentration_p_eq}
		\end{figure}

		Figs. \ref{fig:Concentration_p_eq} and  \ref{fig:Pump_p_eq} show the equilibrium states of concentrations and pump efficiency at different proton concentrations  with difference leak conductance $g_m$.  
		
		First, Fig. \ref{fig:Pump_p_eq} a. illustrates the reaction rate with different $[H]_P$ keeps a constant since $\mathcal{R} = \frac{-S_v}{2F}I_e$ at equilibrium.
		When the leak conductance is zero, the complex IV efficiency does not change, i.e. $\frac{I_{X2P}}{I_e} =1$ and the flow pattern is shown in Fig.\ref{fig:protonflow}a.  When the shunt conductance $g_m$ is larger than zero,  the pump resistance increases with the outside proton concentration. This produces a decrease of the complex IV efficiency all the way to zero after the threshold. Then the proton pattern is the same as  in Fig.\ref{fig:protonflow}c, where all the protons pumped from inside through D and K channels are consumed by the reaction at BNC. 
		
		The concentrations of electrons and protons at $E$ and $B$ sites  are small perturbations in all cases. The concentration in the PLS is almost a constant with different $[H]_P$ when the leak conductance is large.  However, it  increases tremendously when the leak conductance is small. Large leak conductance simulates voltage clamp conditions, which do not describe the normal functional state of the mitochondrion. Small leak conductance presumably corresponds to the natural state in which the sum of all currents across the mitochondrion is `clamped' to zero, by Kirchhoff's current law, because there is nowhere else for the current to flow.

		\subsection{Kirchhoff clamp }
		
		
		Most of this paper describes cytochrome c oxidase embedded in a mitochondrion approximating a preparation without other members of the respiratory chain, but with otherwise normal properties. The mitochondrion is a small cell, as it were, in which the interior potential is unlikely to vary substantially with macroscopic location, on the micron scale, because the cell is much smaller than the length constant of cable theory. In such a system, Kirchhoff’s current law ensures that the sum of all currents across the mitochondrial membrane is zero. The currents are necessarily coupled by electrodynamics, whether or not they are coupled by chemistry. If one current increases, the sum of the others must decrease. A graph of one current against another will give a definite ratio, a coupling ratio, if you will, when other variables are held constant. 
		The various currents are coupled by the electric potential. The electrical potential contributes to the forces that drive the currents. Chemical reactions may contribute as well. But even without chemical reactions, coupling can occur through the electric field. While this may seem strange in context of classical transport biophysics, it is well precedent ed and understood in channel biophysics, that presumably follows the same laws of physics. The coupling of sodium and potassium conductances that allow the action potential to propagate are an example of coupling by the electric field, without chemical interaction of the underlying protein molecules, as we have discussed previously in this paper.
		
		Coupling occurs because the electric field adopts the values that conserve current. That is easy to prove from the Maxwell equations \cite{RN28898,RN30138}. 
		In fact, the conservation of current is a form of Kirchhoff’s law, so currents are clamped to one another (i.e., coupled) in a “Kirchhoff clamp” if we want to coin a phrase for what is really just the UNclamped, natural situation. 
		
		If the electrical potential is controlled, so it is not free to adopt the value that conserves current, a different situation occurs altogether. This situation is called a voltage clamp in electrophysiology, and was invented by Cole and used by Hodgkin and Huxley to understand the mechanism of the action potential. The voltage clamp loosens the Kirchhoff clamp because it has an amplifier (that is outside the biological system) to supply current and energy. Indeed, the Kirchhoff clamp of the natural mitochondrion is entirely removed by the currents supplied by the voltage clamp amplifier.
		
		In the voltage clamp, one current is not coupled to another current by the voltage. They cannot be, because the voltage does not vary with the current. The result is that coupling and flux ratios reflect chemical coupling, not voltage coupling, in the voltage clamp setup. The result is that nearly every experimental result is different in a voltage clamp and the natural UNclamped situation. 
		
		The voltage clamp was invented to gain experimental control of currents so they can be studied as Cole made abundantly clear, followed by Hodgkin and Huxley. But \textbf{the voltage clamp is not natural}. It removes a natural form of flux coupling. Flux coupling by the electric field is absent. Flux coupling by the electric field is natural, just as natural in the mitochondrion as in the nerve, just as natural in the generation of ATP as it is in the generation of the nerve signal. 
		
		It is difficult to voltage clamp mitochondria, and preparations reconstituted into bilayers (that can be voltage clamped) have other difficulties that experimentalists often wish to avoid. Other methods are used to simulate a voltage clamp, quite well, as it turns out. 
		
		In work on mitochondria, voltage clamp is usually produced indirectly, by artificially increasing the leak conductance. An effective carrier of potassium current like valinomycin is often added to solutions. When valinomycin is present in large enough concentrations, it partitions into the mitochnondrial membrane, and the leak conductance dominates. The potential across the mitochondrial membrane is set by the equilibrium  potential of the leak conductance. If valinomycin is used to increase the leak conductance, the potential is in fact close to the potassium equilibrium potential, independent of current because valinomycin is remarkably selective for potassium ions. Valinomycin clamps the potential to the potassium equilibrium potential.  
		
\begin{figure}[!ht]
			\centering
			\begin{subfigure}[]{
					\includegraphics[width=3.in]{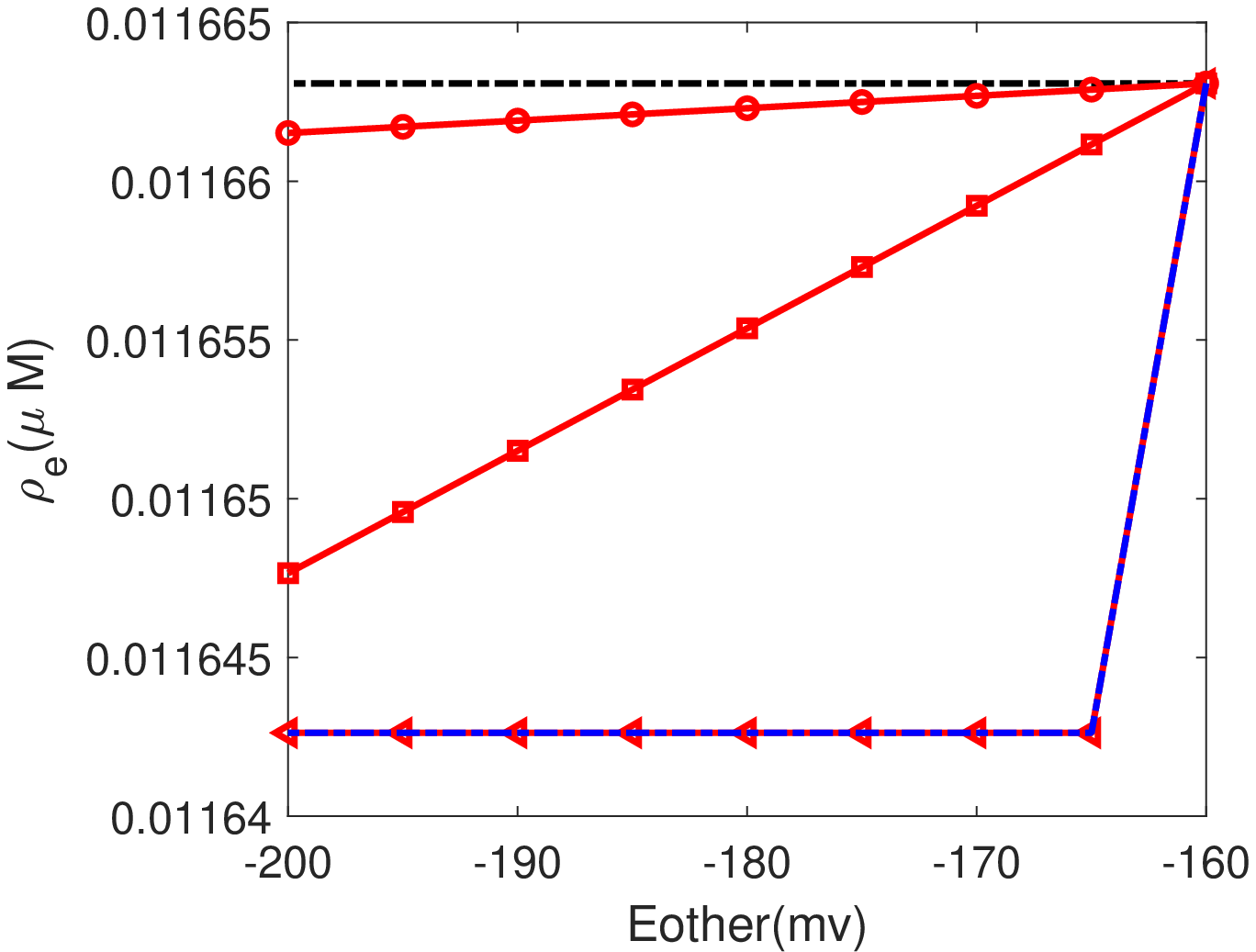}}
			\end{subfigure}
			\begin{subfigure}[]{
					\includegraphics[width=3.in]{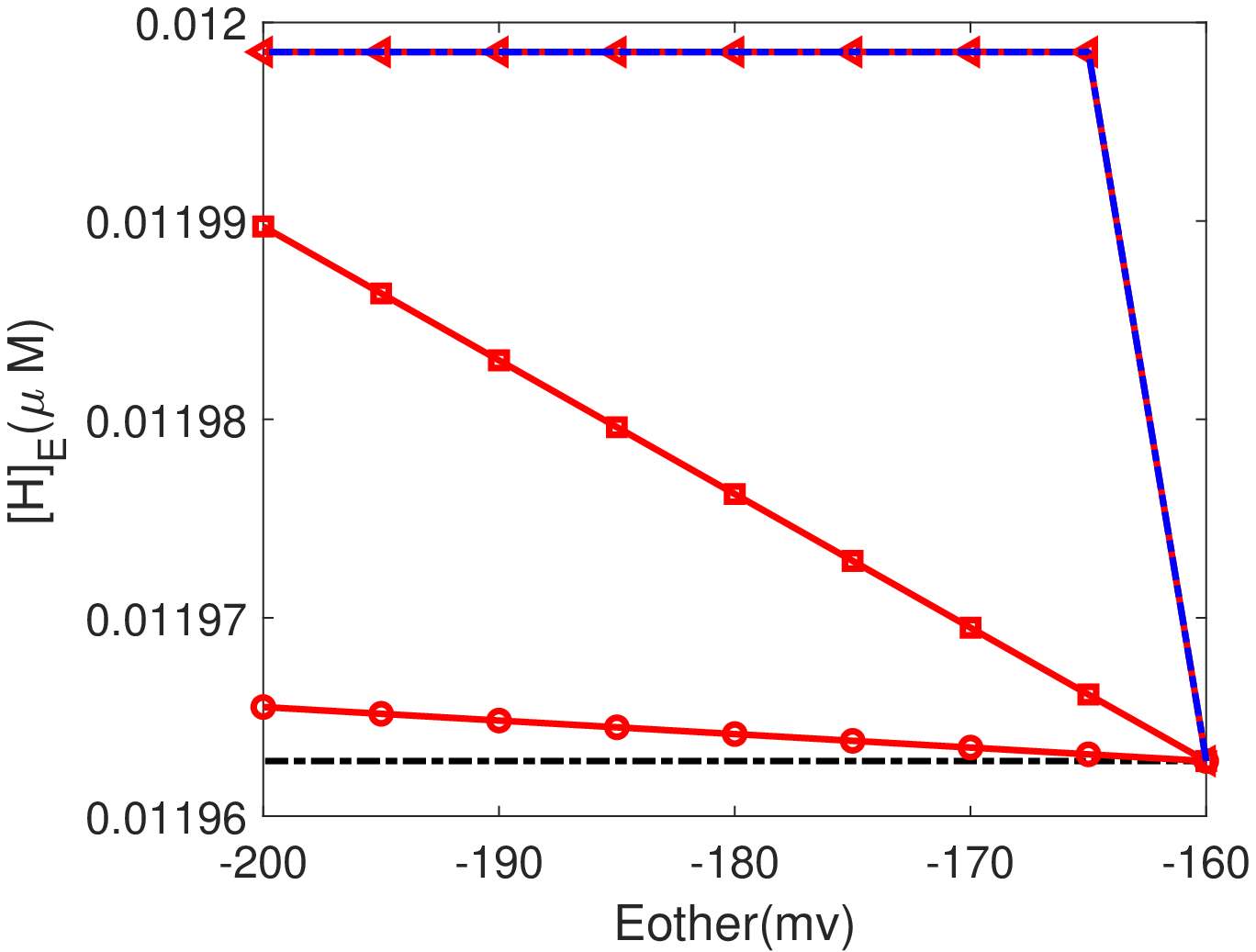}}
			\end{subfigure}
			\begin{subfigure}[]{
					\includegraphics[width=3.in]{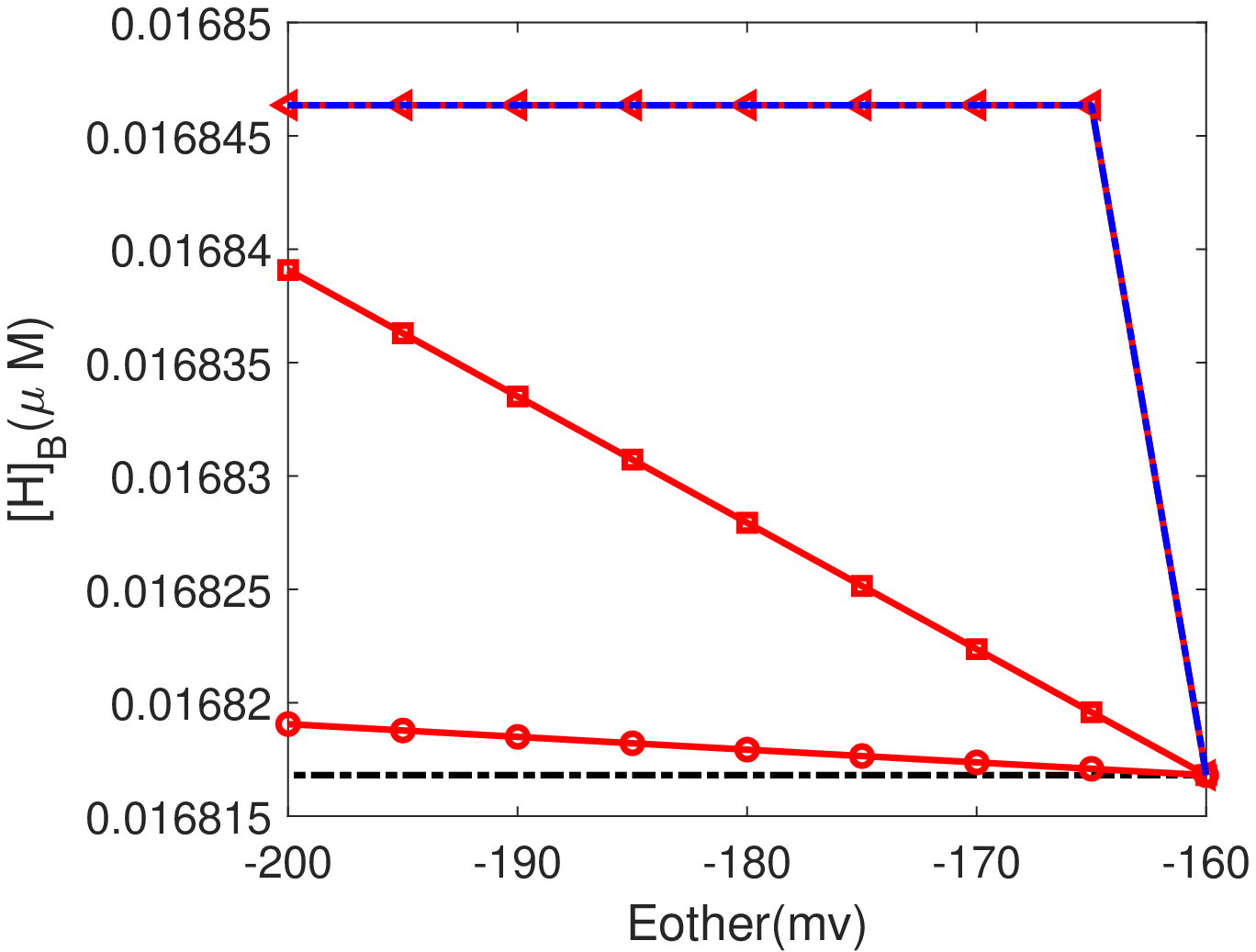}}
			\end{subfigure}
			\begin{subfigure}[]{
					\includegraphics[width=3.in]{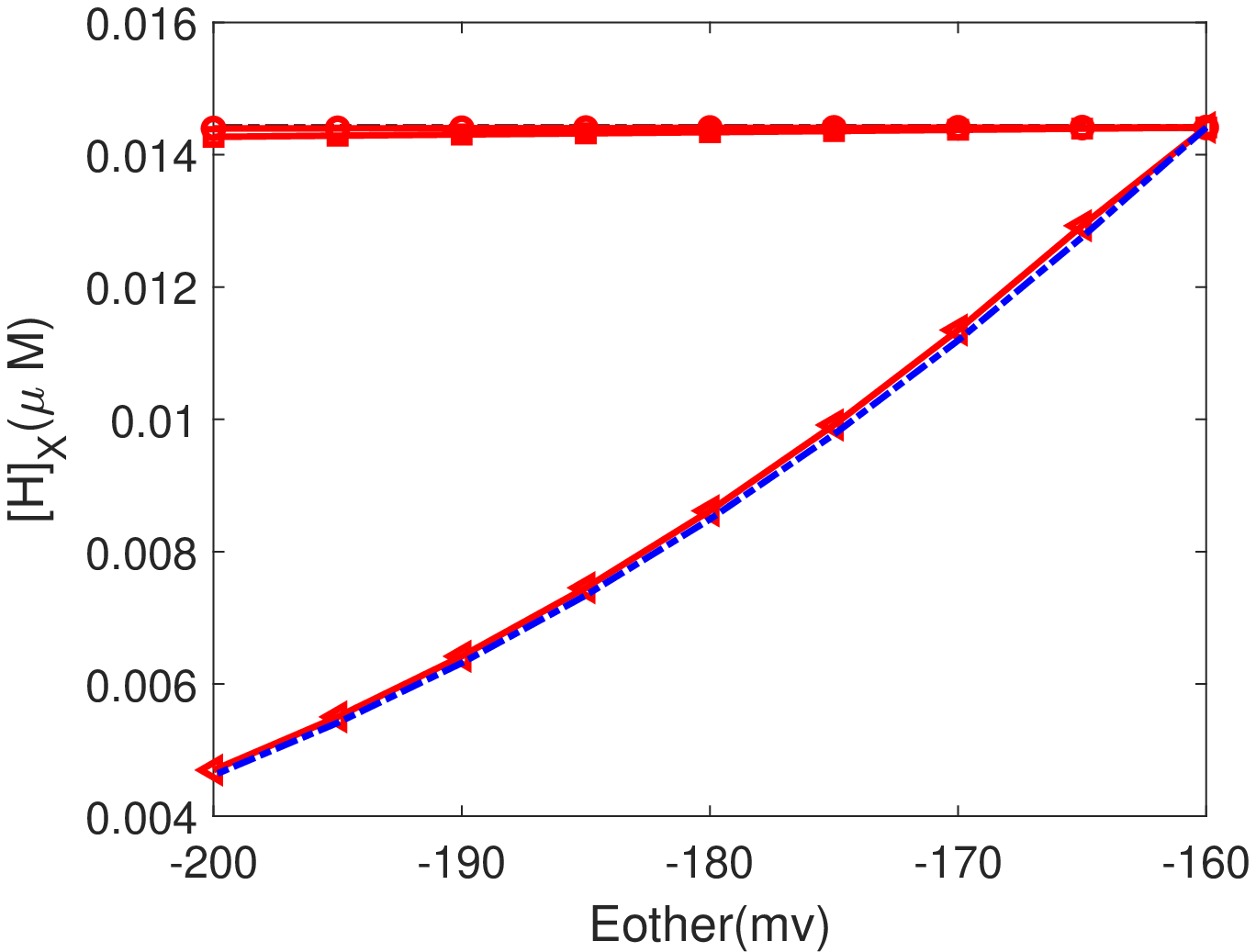}}
			\end{subfigure}
			\caption{Concentration at equilibrium states with different $E_{other} $and   $g_m$. (a) Electron concentration $\rho_e$;  (b) $[H]_E$; (c) $[H]_B$;(d) $[H]_X$.  Black dash line: $g_m=0$; Red line with circles: $g_m =10^{-6}$; Red line with squares: $g_m =10^{-5}$; Red line with triangles: $g_m =10^{-3}$; Blue dash lines: Voltage clamp.}
			\label{fig:Concentration_Eother_gm_eq}
		\end{figure}

\begin{figure}[!ht]
			\centering
			\begin{subfigure}[]{
					\includegraphics[width=3.in]{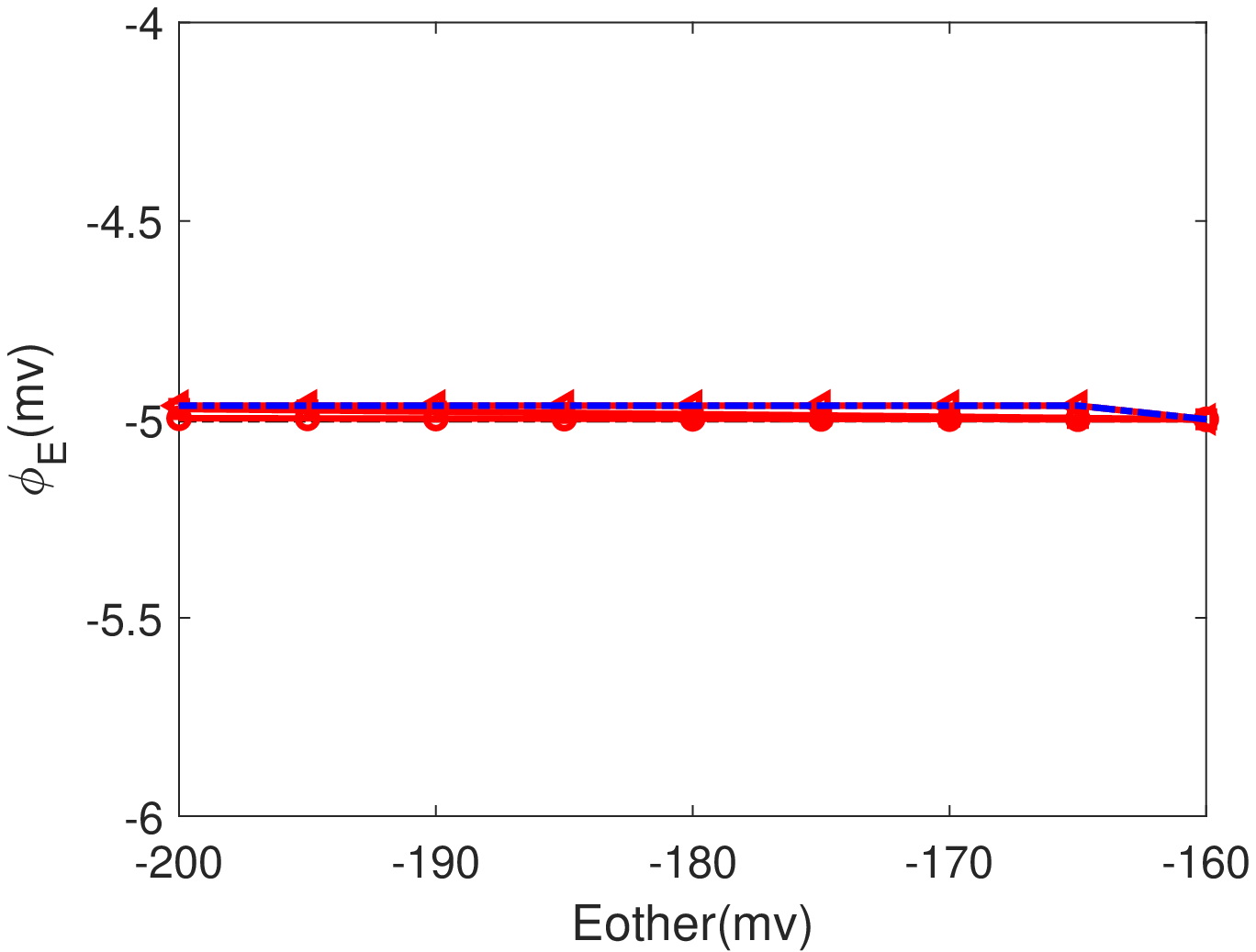}}
			\end{subfigure}
			\begin{subfigure}[]{
					\includegraphics[width=3.in]{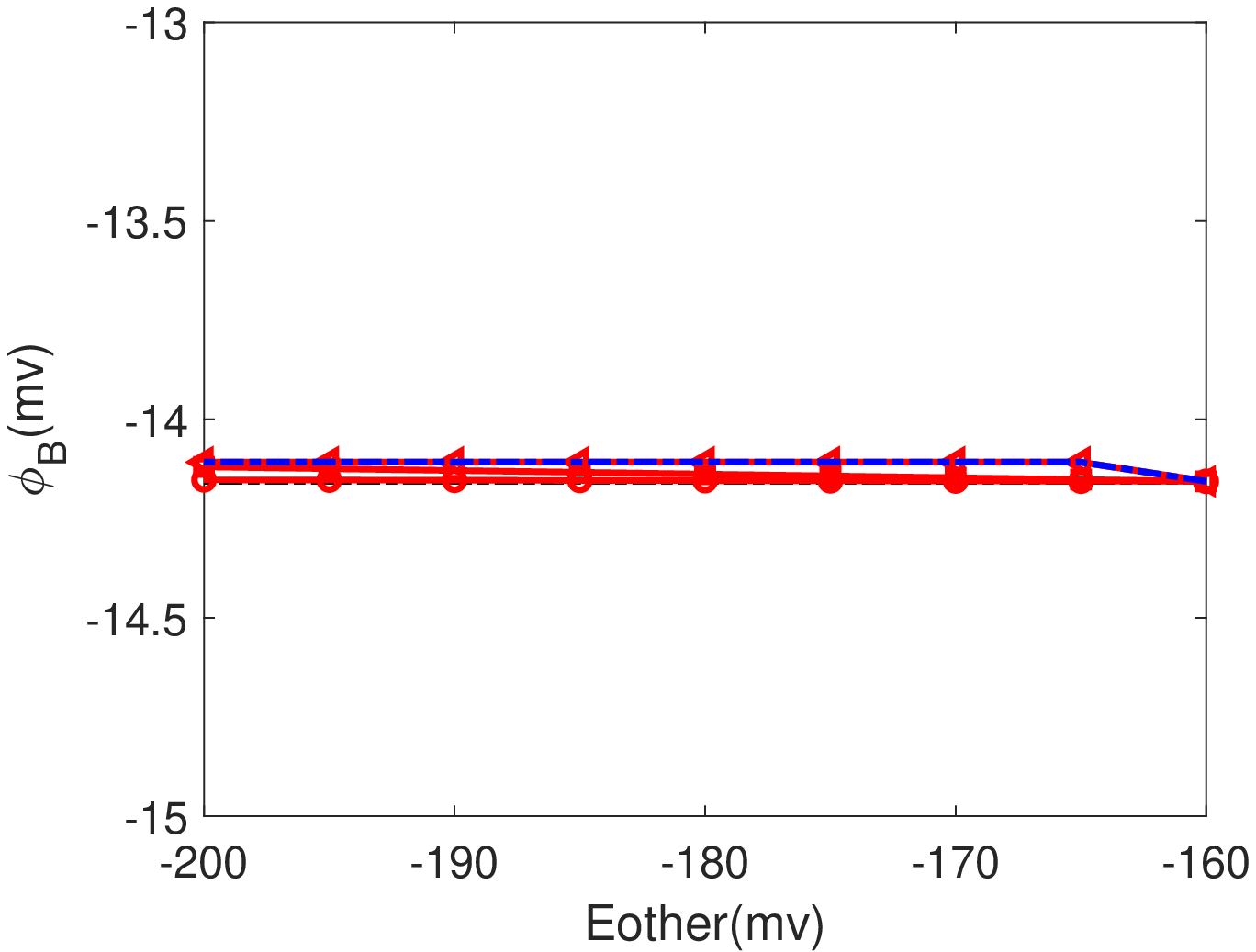}}
			\end{subfigure}
			\begin{subfigure}[]{
					\includegraphics[width=3.in]{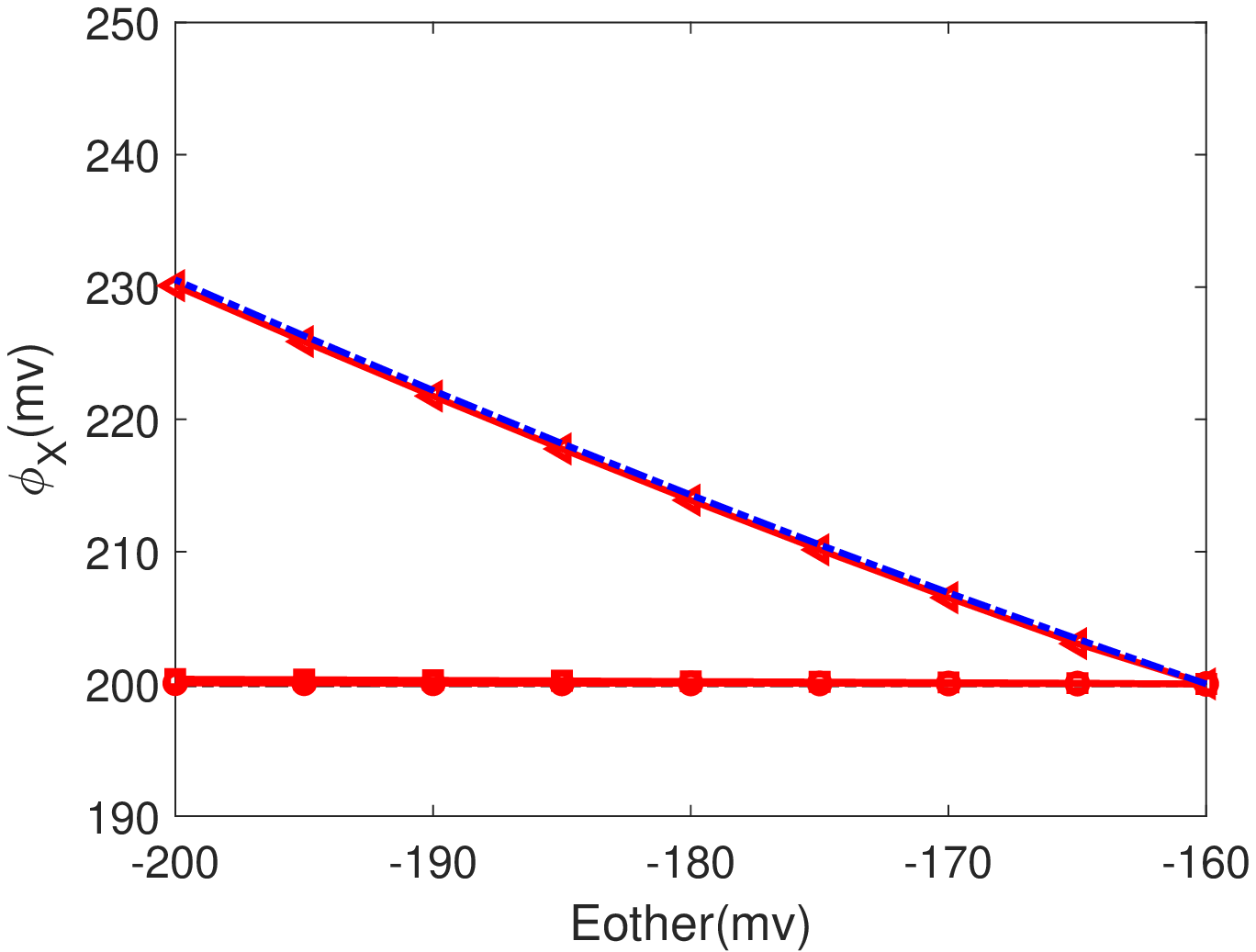}}
			\end{subfigure}
			\begin{subfigure}[]{
					\includegraphics[width=3.in]{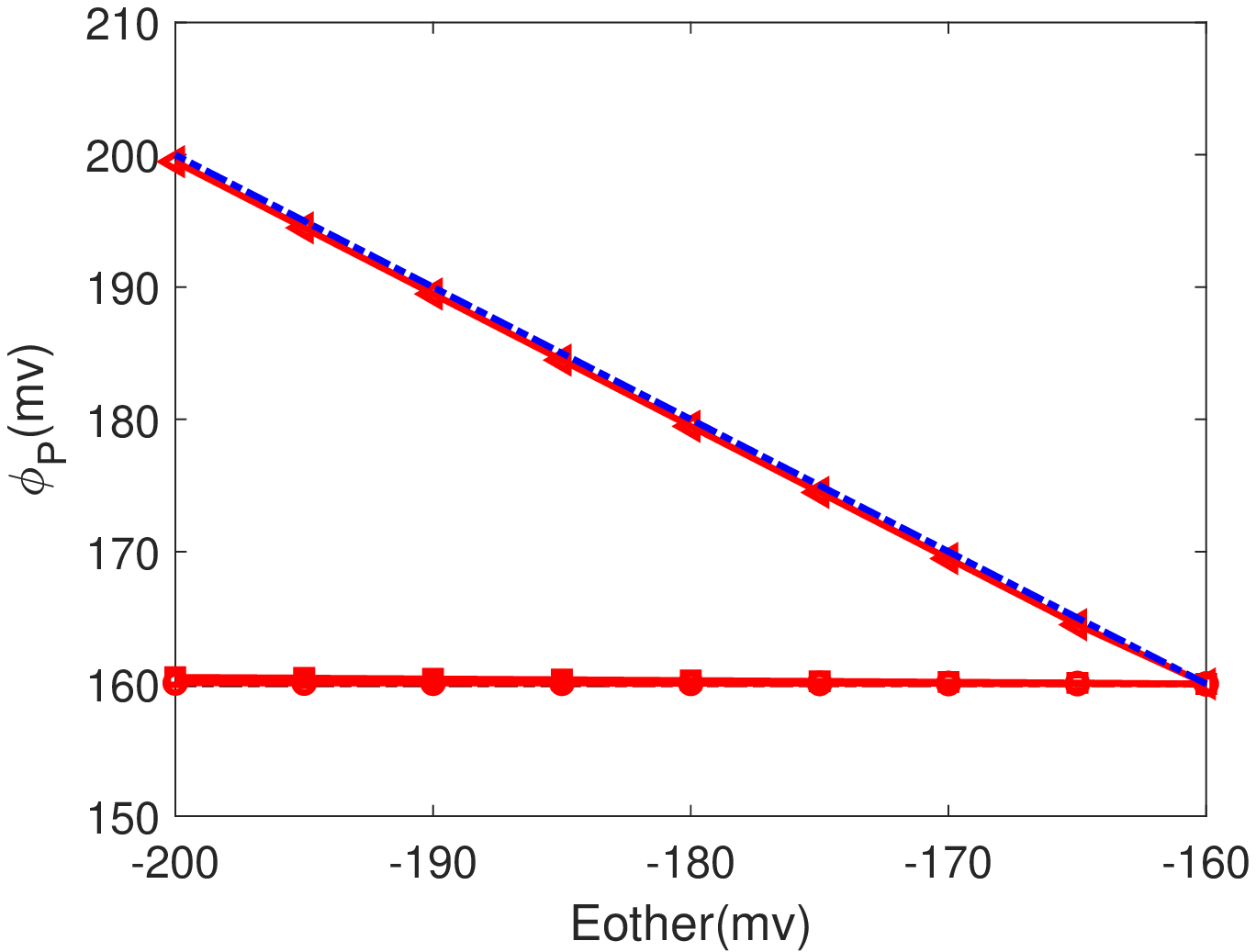}}
			\end{subfigure}
			\caption{Electric potential at equilibrium states with different $E_{other} $and   $g_m$.  (a)   $\phi_E$;  (b) $\phi_B$; (c) $\phi_X$;(d) $\phi_P$. Red line with squares: $g_m =10^{-5}$; Red line with triangles: $g_m =10^{-3}$; Blue dash lines: Voltage clamp.}
			\label{fig:phi_Eother_gm_eq}
		\end{figure}

\begin{figure}[!ht]
			\centering
			\begin{subfigure}[]{
					\includegraphics[width=3.in]{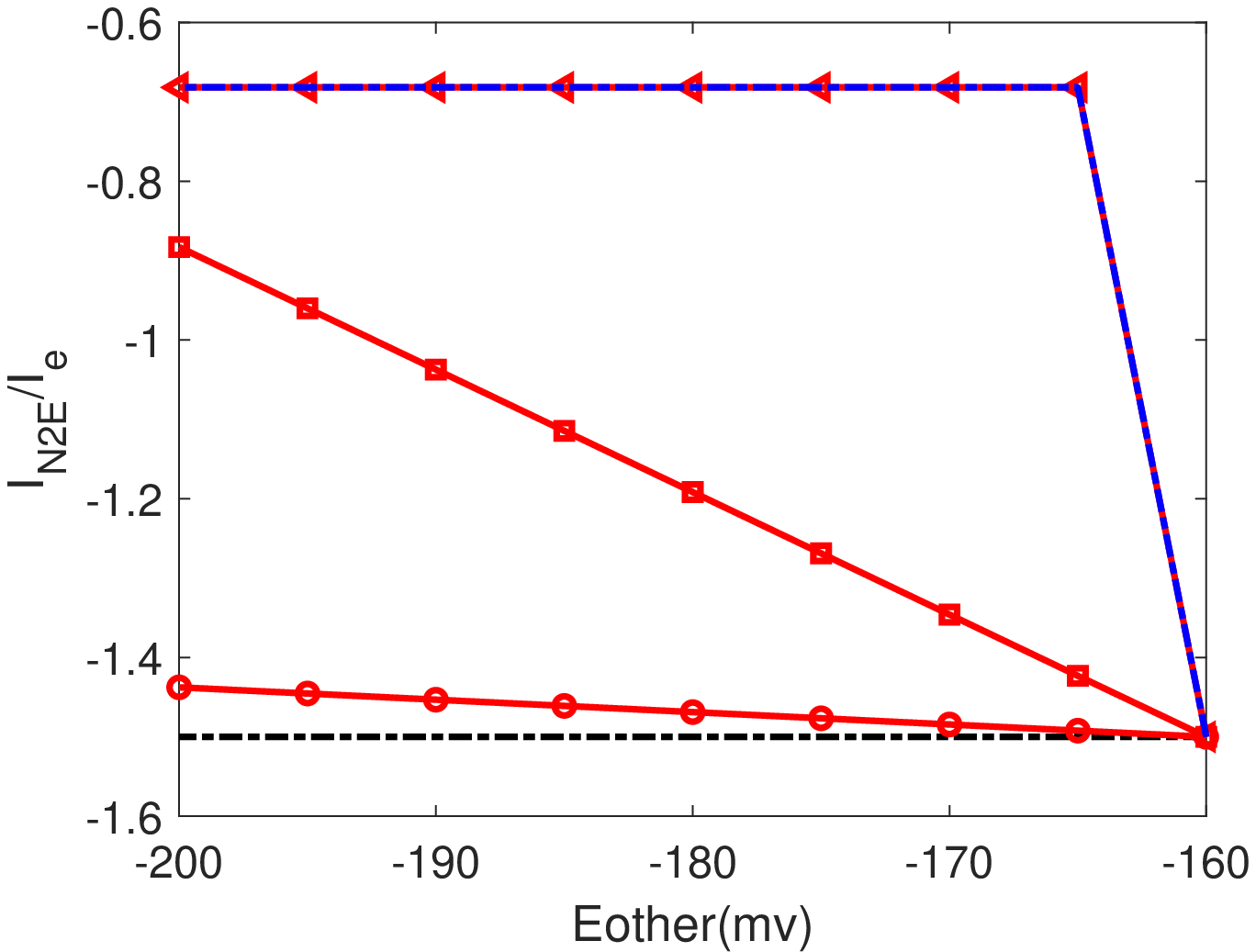}}
			\end{subfigure}
			\begin{subfigure}[]{
					\includegraphics[width=3.in]{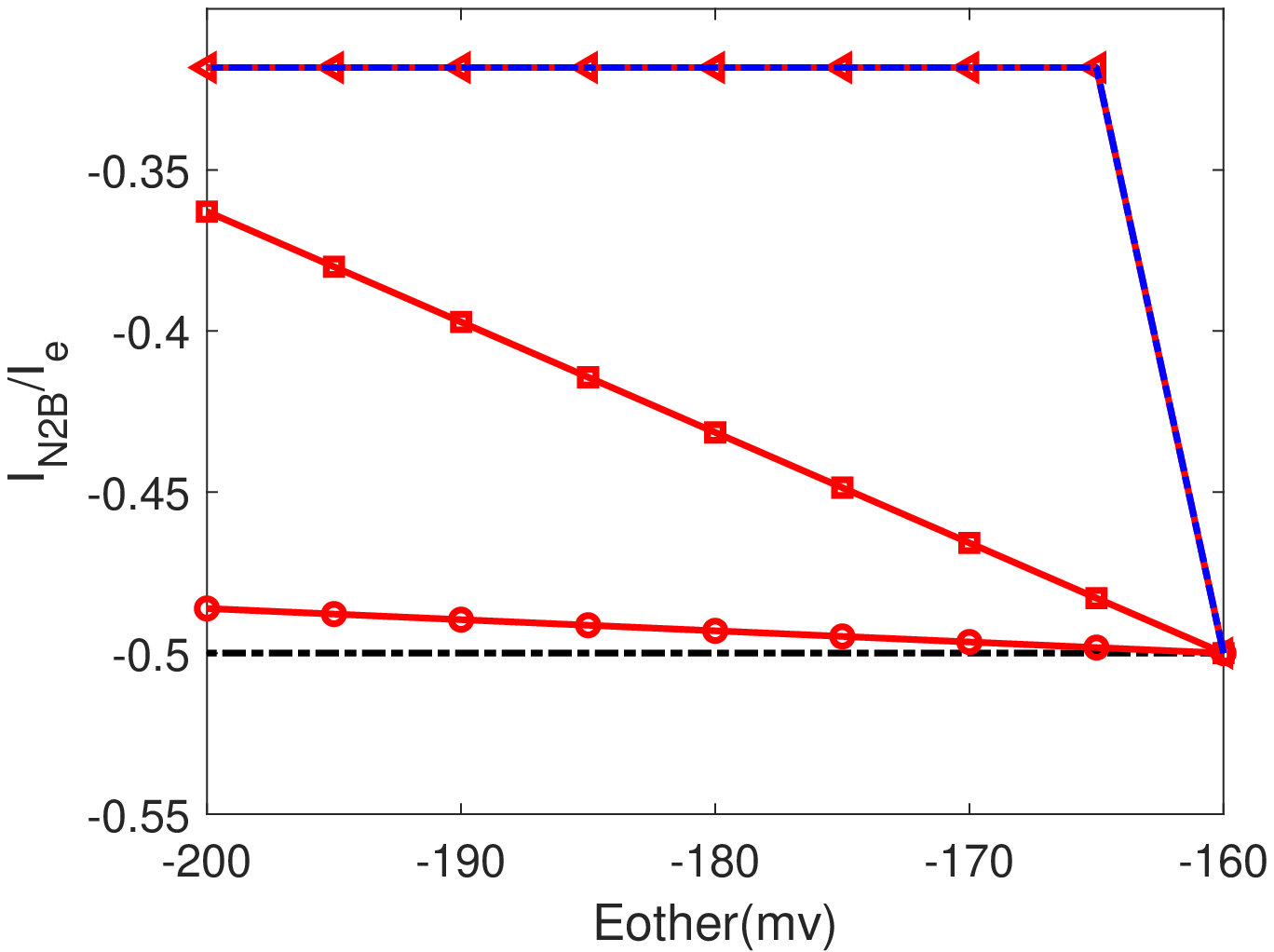}}
			\end{subfigure}
			\begin{subfigure}[]{
					\includegraphics[width=3.in]{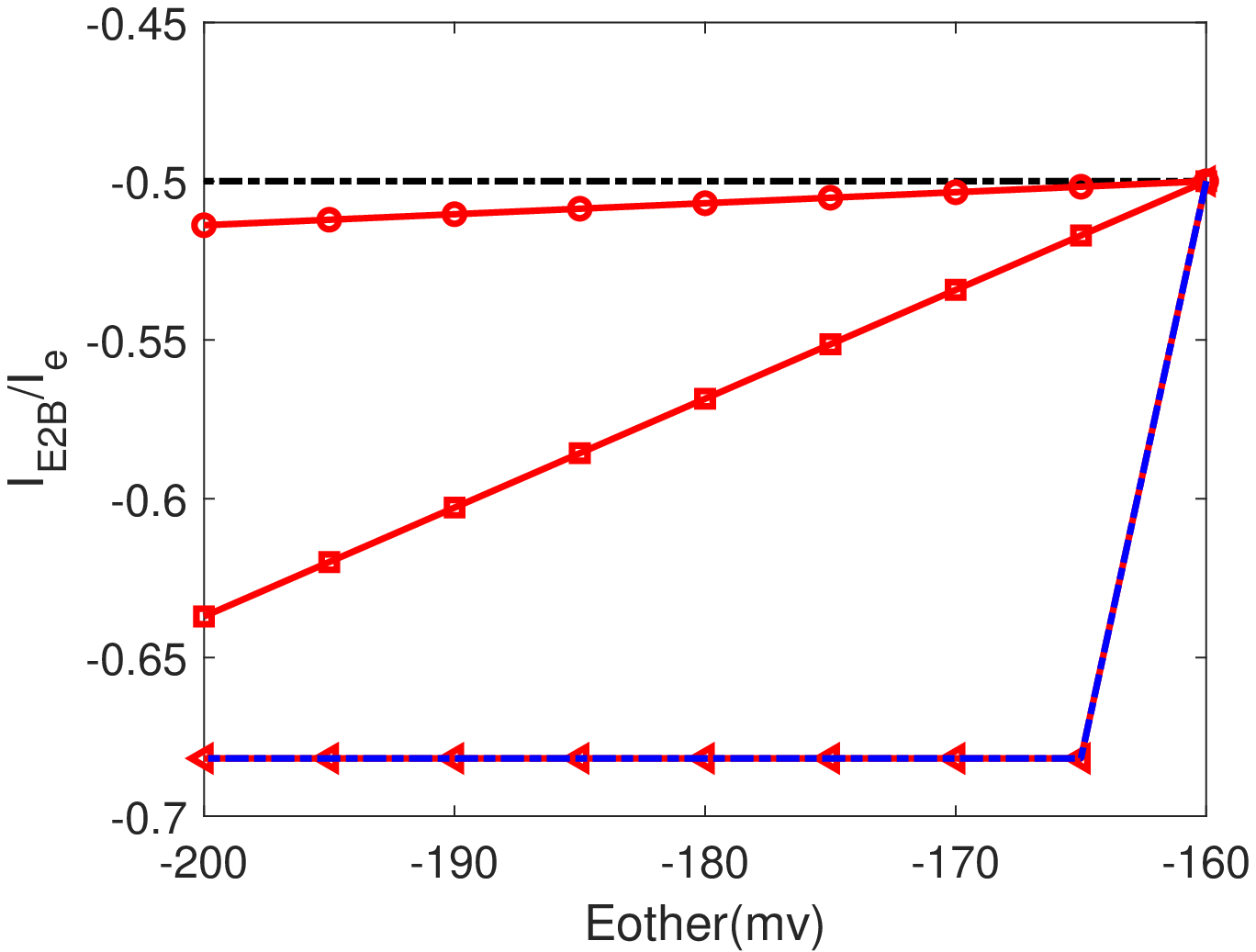}}
			\end{subfigure}
			\begin{subfigure}[]{
					\includegraphics[width=3.in]{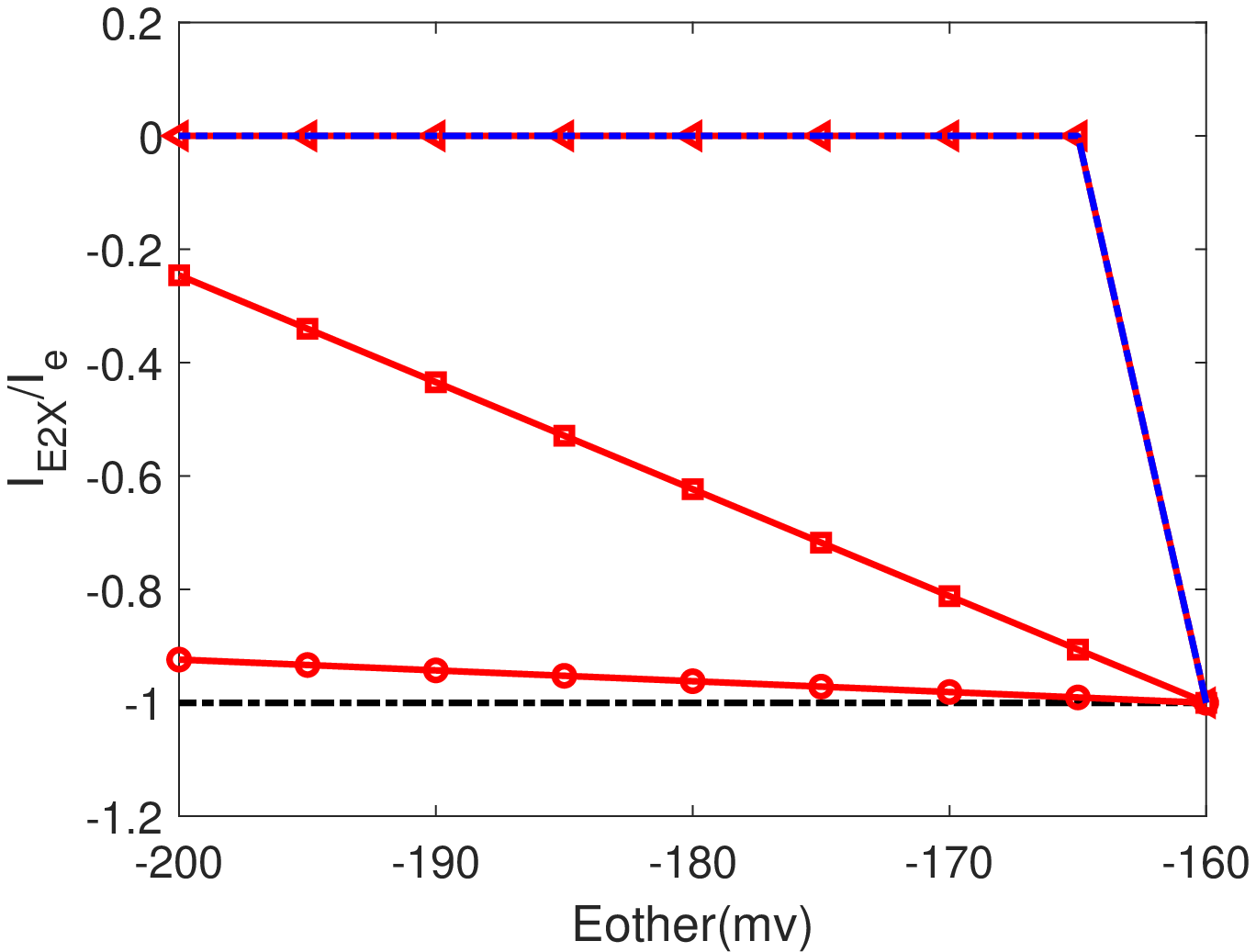}}
			\end{subfigure}
			\begin{subfigure}[]{
					\includegraphics[width=3.in]{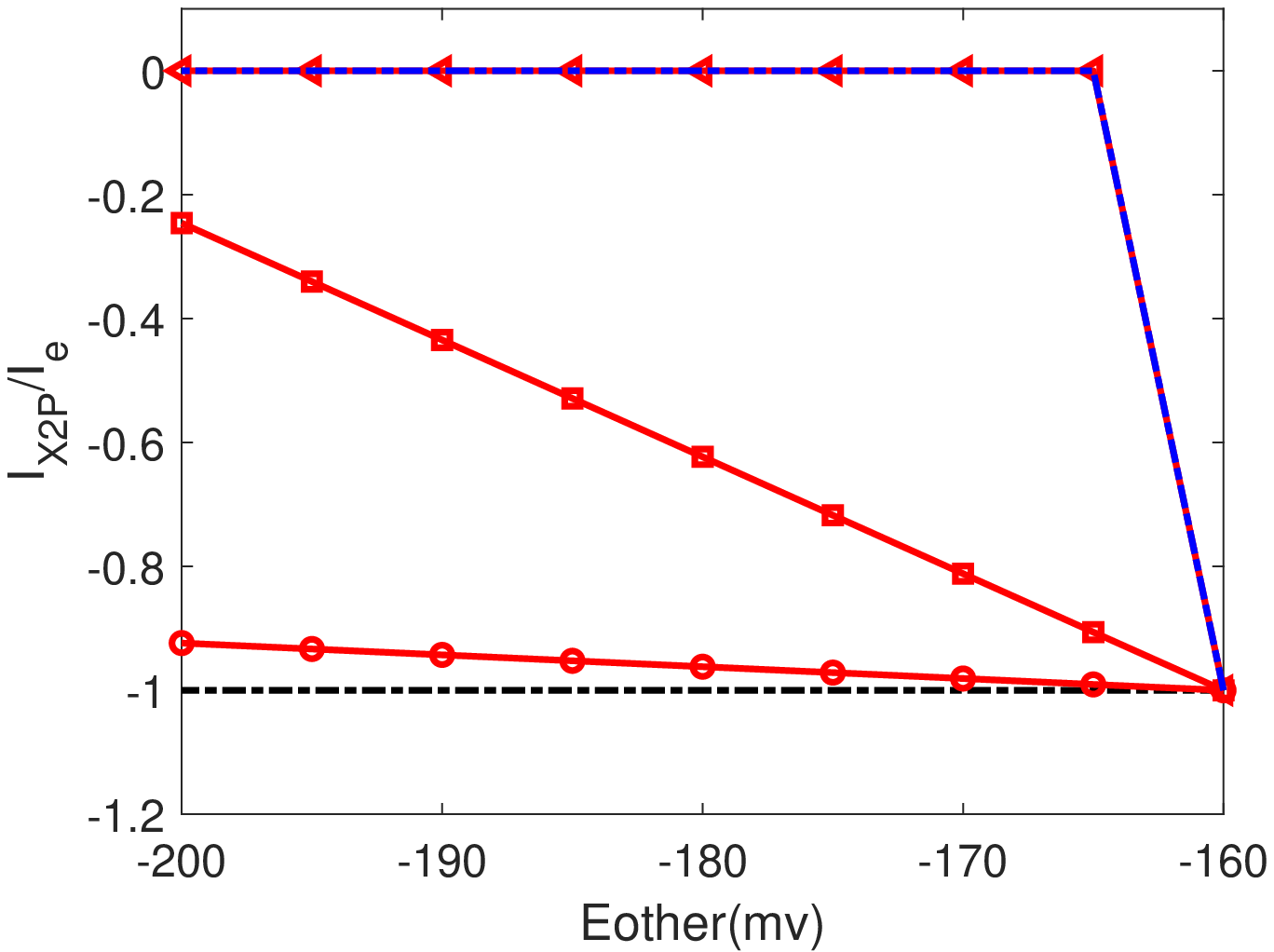}}
			\end{subfigure}
			\caption{Evolution of current   at equilibrium states with different $E_{other} $and   $g_m$.   (a) $I_{N2E}/I_e$; (b) $I_{N2B}/I_e$;  (c) $I_{E2B}/I_e$;  (d) $I_{E2X}/I_e$. (e) $I_{X2P}/I_e$. Red line with squares: $g_m =10^{-5}$; Red line with triangles: $g_m =10^{-3}$; Blue dash lines: Voltage clamp.}
			\label{fig:ratio_Eother_gm_eq}
		\end{figure}

		Figs. \ref{fig:Concentration_Eother_gm_eq}-\ref{fig:ratio_Eother_gm_eq} illustrate  the effect of Nernst potential $E_{other}$ on concentrations, electric potentials and currents under different leak conductance. $E_{other}$ is an approximation to the potassium equilibrium potential  The red lines with circles, squares and triangles are denotes the different leak conductance $g_m = 10^{-6}, 10^{-5}, 10^{-3}$, correspondingly. The black dash lines are the results by setting zero leakage, i.e. $g_m =0$.  And the blue dash lines denotes voltage clamp results  where the electric potential at inside $\phi_N$ is set to be zero and electric potential at outside $\phi_P = -E_{other} $ in system \eqref{current_model}. 
		
		First, consider the natural case, when shunt conductance $g_m=0$. Cytochrome $c$ oxidase is not affected. The efficiency of complex IV is not changed by the Nernst potential because $I_{leak}$ is always zero in this case. 
		When $g_m>0$, as the Nernst potential become more negative, the resistance for proton pumping increases  which leads to the decrease of proton pump efficiency. Of course, when the leak conductance is large enough, the system is nearly voltage clamped to the equilibrium potential for the leak. The Kirchhoff clamp (red lines with triangles) is unlocked, removed by the large leak conductance. The results are the same as the results of the voltage clamp (blue dash lines).
		Fig. \ref{fig:Concentration_Eother_gm_eq}(d),   Fig. \ref{fig:phi_Eother_gm_eq}(c)-(d),  and   Fig. \eqref{fig:ratio_Eother_gm_eq} show the difference in various quantities between voltage and UNclamped natural situations.

		The dramatic effect of antibiotics on membrane properties was studied by \cite{RN938}.
		Valinomycin has been studied in many other papers, which include \cite{RN46023,RN45770,RN27722,RN46059,RN46060,RN46061}.

		\newpage

			\section{Discussion and Conclusion}
			
			As history separated chemistry and field theory, so it separated chemical theory from devices. Almost all of chemical theory devalues the significance of boundary conditions and flows. Almost none of chemical theory allows flows from boundary to boundary. These bald statements are easy to confirm. Most monographs and texts of chemical theory barely mention flow from boundaries, and most deal with equilibrium zero flow systems, using those results to discuss what happens when flow is not zero with certain inherent difficulties (given the obvious inconsistencies involved). References are hard to find that discuss spatially nonuniform boundary conditions or flows from boundary to boundary driven by external sources. Power supplies are essential for most engineering devices and they require different locations on boundaries to have different potential. That is to say, they require spatially nonuniform boundary conditions for potential with nonzero flows on the boundary.
			
			Devices are important. Our entire electronic technology is built from devices that function more or less the same way no matter where they are located (within reasonable limits, it goes without saying. Nothing in engineering or technology is true in general. Everything exists and functions only within reasonable limits). It would obviously be useful if chemical systems could be easily and routinely shaped into devices.
			
			Devices depend on spatially complex boundary conditions. A device has inputs and outputs with different locations and different boundary equations. If inputs and outputs are at the same location, and have the same properties, there is no device! Most devices have power supplies as well as inputs and outputs. These supply flows of energy that allow the device to have well defined input output relations that are robust, quite independent of what is connected to the input or output of the device. A transfer function relates input and output when they are related by a constant coefficient causal ordinary differential equation in time. 
			
			Devices maintain these properties almost entirely by using electricity and energy from power supplies. They are fundamentally nonequilibrium systems with spatially nonuniform Dirichlet boundary conditions for the electrical potential.
			
			Electricity is used to make engineering devices for good reason. Electrical potentials, currents, and electrical energy are described by the Maxwell equations with greater precision over a wider range of conditions than almost any other physical phenomenon. 
			
			The generality of the Maxwell equations in classical form is obscured because they embody an outdated and seriously inadequate representation of dielectrics and polarization. We say this with no disrespect for the enormous contributions of Maxwell and Heaviside, et al. But a representation that described dielectric measurements slower than some 0.05 sec (in the  1890's) cannot be expected to describe systems studied in the 2020's that function on atomic time scales of femtoseconds, let alone the much faster time scales of visible and ultraviolet light ($3x10^{12}$ - $7x10^{12}$) to $10^{16}Hz$ and even higher energy radiation like x-rays ($10^{16}$ - $10^{20}Hz$) and gamma rays($>10^{20}Hz$).
			
			The classical Maxwell equations use a single real number to describe how charge moves when an electric field is changed. Charge moves (and is said to polarize) on an enormous range of times scales, when electric fields are applied to matter, that cannot usefully be described by the single dielectric constant. The classical formulation  of the Maxwell equations embody---as well as depend on--- the classical, but crude representation of dielectrics. They use a dielectric constant in the very definition of key variables. Those variables in fact depend on an out of date constitutive model. If time dependent, complicated charge movements are present, as is always the case in liquids, and in solids over the time scale of modern technology, the Maxwell equations need to be rewritten to isolate the movements of material charge (with mass) from other types of current (e.g., from the displacement current found throughout space $\varepsilon_0 \partial E/\partial t$). The rewritten ``Core Maxwell Equations'' must then be joined to a description of polarization showing how charge moves when electric and magnetic forces are applied \cite{wang2021variational}. That description is not very different from the stress strain description of how mass moves when a force is applied. Mathematicians and physicists have dealt with such stress strain relations in solids and complex fluids with many types of flow (migration, convection, diffusion) and those methods (chiefly of the theory of complex fluids, in its energetic variational flavor)  can be applied to the polarization phenomena of charge.
			
			The Core Maxwell Equations have a special property not found in many other field equations because of the displacement current term $\varepsilon_0 \partial E/\partial t$  that is universal, present everywhere including inside atoms and in empty space, wherever the Maxwell equations are valid.
			
			The Core Maxwell Equations are of use even when the spatial and temporal location of charge is not known because the displacement current term $\varepsilon_0 \partial E/\partial t$ is universal, It is present everywhere, inside atoms and between stars. For this reason, the Core Maxwell Equations (surprisingly) are very useful for any description of polarization and for any type of charge movement, whether created or driven by electrodynamics, diffusion, convection, sunlight (in solar cells), or even heat flow even though they themselves do not describe polarization phenomena.  
			
			The Core Maxwell Equations have this special property because of Maxwell’s Ampere law. This law has no counterpart in the field theories of mechanics. This law allows electric and magnetic fields to flow in perfect vacuum so they can create propagated waves we call `the light of the sun', even though the vacuum contains no charge (with mass). 
			
			The Maxwell Ampere law has a corollary: the divergence of total current is always zero, everywhere to the same accuracy and in the same domain that the Maxwell equations are valid. The Maxwell Ampere law implies a unique definition of total current as the entire source (i.e., right hand side) of\textbf{ curl B} in Maxwell’s Ampere law {\textbf{no matter how material charge moves.}} In fact, Maxwell’s Ampere law  defines total current everywhere, including in a perfect vacuum where there is no mass or charge with mass or movement of charge with mass whatsoever. In a total vacuum, total current is $\varepsilon_0 \partial E/\partial t$). In the presence of matter total current is  $J_{total-  current} = \ J+ \ \varepsilon_0 \partial E/\partial t$) where $J$ described the movement of charge with mass no matter how fast, transient, or small it is. This fundamental property of electricity was well known to Maxwell and his followers, as is made clear on p.155 and p. 511-512 of \cite{RN28696}.
			and is central to the discussion of current flow in
			the classic text of \cite{RN45998}
			and the work of Landauer \cite{RN45927,RN26340}.
			The idea of total current is defined to focus attention on these issues and is discussed in many papers cited in  \cite{eisenberg2021maxwell}.
			
			If total current is confined to a single one dimensional path, conservation of total current becomes equality of total current everywhere (see Fig. 2 of  \cite{RN25303}
			for an extensive physical discussion with examples). If that current is confined to a circuit, conservation of total current becomes Kirchhoff’s current law \cite{RN46001}. In a particular system, it is easy to verify whether the one dimensional approximation for current flow is accurate (enough). Just measure the currents and see if they are equal (in unbranched systems) or whether they sum to zero as Kirchhoff’s law requires at nodes in circuits in general as Kirchhoff’s law requires. It would be interesting to learn to specify both the necessary and sufficient conditions in abstract mathematical terms (that can be evaluated before a particular system is specified) under which Kirchhoff's law is accurate (enough).
			
			The Kirchhoff’s law just described is not quite the Kirchhoff’s law of textbooks of electrodynamics or engineering. The textbook law is derived and presented as valid for long times many orders of magnitude longer than the time scales of electronic devices, let alone atomic motion. There should be no misunderstanding of this crucial point.  The literature of circuit design shows that Kirchhoff's law is used as an essential design tool or analog and digital circuits and integrated circuits   \cite{RN25358,RN4927,RN45662,RN4927,RN26358,RN28108,RN25161,RN26575} that can even function on the $10^{-12}$ second time scale \cite{RN46002}.
			It seems clear to us that the generalization of Kirchhoff’s law to total current either solves this problem, or makes it moot, as you wish \cite{eisenberg2019kirchhoff,eisenberg2018current}, following the practice of Maxwell himself, and his followers, according to p.155 and p. 511-512 of \cite{RN28696}.
			
			This paper combines Kirchhoff’s law for total current with a quite general description of chemical reactions with a general description of ion and water flow, using the EnVarA (energy variational) approach in the tradition of the theory of complex fluids. We use this electro-osmotic framework to analyze a coarse grained description of cytochrome $c$ oxidase in the tradition of `master equations'. Our model is built on the carefully constructed and well analyzed models of many others, but here we analyze the master equations using currents defined as in Kirchhoff's law for total current.  
			
			Our analysis is in the tradition of engineering. It does not depend on details of electron current flow. It does not compute properties of charges and their interactions, except implicitly as defined by the Maxwell equations, particularly Maxwell's Ampere's law $curl \ B=\mu_0 \ J_{total-current} =\mu_0 \  J+ \  \mu_0\varepsilon_0 \partial E/\partial t$ 
			
			It seems obvious to us that each of the systems of oxidative phosphorylation and photosynthesis require a circuit analysis embedded in the theory of complex fluids. Evolution has built structures that conduct electron currents, as in our electronic technology and so should be analyzed by the extraordinarily successful methods of electronic circuit analysis. Evolution has used ion current flow and chemical reactions in addition to electron flow so the circuit approach is embedded here using the theory of complex fluids. Complex fluid theory is designed to combine a wide variety of flows and the forces and energies that drive them in a mathematically consistent way. The structures evolution uses to control these flows form the geometry---or anatomy or histology, depending on the length scale---of the system. The channels and transporters (and electron transport pathways) built by evolution form boundary conditions that decorate (i.e., are located on) these structures and thus describe how they work.
			
			Cytochrome $c$ oxidase is called Complex IV for a reason. It is embedded in a lipid bilayer, connected to electron pathways we approximate as wires (in the engineering tradition), and is  surrounded by electrolytes that store energy in their electric and chemical potential fields that form a complex fluid. Complex IV includes the cytochrome $c$ oxidase enzyme, electron pathways, channels for protons and potassium ions, pathways for oxygen diffusion, and the membrane that encloses it and ions that surround it. Each subsystem stores energy and responds to energy gradients with different types of flows. Complex fluid mathematics is designed to handle systems of this complexity, although biological applications involve more preset structural complexity than in many physical systems of fluids. Complex fluid theory treats all fields, flows and boundary conditions---including spatially nonuniform conditions that power the system as they power electronic devices---consistently. The electro-osmotic extension of the theory hopefully joins the forces, flows and energies of chemical reactions into this formulation, while preserving the mathematical consistency of the original theory, without violating the traditions of chemistry.
			
			Our work is significantly incomplete and limited. We over-approximate several important biophysical mechanisms, including the water-gate switch, and the oxygen reduction mechanism. We are more than aware of the need for higher resolution in later work, with specific atomic scale models that compute the electric field and flows from underlying structures and chemical reactions on time scales of displacement (capacitive) currents that have been so well resolved in experiments of great difficulty. These currents are important in understanding the switches and mechanisms by which cytochrome c oxidase couples electron flow, oxidative chemical reactions, and proton flow to make oxidative phosphorylation possible in mitochondria.

			\section*{Acknowledgement}
			This work was partially supported by the 
			National Natural Science Foundation of China  no. 12071190  and Natural Sciences and Engineering Research Council of Canada (NSERC). 
			We also would like to thank  American Institute of Mathematics where this project  initiated.

			\newpage
			\printbibliography

@book{garber1995maxwell,
  title={Maxwell on Heat and Statistical Mechanics: On" Avoiding All Personal Enquiries" of Molecules},
  author={Garber, Elizabeth and Brush, Stephen G and Everitt, CWF},
  year={1995},
  publisher={Lehigh University Press}
}

@book{brush1976kind,
  title={The kind of motion we call heat},
  author={Brush, Stephen G},
  volume={2},
  year={1976},
  publisher={North-Holland Amsterdam}
}

@article{eisenberg2020electrodynamics,
  title={Electrodynamics Correlates Knock-on and Knock-off: Current is Spatially Uniform in Ion Channels},
  author={Eisenberg, Robert S},
  journal={arXiv preprint arXiv:2002.09012},
  year={2020}
}

@article{eisenberg2019kirchhoff,
  title={Kirchhoff's Law Can Be Exact},
  author={Eisenberg, Robert S},
  journal={arXiv preprint arXiv:1905.13574},
  year={2019}
}

@article{eisenberg2018current,
  title={What current flows through a resistor?},
  author={Eisenberg, Bob and Gold, Nathan and Song, Zilong and Huang, Huaxiong},
  journal={arXiv preprint arXiv:1805.04814},
  year={2018}
}

@misc{simpson1998maxwell,
  title={Maxwell on the Electromagnetic Field: A Guided Study.},
  author={Simpson, Thomas K},
  year={1998},
  publisher={American Association of Physics Teachers}
}

@article{chree1908mathematical,
  title={The Mathematical Theory of Electricity and Magnetism},
  author={Chree, C},
  journal={Nature},
  volume={78},
  number={2031},
  pages={537--538},
  year={1908},
  publisher={Nature Publishing Group}
}

@book{zangwill2013modern,
  title={Modern electrodynamics},
  author={Zangwill, Andrew},
  year={2013},
  publisher={Cambridge University Press}
}

@article{eisenberg2010energy,
  title={Energy variational analysis of ions in water and channels: Field theory for primitive models of complex ionic fluids},
  author={Eisenberg, Bob and Hyon, Yunkyong and Liu, Chun},
  journal={The Journal of Chemical Physics},
  volume={133},
  number={10},
  pages={104104},
  year={2010},
  publisher={American Institute of Physics}
}

@misc{giga2017variational,
  title={Variational modeling and complex fluids},
  author={Giga, Mi-Ho and Kirshtein, Arkadz and Liu, Chun},
  journal={Handbook of mathematical analysis in mechanics of viscous fluids},
  pages={1--41},
  year={2017},
  publisher={Springer International Publishing, Cham}
}

@article{chen2020differential,
  title={Differential Capacitance of Electric Double Layers: A Poisson-Bikerman Formula},
  author={Chen, Ren-Chuen and Li, Chin-Lung and Chen, Jen-Hao and Eisenberg, Bob and Liu, Jinn-Liang},
  journal={arXiv preprint arXiv:2012.13141},
  year={2020}
}

@article{li2020generalized,
  title={Generalized Debye--Hückel Equation From Poisson--Bikerman Theory},
  author={Li, Chin-Lung and Liu, Jinn-Liang},
  journal={SIAM Journal on Applied Mathematics},
  volume={80},
  number={5},
  pages={2003--2023},
  year={2020},
  publisher={SIAM}
}

@article{liu2020molecular,
  title={Molecular mean-field theory of ionic solutions: A Poisson-Nernst-Planck-Bikerman model},
  author={Liu, Jinn-Liang and Eisenberg, Bob},
  journal={Entropy},
  volume={22},
  number={5},
  pages={550},
  year={2020},
  publisher={Multidisciplinary Digital Publishing Institute}
}

@article{wang2020field,
  title={Field theory of reaction-diffusion: Law of mass action with an energetic variational approach},
  author={Wang, Yiwei and Liu, Chun and Liu, Pei and Eisenberg, Bob},
  journal={Physical Review E},
  volume={102},
  number={6},
  pages={062147},
  year={2020},
  publisher={APS}
}

@article{eisenberg2021maxwell,
  title={Maxwell Equations Without a Polarization Field, Using a Paradigm from Biophysics},
  author={Eisenberg, Robert S},
  journal={Entropy},
  volume={23},
  number={2},
  pages={172},
  year={2021},
  publisher={Multidisciplinary Digital Publishing Institute}
}

@article{kim2009kinetic,
  title={Kinetic gating of the proton pump in cytochrome c oxidase},
  author={Kim, Young C and Wikstr{\"o}m, M{\aa}rten and Hummer, Gerhard},
  journal={Proceedings of the National Academy of Sciences},
  volume={106},
  number={33},
  pages={13707--13712},
  year={2009},
  publisher={National Acad Sciences}
}

@article{eisenberg1975electrophysiology,
  title={Electrophysiology: Electric Current Flow in Excitable Cells. JJB Jack, D. Noble, and RW Tsien. Clarendon (Oxford University Press)},
  author={Eisenberg, Robert S},
  journal={Science},
  volume={190},
  number={4219},
  pages={1087--1087},
  year={1975},
  publisher={American Association for the Advancement of Science}
}

@inproceedings{thompson1855theory,
  title={On the theory of the electric telegraph},
  author={Thompson, William and Kelvin, L},
  booktitle={Proc. Royal Soc. London},
  volume={7},
  pages={382--399},
  year={1855}
}

@article{wang2021variational,
  title={On variational principles for polarization in electromechanical systems},
  author={Wang, Yiwei and Liu, Chun and Eisenberg, Bob},
  journal={arXiv e-prints},
  pages={arXiv--2108},
  year={2021}
}

@article{bush1929operational,
  title={Operational circuit analysis},
  author={Bush, Vannevar},
  journal={New York},
  pages={197},
  year={1929}
}

@book{RN45733,
   author = {Guillemin, E.A.},
   title = {Communications Networks Vol. 1 The Classical Theory of Lumped Constant Networks},
   publisher = {John Wiley},
   year = {1931},
   type = {Book}
}

@article{sugitani2008theoretical,
  title={Theoretical and computational analysis of the membrane potential generated by cytochrome c oxidase upon single electron injection into the enzyme},
  author={Sugitani, Ryogo and Medvedev, Emile S and Stuchebrukhov, Alexei A},
  journal={Biochimica et Biophysica Acta (BBA)-Bioenergetics},
  volume={1777},
  number={9},
  pages={1129--1139},
  year={2008},
  publisher={Elsevier}
}

@article{belevitch1962summary,
  title={Summary of the history of circuit theory},
  author={Belevitch, Vitold},
  journal={Proceedings of the IRE},
  volume={50},
  number={5},
  pages={848--855},
  year={1962},
  publisher={IEEE}
}

@article{darlington1984history,
  title={A history of network synthesis and filter theory for circuits composed of resistors, inductors, and capacitors},
  author={Darlington, Sidney},
  journal={IEEE transactions on circuits and systems},
  volume={31},
  number={1},
  pages={3--13},
  year={1984},
  publisher={IEEE}
}

@article{miedema2007biological,
  title={A biological porin engineered into a molecular, nanofluidic diode},
  author={Miedema, Henk and Vrouenraets, Maarten and Wierenga, Jenny and Meijberg, Wim and Robillard, George and Eisenberg, Bob},
  journal={Nano letters},
  volume={7},
  number={9},
  pages={2886--2891},
  year={2007},
  publisher={ACS Publications}
}

@book{colinge2005physics,
  title={Physics of semiconductor devices},
  author={Colinge, J-P and Colinge, Cynthia A},
  year={2005},
  publisher={Springer Science \& Business Media}
}

@book{pierret1996semiconductor,
  title={Semiconductor device fundamentals},
  author={Pierret, Robert F},
  year={1996},
  publisher={Pearson Education India}
}

@article{laux1999revisiting,
  title={Revisiting the analytic theory of pn junction impedance: Improvements guided by computer simulation leading to a new equivalent circuit},
  author={Laux, Steven E and Hess, Karl},
  journal={IEEE Transactions on Electron Devices},
  volume={46},
  number={2},
  pages={396--412},
  year={1999},
  publisher={IEEE}
}

@article{haggag2000analytical,
  title={Analytical theory of semiconductor pn junctions and the transition between depletion and quasineutral region},
  author={Haggag, Amr and Hess, Karl},
  journal={IEEE Transactions on Electron Devices},
  volume={47},
  number={8},
  pages={1624--1629},
  year={2000},
  publisher={IEEE}
}

@article{kim2012proton,
  title={Proton-pumping mechanism of cytochrome c oxidase: A kinetic master-equation approach},
  author={Kim, Young C and Hummer, Gerhard},
  journal={Biochimica et Biophysica Acta (BBA)-Bioenergetics},
  volume={1817},
  number={4},
  pages={526--536},
  year={2012},
  publisher={Elsevier}
}

@article{kim2007kinetic,
  title={Kinetic models of redox-coupled proton pumping},
  author={Kim, Young C and Wikstr{\"o}m, M{\aa}rten and Hummer, Gerhard},
  journal={Proceedings of the National Academy of Sciences},
  volume={104},
  number={7},
  pages={2169--2174},
  year={2007},
  publisher={National Acad Sciences}
}

@article{shen2020energy,
  title={An energy stable C0 finite element scheme for a quasi-incompressible phase-field model of moving contact line with variable density},
  author={Shen, Lingyue and Huang, Huaxiong and Lin, Ping and Song, Zilong and Xu, Shixin},
  journal={Journal of Computational Physics},
  volume={405},
  pages={109179},
  year={2020},
  publisher={Elsevier}
}

@article{shen2022energy,
  title={An Energy Stable C\^{}0 Finite Element Scheme for A Phase-Field Model of Vesicle Motion and Deformation},
  author={Shen, Lingyue and Xu, Zhiliang and Lin, Ping and Huang, Huaxiong and Xu, Shixin},
  journal={SIAM Journal on Scientific Computing},
  volume={44},
  number={1},
  pages={B122--B145},
  year={2022},
  publisher={SIAM}
}

@article{xu2018osmosis,
  title={Osmosis through a semi-permeable membrane: a consistent approach to interactions},
  author={Xu, Shixin and Eisenberg, Bob and Song, Zilong and Huang, Huaxiong},
  journal={arXiv preprint arXiv:1806.00646},
  year={2018}
}

@article{RN4995,
   author = {Wikstrom, M. and Verkhovsky, M. I. and Hummer, G.},
   title = {Water-gated mechanism of proton translocation by cytochrome c oxidase},
   journal = {Biochim Biophys Acta},
   volume = {1604},
   number = {2},
   pages = {61-5},
   url = {http://www.ncbi.nlm.nih.gov/entrez/query.fcgi?cmd=Retrieve&db=PubMed&dopt=Citation&list_uids=12765763 },
   year = {2003},
   type = {Journal Article}
   }

@article{RN45750,
   author = {Wikström, Mårten and Sharma, Vivek},
   title = {Proton pumping by cytochrome c oxidase–a 40 year anniversary},
   journal = {Biochimica et Biophysica Acta (BBA)-Bioenergetics},
   volume = {1859},
   number = {9},
   pages = {692-698},
   ISSN = {0005-2728},
   year = {2018},
   type = {Journal Article}}

@article{RN938,
   author = {Leung, J. and Eisenberg, R. S.},
   title = {The effects of the antibiotics gramicidin A, amphotericin B, and nystatin on the electrical properties of frog skeletal muscle},
   journal = {Biochim Biophys Acta},
   volume = {298},
   number = {3},
   pages = {718-23},
   url = {http://www.ncbi.nlm.nih.gov/entrez/query.fcgi?cmd=Retrieve&db=PubMed&dopt=Citation&list_uids=4541500},
   year = {1973},
   type = {Journal Article}
}

@article{RN24,
   author = {Eisenberg, R.S.},
   title = {Computing the field in proteins and channels.},
   journal = {Journal of Membrane Biology},
   volume = {150},
   pages = {1–25.  Preprint available on physics arXiv as document 1009.2857},
   year = {1996},
   type = {Journal Article}
}

@inbook{RN15949,
   author = {Eisenberg, R.S.},
   title = {Atomic Biology, Electrostatics and Ionic Channels.},
   booktitle = {New Developments and Theoretical Studies of Proteins},
   editor = {Elber, Ron},
   publisher = {World Scientific},
   address = {Philadelphia},
   volume = {7},
   pages = {269-357.  Published in the Physics ArXiv as arXiv:0807.0715},
   year = {1996},
   type = {Book Section}
}

@article{RN6246,
   author = {Miedema, H. and Vrouenraets, M. and Wierenga, J. and Meijberg, W. and Robillard, G. and Eisenberg, B.},
   title = {A Biological Porin Engineered into a Molecular, Nanofluidic Diode},
   journal = {Nano Lett.},
   volume = {7},
   number = {9},
   pages = {2886-2891},
   ISSN = {1530-6984},
   url = {http://pubs3.acs.org/acs/journals/doilookup?in_doi=10.1021/nl0716808 },
   year = {2007},
   type = {Journal Article}
}

@article{yamashita2012insights,
  title={Insights into the mechanism of proton transport in cytochrome c oxidase},
  author={Yamashita, Takefumi and Voth, Gregory A},
  journal={Journal of the American Chemical Society},
  volume={134},
  number={2},
  pages={1147--1152},
  year={2012},
  publisher={ACS Publications}
}

@article{han1993superconducting,
  title={SUPERCONDUCTING QUANTUM INTERFERENCE DEVICES},
  author={Han, Siyuan and Lapointe, J and Lukens, JE},
  journal={Activated Barrier Crossing: Applications in Physics, Chemistry and Biology},
  volume={4},
  pages={241},
  year={1993},
  publisher={World Scientific}
}

@article{eisenberg2010computing,
  title={Computing the field in proteins and channels},
  author={Eisenberg, Bob},
  journal={arXiv preprint arXiv:1009.2857},
  year={2010}
}

@article{eisenberg1998ionic,
  title={Ionic channels in biological membranes-electrostatic analysis of a natural nanotube},
  author={Eisenberg, Bob},
  journal={Contemporary Physics},
  volume={39},
  number={6},
  pages={447--466},
  year={1998},
  publisher={Taylor \& Francis}
}

@article{RN23400,
   author = {Finkelstein, Alan and Mauro, Alexander},
   title = {Equivalent Circuits as Related to Ionic Systems},
   journal = {Biophysical Journal},
   volume = {3},
   number = {3},
   pages = {215-237},
   ISSN = {0006-3495},
   DOI = {http://dx.doi.org/10.1016/S0006-3495(63)86817-4},
   url = {http://www.sciencedirect.com/science/article/pii/S0006349563868174},
   year = {1963},
   type = {Journal Article}
}

@article{RN23397,
   author = {Mauro, Alexander},
   title = {Space Charge Regions in Fixed Charge Membranes and the Associated Property of Capacitance},
   journal = {Biophysical Journal},
   volume = {2},
   number = {2, Part 1},
   pages = {179-198},
   ISSN = {0006-3495},
   DOI = {http://dx.doi.org/10.1016/S0006-3495(62)86848-9},
   url = {http://www.sciencedirect.com/science/article/pii/S0006349562868489},
   year = {1962},
   type = {Journal Article}
}

@article{RN45776,
   author = {Riza Putra, Budi and Tshwenya, Luthando and Buckingham, Mark A. and Chen, Jingyuan and Jeremiah Aoki, Koichi and Mathwig, Klaus and Arotiba, Omotayo A. and Thompson, Abigail K. and Li, Zhongkai and Marken, Frank},
   title = {Microscale Ionic Diodes: An Overview},
   journal = {Electroanalysis},
   volume = {33},
   number = {6},
   pages = {1398-1418},
   ISSN = {1040-0397},
   DOI = {10.1002/elan.202060614},
   url = {https://dx.doi.org/10.1002/elan.202060614},
   year = {2021},
   type = {Journal Article}
}

@article{RN45777,
   author = {Sun, Gongchen and Senapati, Satyajyoti and Chang, Hsueh-Chia},
   title = {High-flux ionic diodes, ionic transistors and ionic amplifiers based on external ion concentration polarization by an ion exchange membrane: a new scalable ionic circuit platform},
   journal = {Lab on a Chip},
   volume = {16},
   number = {7},
   pages = {1171-1177},
   ISSN = {1473-0197},
   DOI = {10.1039/c6lc00026f},
   url = {https://dx.doi.org/10.1039/c6lc00026f},
   year = {2016},
   type = {Journal Article}
}

@article{RN30417,
   author = {Catacuzzeno, Luigi and Sforna, Luigi and Franciolini, Fabio and Eisenberg, Robert S.},
   title = {Multiscale modeling shows that dielectric differences make NaV channels faster than KV channels},
   journal = {Journal of General Physiology},
   volume = {153  DOI: 10.1085/jgp.202012706},
   number = {2},
   ISSN = {0022-1295},
   DOI = {10.1085/jgp.202012706},
   url = {https://dx.doi.org/10.1085/jgp.202012706},
   year = {2021},
   type = {Journal Article}
}

@book{RN25358,
   author = {Horowitz, Paul and Hill, Winfield},
   title = {The Art of Electronics},
   publisher = {Cambridge University Press},
   edition = {Third Edition},
   pages = {1224 },
   ISBN = {ISBN-10: 0521809266
ISBN-13: 978-0521809269},
   year = {2015},
   type = {Book}
}

@book{RN45662,
   author = {Lienig, Jens and Scheible, Juergen},
   title = {Fundamentals of layout design for electronic circuits},
   publisher = {Springer Nature},
   ISBN = {3030392848},
   year = {2020},
   type = {Book}
}

@book{RN4927,
   author = {Gray, P. R. and Hurst, P. J. and Lewis, S. H. and Meyer, R. G.},
   title = {Analysis and Design of Analog Integrated Circuits},
   publisher = {John Wiley},
   address = {New York},
   edition = {4th Edition},
   pages = {875},
   year = {2001},
   type = {Book}
}

@book{RN26358,
   author = {Ghausi, Mohammed Shuaib and Kelly, John Joseph},
   title = {Introduction to distributed-parameter networks: with application to integrated circuits},
   publisher = {Holt, Rinehart and Winston},
   year = {1968},
   type = {Book}
}

@book{RN28108,
   author = {Muller, R.S. and Chan, M. and Kamins, T.I.},
   title = {Device Electronics For Integrated Circuits, 3rd Ed},
   publisher = {Wiley India Pvt. Limited},
   ISBN = {9788126510962},
   url = {https://books.google.com/books?id=X9Q-QQAACAAJ},
   year = {2003},
   type = {Book}
}

@book{RN25161,
   author = {Joffe, Elya B. and Lock, Kai-Sang},
   title = {Grounds for Grounding},
   publisher = { Wiley-IEEE Press},
   address = {NY},
   pages = {1088},
   ISBN = {ISBN-13: 978-0471660088  ISBN-10: 0471660086  },
   year = {2010},
   type = {Book}
}

@book{RN26575,
   author = {Johnson, Howard W and Graham, Martin},
   title = {High-speed signal propagation: advanced black magic},
   publisher = {Prentice Hall Professional},
   ISBN = {013084408X},
   year = {2003},
   type = {Book}
}

@article{RN25303,
   author = {Eisenberg, Robert S.},
   title = {Mass Action and Conservation of Current},
   journal = {Hungarian Journal of Industry and Chemistry  Posted on arXiv.org with paper ID arXiv:1502.07251},
   volume = {44},
   number = {1},
   pages = {1-28},
   DOI = {10.1515/hjic-2016-0001},
   year = {2016},
   type = {Journal Article}
}

@book{RN267,
   author = {Hodgkin, A.L.},
   title = {Chance and Design},
   publisher = {Cambridge University Press},
   address = {New York},
   pages = {401},
   year = {1992},
   type = {Book}
}

@article{RN309,
   author = {Hodgkin, A.L. and Huxley, A.F. and Katz, B.},
   title = {Measurement of current- voltage relations in the membrane of the giant axon of Loligo},
   journal = {J. Physiol. (London)},
   volume = {116},
   pages = {424-448},
   year = {1952},
   type = {Journal Article}
}

@book{RN28957,
   author = {Huxley, Andrew F},
   title = {The quantitative analysis of excitation and conduction in nerve},
   volume = {1963},
   series = {Les Prix Nobel},
   pages = {242-260},
   year = {1963},
   type = {Book}
}

@article{RN12551,
   author = {Huxley, A.F.},
   title = {From overshoot to voltage clamp},
   journal = {Trends in Neurosciences },
   volume = {25 },
   number = {11},
   pages = {553-558},
   year = {2002},
   type = {Journal Article}
}

@article{RN28654,
   author = {Cole, Kenneth Stewart},
   title = {Dynamic electrical characteristics of the squid axon membrane},
   journal = {Archives des sciences physiologiques},
   volume = {3},
   number = {2},
   pages = {253-258},
   ISSN = {0003-9713},
   year = {1949},
   type = {Journal Article}
}

@article{RN291,
   author = {Hodgkin, Alan and Huxley, Andrew and Katz, Bernhard},
   title = {Ionic Currents underlying activity in the giant axon of the squid},
   journal = {Arch. Sci. physiol.},
   volume = {3},
   pages = {129-150},
   year = {1949},
   type = {Journal Article}
}

@article{RN45995,
   author = {Caldwell, PC and Hodgkin, AL and Keynes, RD and Shaw, TI},
   title = {Partial inhibition of the active transport of cations in the giant axons of Loligo},
   journal = {The Journal of Physiology},
   volume = {152},
   number = {3},
   pages = {591},
   year = {1960},
   type = {Journal Article}
}

@article{RN9363,
   author = {Caldwell, P. C. and Hodgkin, A. L. and Keynes, R. D. and Shaw, T. I.},
   title = {The effects of injecting `energy-rich' phosphate compounds on the active transport of ions in the giant axons of Loligo},
   journal = {The Journal of Physiology},
   volume = {152},
   number = {3},
   pages = {561-590},
   url = {http://jp.physoc.org/content/152/3/561.short},
   year = {1960},
   type = {Journal Article}
}

@article{RN10470,
   author = {Caldwell, P. C. and Hodgkin, A. L. and Keynes, R. D. and Shaw, T. I.},
   title = {The Rate of Formation and Turnover of Phosphorus Compounds in Squid Giant Axons},
   journal = {J Physiol},
   volume = {171},
   pages = {119-31},
   ISSN = {0022-3751 (Print)
0022-3751 (Linking)},
   url = {http://www.ncbi.nlm.nih.gov/entrez/query.fcgi?cmd=Retrieve&db=PubMed&dopt=Citation&list_uids=14170137},
   year = {1964},
   type = {Journal Article}
}

@article{RN29611,
   author = {Gadsby, D. C.},
   title = {Structural biology: ion pumps made crystal clear},
   journal = {Nature},
   volume = {450},
   number = {7172},
   pages = {957-9},
   ISSN = {1476-4687 (Electronic)
0028-0836 (Linking)},
   DOI = {10.1038/450957a},
   url = {https://www.ncbi.nlm.nih.gov/pubmed/18075569},
   year = {2007},
   type = {Journal Article}
}

@book{RN434,
   author = {Tosteson, Daniel},
   title = {Membrane Transport: People and Ideas},
   address = {Bethesda MD},
   pages = {414},
   year = {1989},
   type = {Book}
}

@inbook{RN155,
   author = {Hille, B.},
   title = {Transport Across Cell Membranes: Carrier Mechanisms, Chapter 2},
   booktitle = {Textbook of Physiology},
   editor = {Patton, H.D. and Fuchs, A.F. and Hille, B. and Scher, A.M. and Steiner, R.D.},
   publisher = {Saunders},
   address = {Philadelphia},
   volume = {1},
   edition = {21},
   pages = {24-47.},
   year = {1989},
   type = {Book Section}
}

@article{RN23,
   author = {Eisenberg, R.S.},
   title = {Channels as enzymes: Oxymoron and Tautology},
   journal = {Journal of Membrane Biology},
   volume = {115},
   pages = {1–12.  Available on arXiv as  http://arxiv.org/abs/1112.2363},
   year = {1990},
   type = {Journal Article}
}

@article{RN29619,
   author = {Gadsby, D. C.},
   title = {Ion transport: spot the difference},
   journal = {Nature},
   volume = {427},
   number = {6977},
   pages = {795-7},
   ISSN = {1476-4687 (Electronic)
0028-0836 (Linking)},
   DOI = {10.1038/427795a},
   url = {https://www.ncbi.nlm.nih.gov/pubmed/14985745},
   year = {2004},
   type = {Journal Article}
}

@article{RN29607,
   author = {Gadsby, D. C.},
   title = {Ion channels versus ion pumps: the principal difference, in principle},
   journal = {Nat Rev Mol Cell Biol},
   volume = {10},
   number = {5},
   pages = {344-52},
   ISSN = {1471-0080 (Electronic)
1471-0072 (Linking)},
   DOI = {10.1038/nrm2668},
   url = {https://www.ncbi.nlm.nih.gov/pubmed/19339978},
   year = {2009},
   type = {Journal Article}
}

@article{RN30714,
   author = {Belevich, Ilya and Bloch, Dmitry A and Belevich, Nikolai and Wikström, Mårten and Verkhovsky, Michael I},
   title = {Exploring the proton pump mechanism of cytochrome c oxidase in real time},
   journal = {Proceedings of the National Academy of Sciences},
   volume = {104},
   number = {8},
   pages = {2685-2690},
   ISSN = {0027-8424},
   year = {2007},
   type = {Journal Article}
}

@article{RN30642,
   author = {Bloch, Dmitry and Belevich, Ilya and Jasaitis, Audrius and Ribacka, Camilla and Puustinen, Anne and Verkhovsky, Michael I. and Wikström, Mårten},
   title = {The catalytic cycle of cytochrome <em>c</em> oxidase is not the sum of its two halves},
   journal = {Proceedings of the National Academy of Sciences of the United States of America},
   volume = {101},
   number = {2},
   pages = {529-533},
   DOI = {10.1073/pnas.0306036101},
   url = {https://www.pnas.org/content/pnas/101/2/529.full.pdf},
   year = {2004},
   type = {Journal Article}
}

@article{RN6359,
   author = {Verkhovsky, M. I. and Belevich, I. and Bloch, D. A. and Wikstrom, M.},
   title = {Elementary steps of proton translocation in the catalytic cycle of cytochrome oxidase},
   journal = {Biochim Biophys Acta},
   volume = {1757},
   number = {5-6},
   pages = {401-7},
   ISSN = {0006-3002 (Print)},
   url = {http://www.ncbi.nlm.nih.gov/entrez/query.fcgi?cmd=Retrieve&db=PubMed&dopt=Citation&list_uids=16829227 },
   year = {2006},
   type = {Journal Article}
}

@article{RN30008,
   author = {Blomberg, Margareta R. A. and Siegbahn, Per E. M.},
   title = {The mechanism for proton pumping in cytochrome c oxidase from an electrostatic and quantum chemical perspective},
   journal = {Biochimica et Biophysica Acta (BBA) - Bioenergetics},
   volume = {1817},
   number = {4},
   pages = {495-505},
   ISSN = {0005-2728},
   DOI = {10.1016/j.bbabio.2011.09.014},
   url = {https://dx.doi.org/10.1016/j.bbabio.2011.09.014},
   year = {2012},
   type = {Journal Article}
}

@article{RN30276,
   author = {Cai, Xiuhong and Haider, Kamran and Lu, Jianxun and Radic, Slaven and Son, Chang Yun and Cui, Qiang and Gunner, M. R.},
   title = {Network analysis of a proposed exit pathway for protons to the P-side of cytochrome c oxidase},
   journal = {Biochimica et Biophysica Acta (BBA) - Bioenergetics},
   volume = {1859},
   number = {10},
   pages = {997-1005},
   ISSN = {0005-2728},
   DOI = {10.1016/j.bbabio.2018.05.010},
   url = {https://dx.doi.org/10.1016/j.bbabio.2018.05.010},
   year = {2018},
   type = {Journal Article}
}

@article{RN45724,
   author = {Verkhovskaya, Marina L and Belevich, Nikolai and Euro, Liliya and Wikström, Mårten and Verkhovsky, Michael I},
   title = {Real-time electron transfer in respiratory complex I},
   journal = {Proceedings of the National Academy of Sciences},
   volume = {105},
   number = {10},
   pages = {3763-3767},
   ISSN = {0027-8424},
   year = {2008},
   type = {Journal Article}
}

@book{RN7109,
   author = {Boron, Walter and Boulpaep, Emile},
   title = {Medical Physiology},
   publisher = {Saunders},
   address = {New York},
   pages = {1352},
   year = {2008},
   type = {Book}
}

@book{RN45615,
   author = {Feher, Joseph J},
   title = {Quantitative human physiology: an introduction},
   publisher = {Academic press},
   ISBN = {0128011548},
   year = {2017},
   type = {Book}
}

@book{RN28910,
   author = {Keener, J. and Sneyd, J.},
   title = {Mathematical Physiology: I: Cellular Physiology},
   publisher = {Springer New York},
   ISBN = {9781489986702},
   url = {https://books.google.com/books?id=Mf7HsgEACAAJ},
   year = {2014},
   type = {Book}
}

@book{RN26033,
   author = {Prosser, Clifford. Ladd and Curtis, Brian A. and Meisami, Esmail},
   title = {A History of Nerve, Muscle and Synapse Physiology},
   publisher = {Stipes Public License},
   pages = {572},
   year = {2009},
   type = {Book}
}

@book{RN45909,
   author = {Silverthorn, D.U. and Johnson, B.R. and Ober, W.C. and Ober, C.E. and Impagliazzo, A. and Silverthorn, A.C.},
   title = {Human Physiology: An Integrated Approach},
   publisher = {Pearson Education, Incorporated},
   ISBN = {9780134605197},
   url = {https://books.google.com/books?id=JOPStAEACAAJ},
   year = {2019},
   type = {Book}
}

@book{RN45950,
   author = {Sperelakis, N. and Sperelakis, N.},
   title = {Cell Physiology Source Book: Essentials of Membrane Biophysics},
   publisher = {Elsevier Science},
   ISBN = {9780123877383},
   url = {https://books.google.com/books?id=n7kxScnqQsQC},
   year = {2012},
   type = {Book}
}

@book{RN134,
   author = {Alberts, B. and Bray, D. and Lewis, J. and Raff, M. and Roberts, K.  and Watson, J.D. },
   title = {Molecular Biology of the Cell},
   publisher = {Garland},
   address = {New York},
   edition = {Third},
   pages = {1294},
   year = {1994},
   type = {Book}
}

@book{RN295,
   author = {Darnell, James and Lodish, Harvey and Baltimore, David},
   title = {Molecular Cell Biology},
   publisher = {Scientific American Books},
   address = {New York},
   edition = {2nd Edition},
   year = {1990},
   type = {Book}
}

@book{RN45998,
   author = {Ramo, S. and Whinnery, J.R. and Van Duzer, T.},
   title = {Fields and Waves in Communication Electronics},
   publisher = {J. Wiley},
   ISBN = {9780471707202},
   url = {https://books.google.com/books?id=ABpRAAAAMAAJ},
   year = {1965},
   type = {Book}
}

@book{RN45934,
   author = {Boylestad, R.L. and Nashelsky, L.},
   title = {Electronic Devices and Circuit Theory: Pearson New International Edition PDF eBook},
   publisher = {Pearson Education},
   ISBN = {9781292038063},
   url = {https://books.google.com/books?id=zyypBwAAQBAJ},
   year = {2013},
   type = {Book}
}

@book{RN21598,
   author = {Howe, Roger T. and Sodini, Charles G.},
   title = {Microelectronics: an integrated approach},
   publisher = {Prentice Hall},
   address = {Upper Saddle River, NJ USA},
   pages = {908},
   year = {1997},
   type = {Book}
}

@book{RN26593,
   author = {Scherz, Paul and Monk, S.},
   title = {Practical electronics for inventors},
   publisher = {McGraw-Hill, Inc.},
   pages = {1056},
   ISBN = {ISBN-13: 978-1259587542
ISBN-10: 1259587541},
   year = {2006},
   type = {Book}
}

@book{RN45935,
   author = {Sedra, A.S. and Smith, K.C. and Chan, T. and Carusone, T.C. and Gaudet, V.},
   title = {Microelectronic Circuits},
   publisher = {Oxford University Press, Incorporated},
   ISBN = {9780190853501},
   url = {https://books.google.com/books?id=qGfRzQEACAAJ},
   year = {2020},
   type = {Book}
}

@book{RN45999,
   author = {Gielen, Georges and Sansen, Willy MC},
   title = {Symbolic analysis for automated design of analog integrated circuits},
   publisher = {Springer Science \& Business Media},
   volume = {137},
   ISBN = {1461539625},
   year = {2012},
   type = {Book}
}

@book{RN28696,
   author = {Jeans, James Hopwood},
   title = {The mathematical theory of electricity and magnetism},
   publisher = {Cambridge University Press},
   year = {1908},
   type = {Book}
}

@article{RN45927,
   author = {Imry, Yoseph and Landauer, Rolf},
   title = {Conductance viewed as transmission},
   journal = {Reviews of Modern Physics},
   volume = {71},
   number = {2},
   pages = {S306},
   year = {1999},
   type = {Journal Article}
}

@article{RN26340,
   author = {Landauer, Rolf},
   title = {Conductance from transmission: common sense points},
   journal = {Physica Scripta},
   volume = {1992},
   number = {T42},
   pages = {110},
   ISSN = {1402-4896},
   url = {http://stacks.iop.org/1402-4896/1992/i=T42/a=020},
   year = {1992},
   type = {Journal Article}
}

@article{RN46001,
   author = {Eisenberg, Robert},
   title = {A Necessary Addition to Kirchhoff’s Current Law of CircuitsVersion  },
   journal = {Engineering Archive EngArXiv},
   volume = {https://doi.org/10.31224/2234},
   DOI = { https://doi.org/10.31224/2234},
   year = {2022},
   type = {Journal Article}
}

@article{RN46002,
   author = {Zou, Lianfeng and Gupta, Shulabh and Caloz, Christophe},
   title = {A Simple Picosecond Pulse Generator Based on a Pair of Step Recovery Diodes},
   journal = {IEEE Microwave and Wireless Components Letters},
   volume = {27},
   number = {5},
   pages = {467-469},
   ISSN = {1531-1309},
   DOI = {10.1109/lmwc.2017.2690880},
   url = {https://dx.doi.org/10.1109/lmwc.2017.2690880},
   year = {2017},
   type = {Journal Article}
}

@article{RN46052,
   author = {Wang, Yiwei and Liu, Chun},
   title = {Some Recent Advances in Energetic Variational Approaches},
   journal = {Entropy},
   volume = {24},
   number = {5},
   pages = {721},
   ISSN = {1099-4300},
   DOI = {10.3390/e24050721},
   url = {https://dx.doi.org/10.3390/e24050721},
   year = {2022},
   type = {Journal Article}
}

@book{RN46053,
   author = {Truesdell, Clifford},
   title = {Rational thermodynamics: a course of lectures on selected topics},
   publisher = {McGraw-Hill},
   ISBN = {0070653003},
   year = {1969},
   type = {Book}
}

@book{RN46056,
   author = {Friedli, Sacha and Velenik, Yvan},
   title = {Statistical mechanics of lattice systems: a concrete mathematical introduction},
   publisher = {Cambridge University Press},
   ISBN = {1107184827},
   year = {2017},
   type = {Book}
}

@book{RN46055,
   author = {Shavitt, Isaiah and Bartlett, Rodney J},
   title = {Many-body methods in chemistry and physics: MBPT and coupled-cluster theory},
   publisher = {Cambridge university press},
   ISBN = {052181832X},
   year = {2009},
   type = {Book}
}

@article{RN28691,
   author = {Zhu, Yi and Xu, Shixin and Eisenberg, Robert S. and Huang, Huaxiong},
   title = {A Bidomain Model for Lens Microcirculation 
},
   journal = {Biophysical Journal},
   volume = {116},
   number = {6},
   pages = {1171-1184 Preprint available at https://arxiv.org/abs/1810.04162},
   ISSN = {0006-3495},
   DOI = {https://doi.org/10.1016/j.bpj.2019.02.007},
   url = {http://www.sciencedirect.com/science/article/pii/S0006349519301341},
   year = {2019},
   type = {Journal Article}
}

@article{RN30228,
   author = {Zhu, Yi and Xu, Shixin and Eisenberg, Robert S. and Huang, Huaxiong},
   title = {A Tridomain Model for Potassium Clearance in Optic Nerve},
   journal = {arXiv:2012.03303},
   DOI = {arxiv:2012.03303},
   url = {https://arxiv.org/abs/2012.03303},
   year = {2020},
   type = {Journal Article}
}

@article{RN30601,
   author = {Zhu, Yi and Xu, Shixin and Eisenberg, Robert S and Huang, Huaxiong},
   title = {Membranes in Optic Nerve Models},
   journal = {arXiv preprint arXiv:2105.14411},
   year = {2021},
   type = {Journal Article}
}

@article{RN30649,
   author = {Zhu, Yi and Xu, Shixin and Eisenberg, Robert S and Huang, Huaxiong},
   title = {Optic nerve microcirculation: Fluid flow and electrodiffusion},
   journal = {Physics of Fluids},
   volume = {33},
   number = {4},
   pages = {041906},
   ISSN = {1070-6631},
   year = {2021},
   type = {Journal Article}
}

@article{RN46054,
   author = {Song, Zilong and Eisenberg, Robert and Xu, Shixin and Huang, Huaxiong},
   title = {A Bubble Model for the Gating of K $ _\mathrm {v} $ Channels},
   journal = {arXiv preprint arXiv:2204.13077},
   year = {2022},
   type = {Journal Article}
}

@article{RN28898,
   author = {Eisenberg, Robert S},
   title = {Updating Maxwell with Electrons, Charge, and More Realistic Polarization},
   journal = {arXiv preprint available at https://arxiv.org/abs/1904.09695},
   year = {2019},
   type = {Journal Article}
}

@article{RN30138,
   author = {Eisenberg, Robert s.},
   title = {Maxwell Equations for Material Systems},
   journal = {doi: 10.20944/preprints202011.0201.v1},
   DOI = {10.20944/preprints202011.0201.v1},
   url = {https://dx.doi.org/10.20944/preprints202011.0201.v1},
   year = {2020},
   type = {Journal Article}
}

@article{RN46023,
   author = {Harris, Eric J and Catlin, Graham and Pressman, Berton C},
   title = {Effect of transport-inducing antibiotics and other agents on potassium flux in mitochondria},
   journal = {Biochemistry},
   volume = {6},
   number = {5},
   pages = {1360-1370},
   ISSN = {0006-2960},
   year = {1967},
   type = {Journal Article}
}

@article{RN45770,
   author = {Pressman, B. C.},
   title = {Induced active transport of ions in mitochondria},
   journal = {Proceedings of the National Academy of Sciences},
   volume = {53},
   number = {5},
   pages = {1076-1083},
   ISSN = {0027-8424},
   DOI = {10.1073/pnas.53.5.1076},
   url = {https://dx.doi.org/10.1073/pnas.53.5.1076},
   year = {1965},
   type = {Journal Article}
}

@article{RN27722,
   author = {Pressman, Berton C},
   title = {Biological applications of ionophores},
   journal = {Annual review of biochemistry},
   volume = {45},
   number = {1},
   pages = {501-530},
   ISSN = {0066-4154},
   year = {1976},
   type = {Journal Article}
}

@article{RN46059,
   author = {Andreoli, Thomas E and Tieffenberg, M and Tosteson, Daniel C},
   title = {The effect of valinomycin on the ionic permeability of thin lipid membranes},
   journal = {The Journal of general physiology},
   volume = {50},
   number = {11},
   pages = {2527-2545},
   ISSN = {1540-7748},
   year = {1967},
   type = {Journal Article}
}

@article{RN46060,
   author = {Tosteson, Daniel C and Cook, Pl and Andreoli, Thomas and Tieffenberg, M},
   title = {The effect of valinomycin on potassium and sodium permeability of HK and LK sheep red cells},
   journal = {The Journal of general physiology},
   volume = {50},
   number = {11},
   pages = {2513-2525},
   ISSN = {1540-7748},
   year = {1967},
   type = {Journal Article}
}

@article{RN46061,
   author = {Su, Zhangfei and Ran, Xueqin and Leitch, J. Jay and Schwan, Adrian L. and Faragher, Robert and Lipkowski, Jacek},
   title = {How Valinomycin Ionophores Enter and Transport K+ across Model Lipid Bilayer Membranes},
   journal = {Langmuir},
   volume = {35},
   number = {51},
   pages = {16935-16943},
   ISSN = {0743-7463},
   DOI = {10.1021/acs.langmuir.9b03064},
   url = {https://dx.doi.org/10.1021/acs.langmuir.9b03064},
   year = {2019},
   type = {Journal Article}
}

@article{belevich2010initiation,
  title={Initiation of the proton pump of cytochrome c oxidase},
  author={Belevich, Ilya and Gorbikova, Elena and Belevich, Nikolai P and Rauhamaki, Virve and Wikstrom, Marten and Verkhovsky, Michael I},
  journal={Proceedings of the National Academy of Sciences},
  volume={107},
  number={43},
  pages={18469--18474},
  year={2010},
  publisher={National Acad Sciences}
}
			
			\newpage
			\appendix
			\section{Simulation results for case1}
			\subsection{Effect of Oxygen}\label{sec: simulation_oxygen}

\begin{figure}[!ht]
				\centering
				\begin{subfigure}[]{
						\includegraphics[width=6.in]{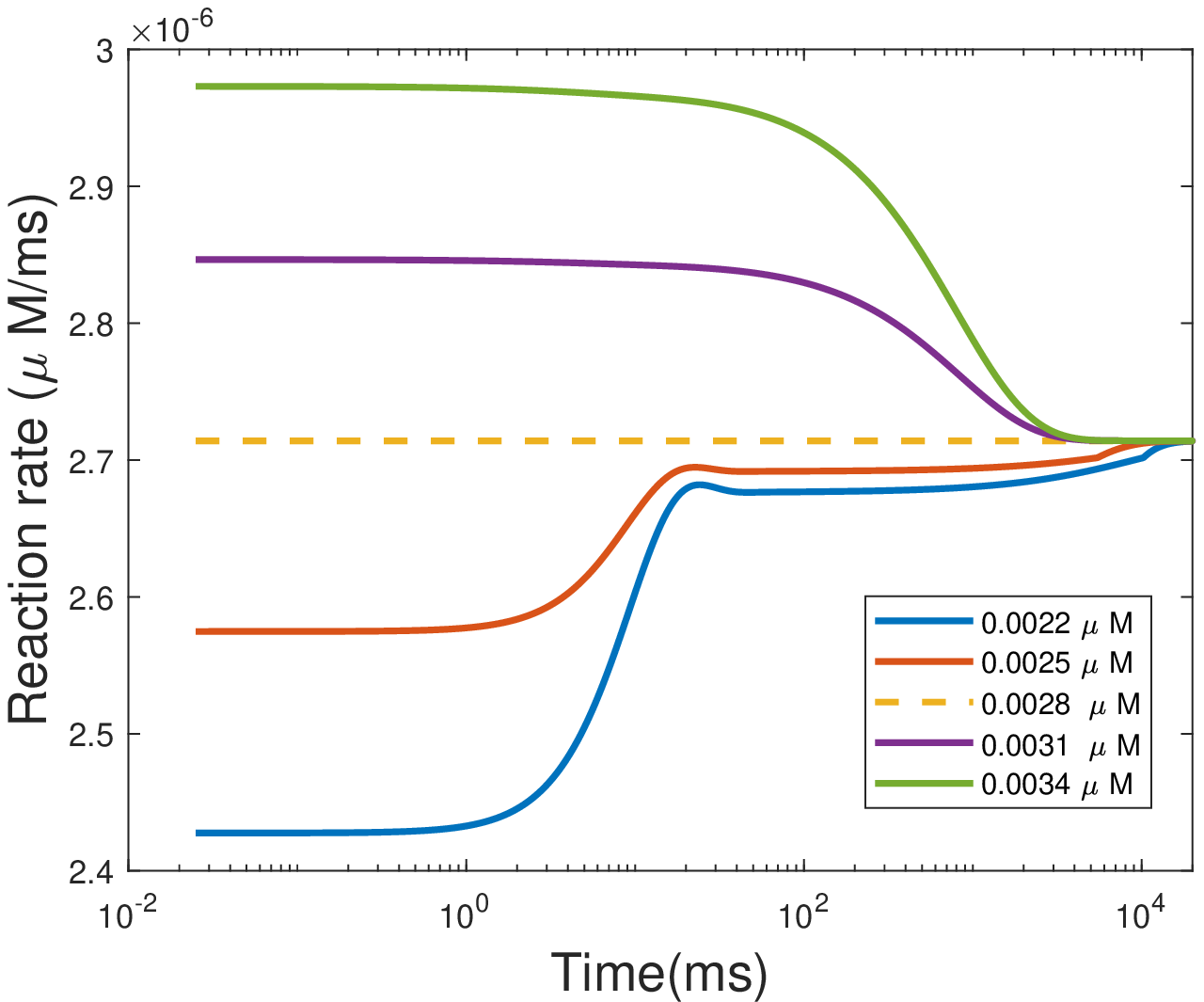}
					}
				\end{subfigure}
				\begin{subfigure}[]{
						\includegraphics[width=3.in]{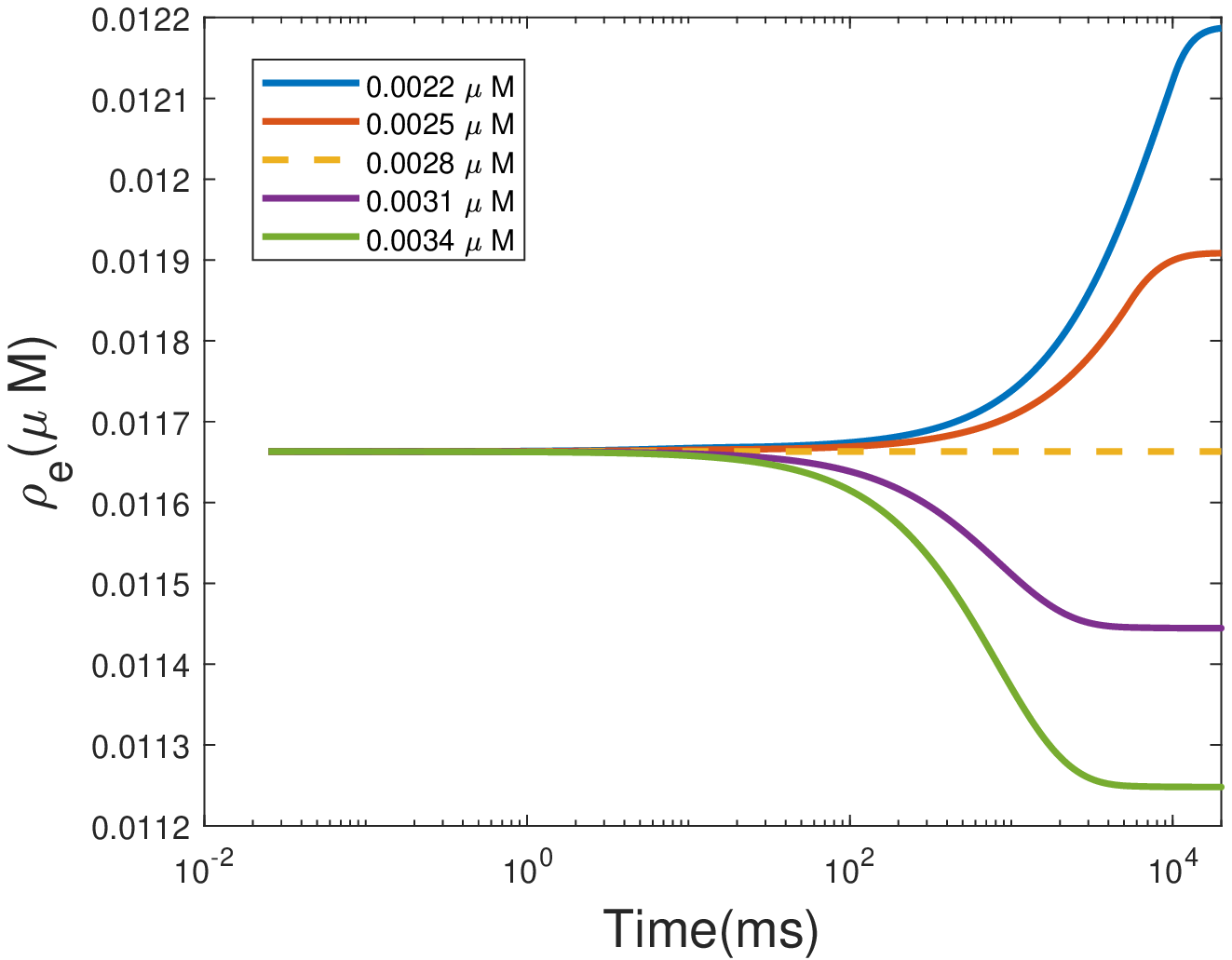}}
				\end{subfigure}
				\begin{subfigure}[]{
						\includegraphics[width=3.in]{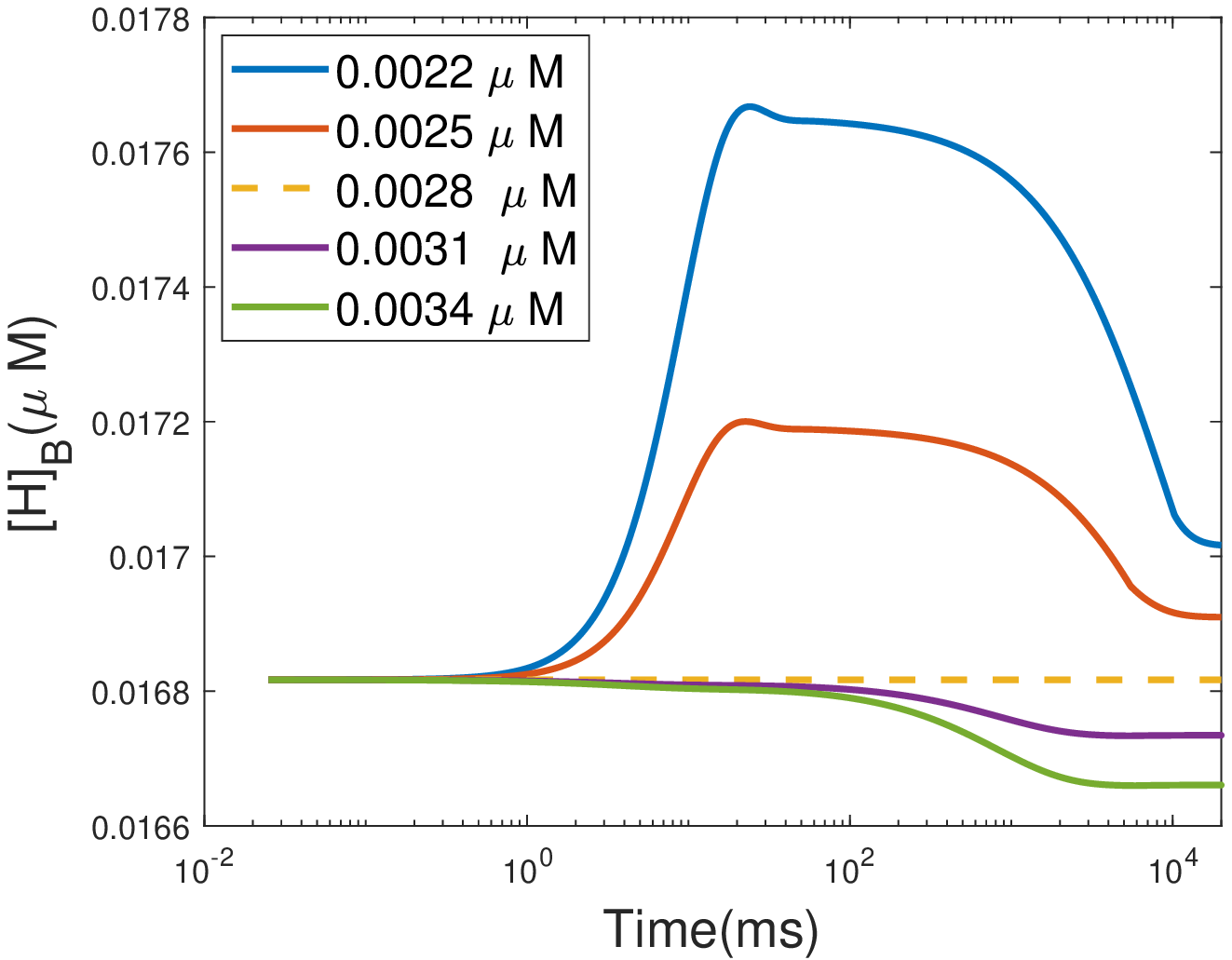}}
				\end{subfigure}
				\begin{subfigure}[]{
						\includegraphics[width=3.in]{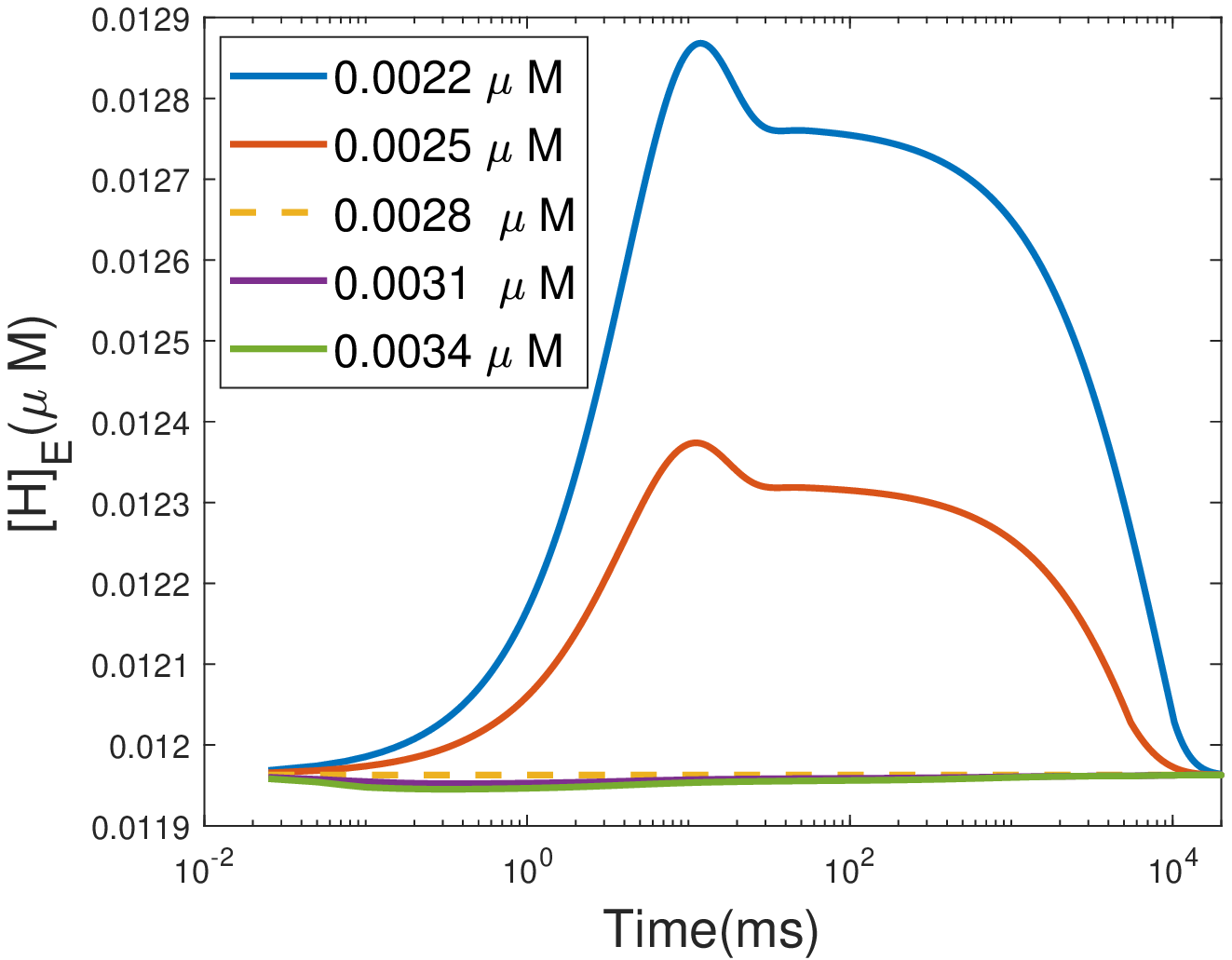}}
				\end{subfigure}
				\begin{subfigure}[]{
						\includegraphics[width=3.in]{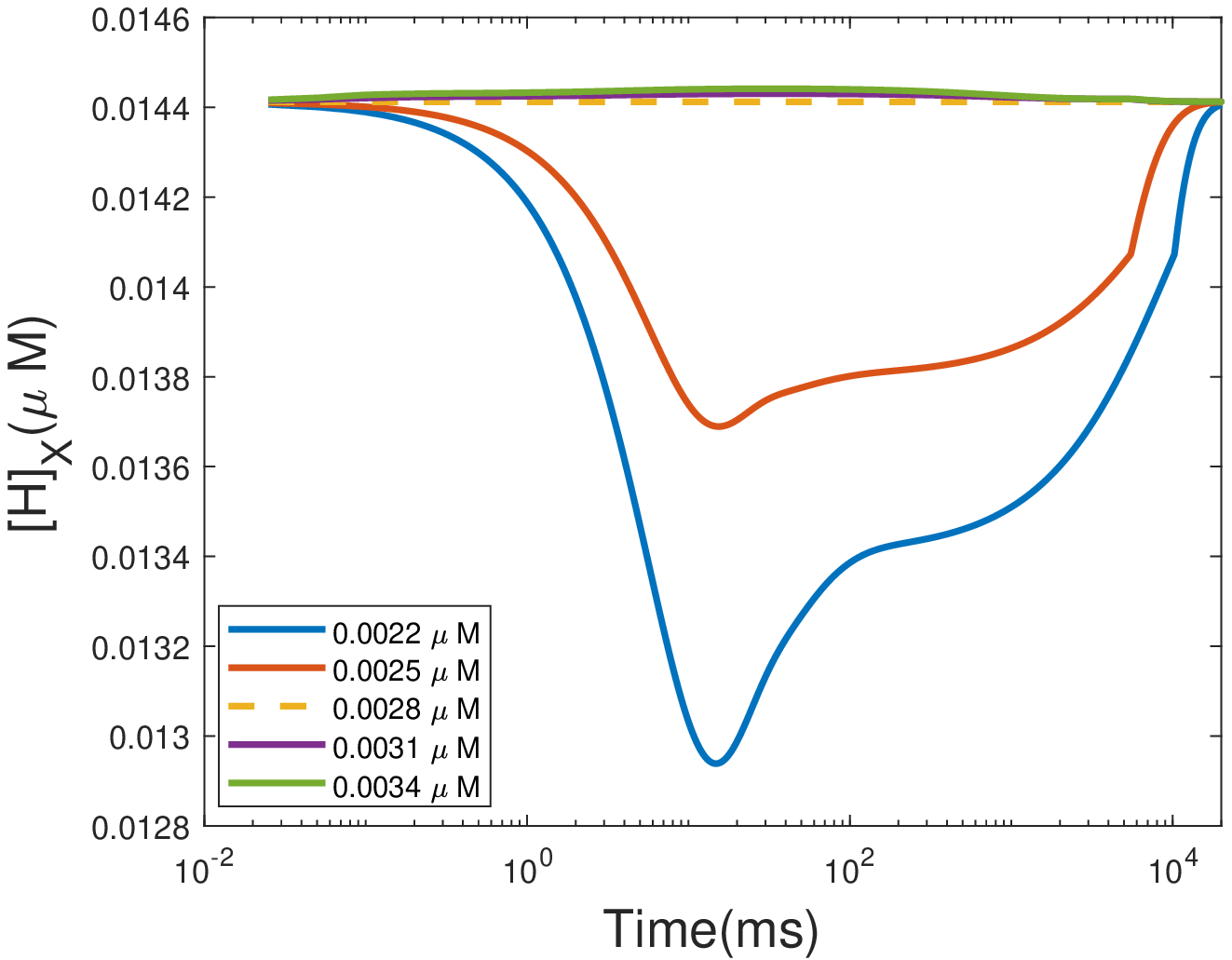}}
				\end{subfigure}
				\caption{Concentration with Different Oxygen concentration. (a) Reaction rate; (b) Electron concentration $\rho_e$; (c) $[H]_B$; (d) $[H]_E$; (e) $[H]_X$. The dash lines are results with default parameters.}
				\label{fig:Concentration_ox}
			\end{figure}

\begin{figure}[!ht]
				\centering
				\begin{subfigure}[]{
						\includegraphics[width=6.in]{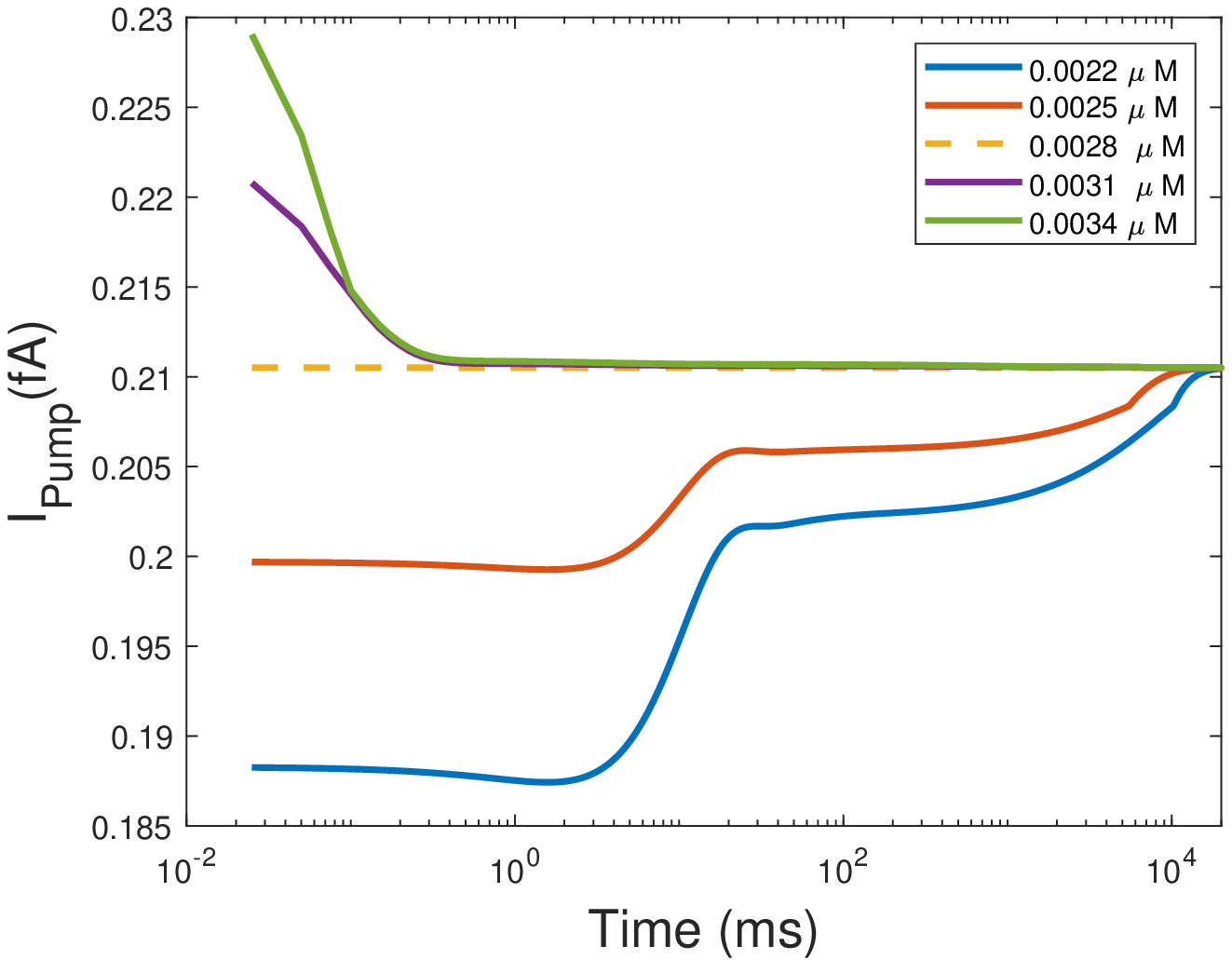}
					}
				\end{subfigure}
				\begin{subfigure}[]{
						\includegraphics[width=2.in]{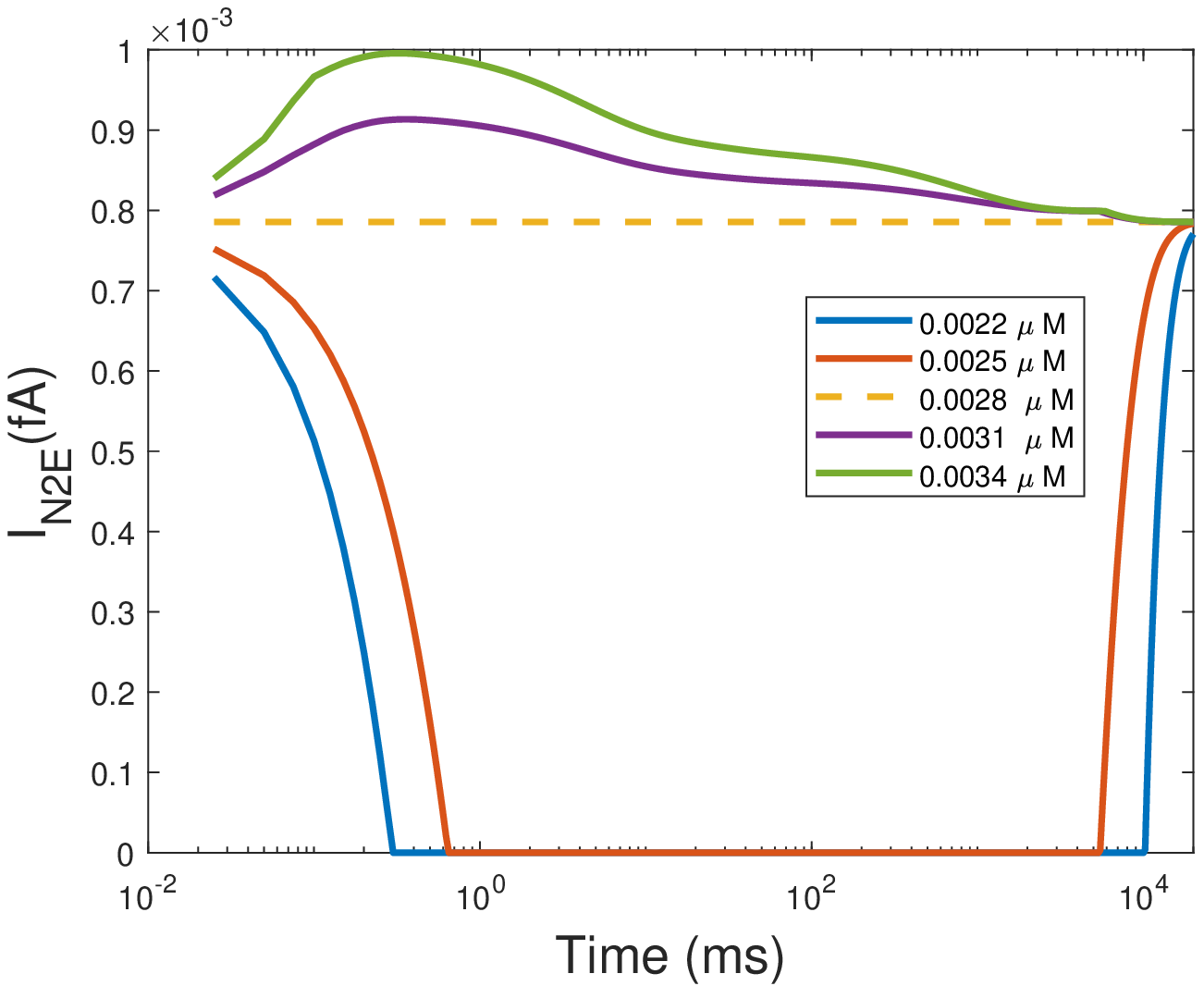}}
				\end{subfigure}
				\begin{subfigure}[]{
						\includegraphics[width=2.in]{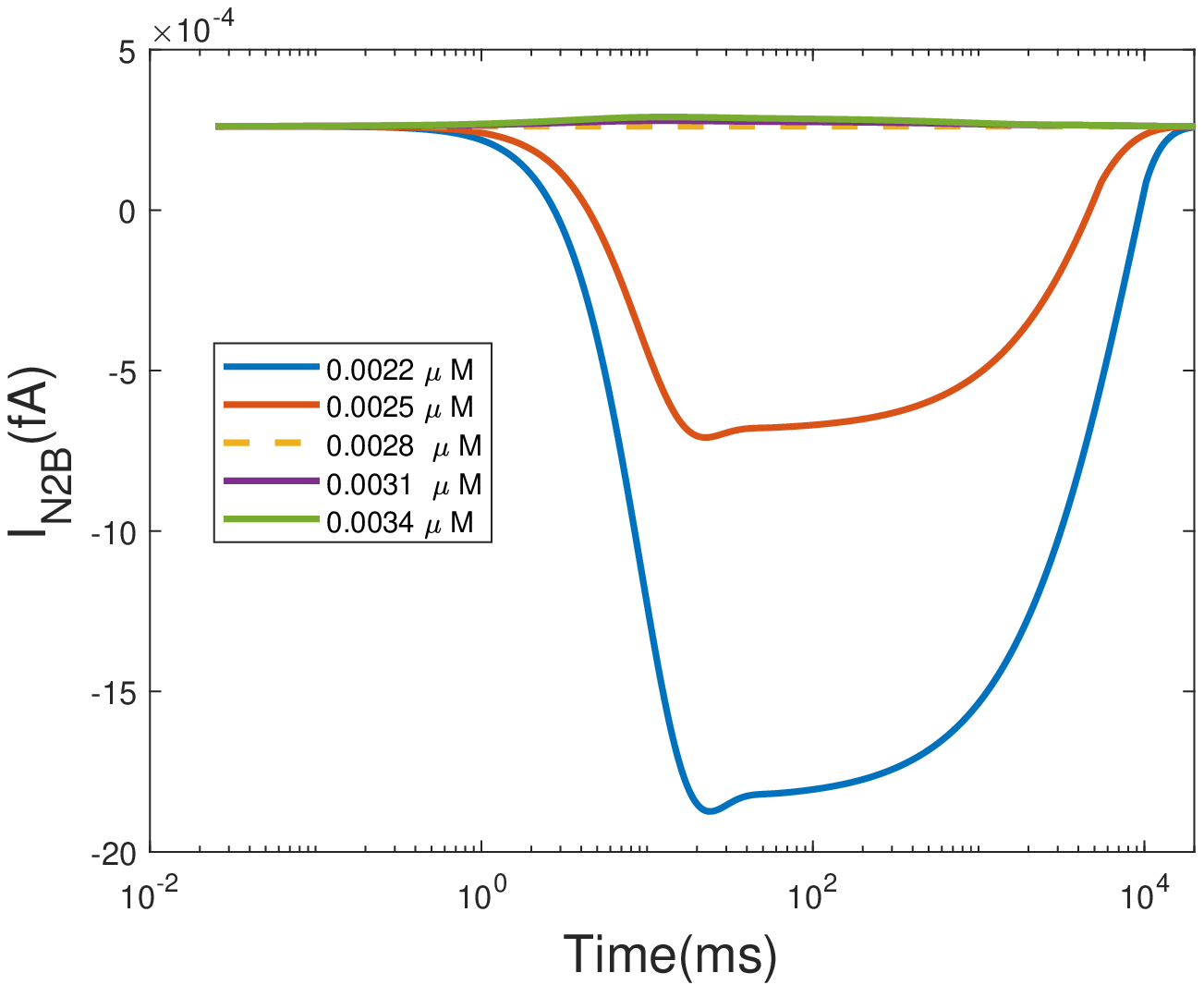}}
				\end{subfigure}
				\begin{subfigure}[]{
						\includegraphics[width=2.in]{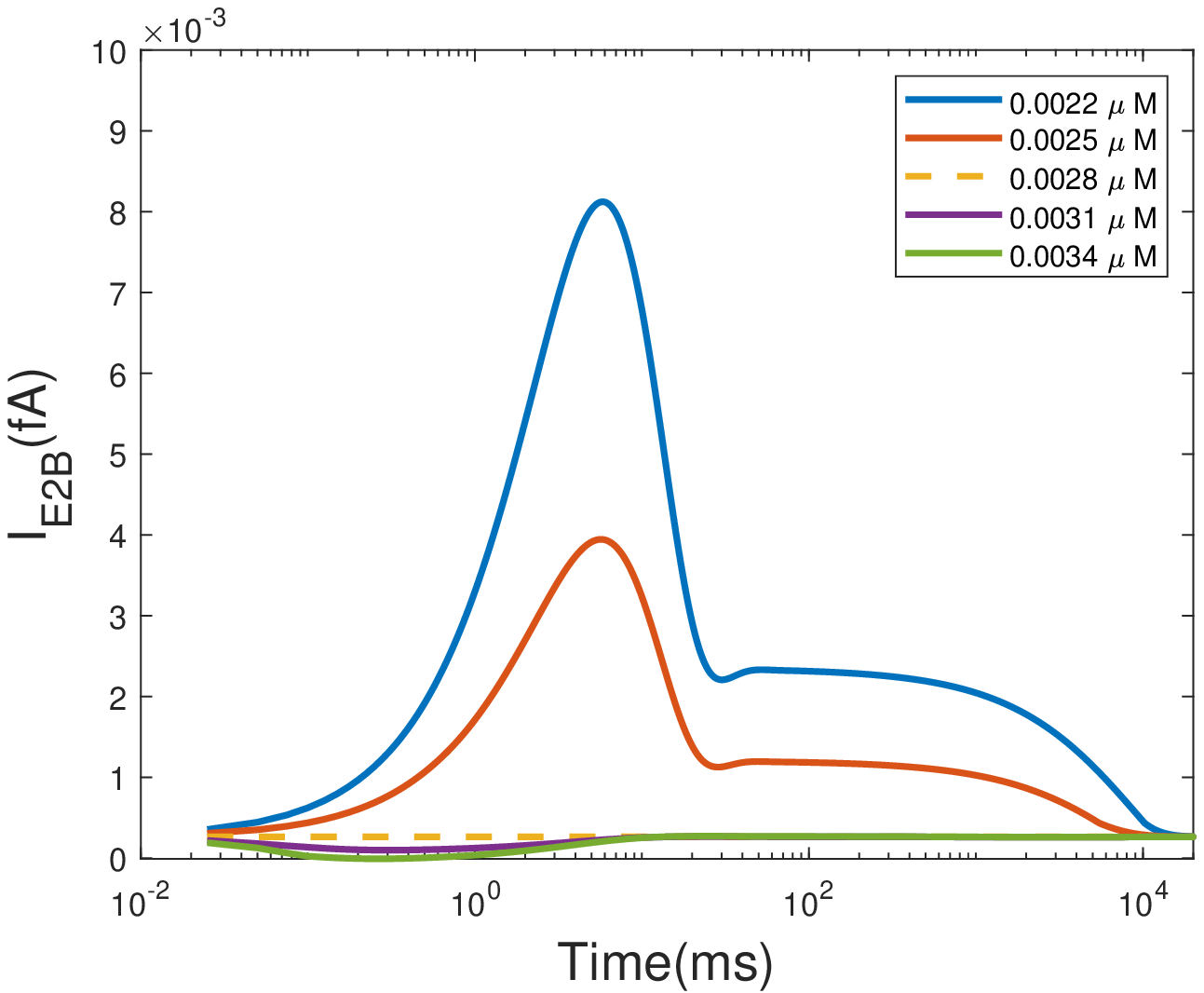}}
				\end{subfigure}
				\begin{subfigure}[]{
						\includegraphics[width=2.in]{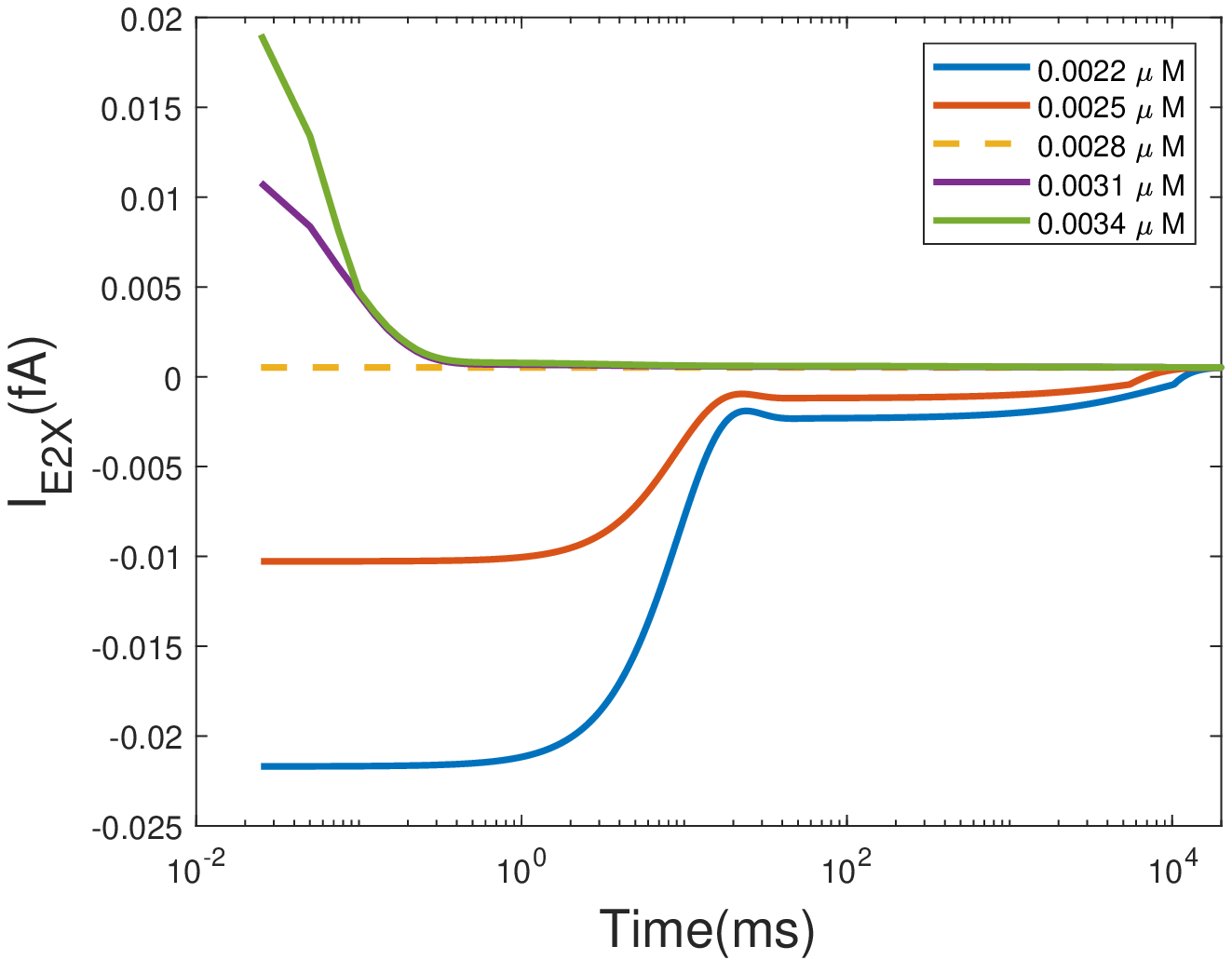}}
				\end{subfigure}
				\begin{subfigure}[]{
						\includegraphics[width=2.in]{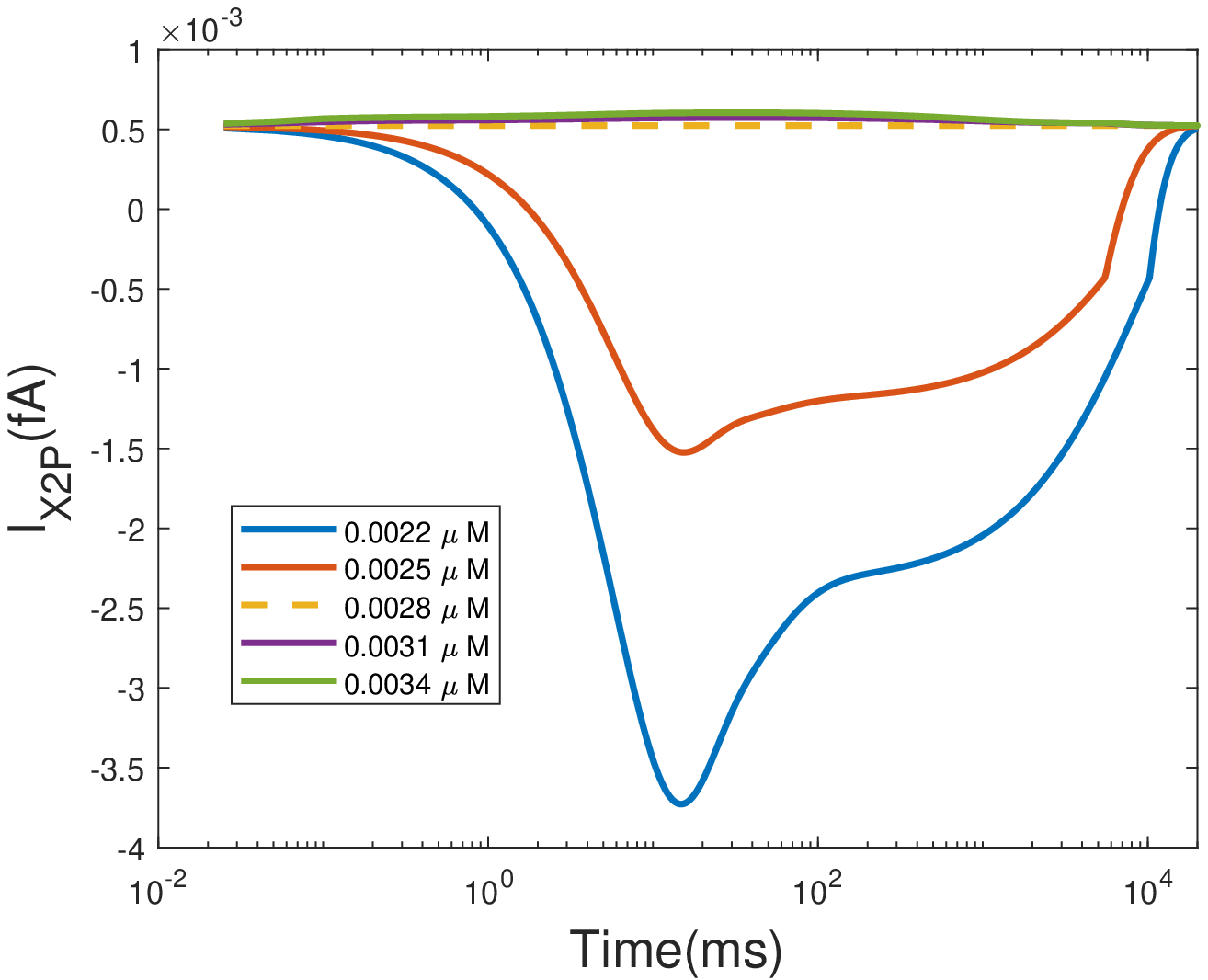}}
				\end{subfigure}
				\begin{subfigure}[]{
						\includegraphics[width=2.in]{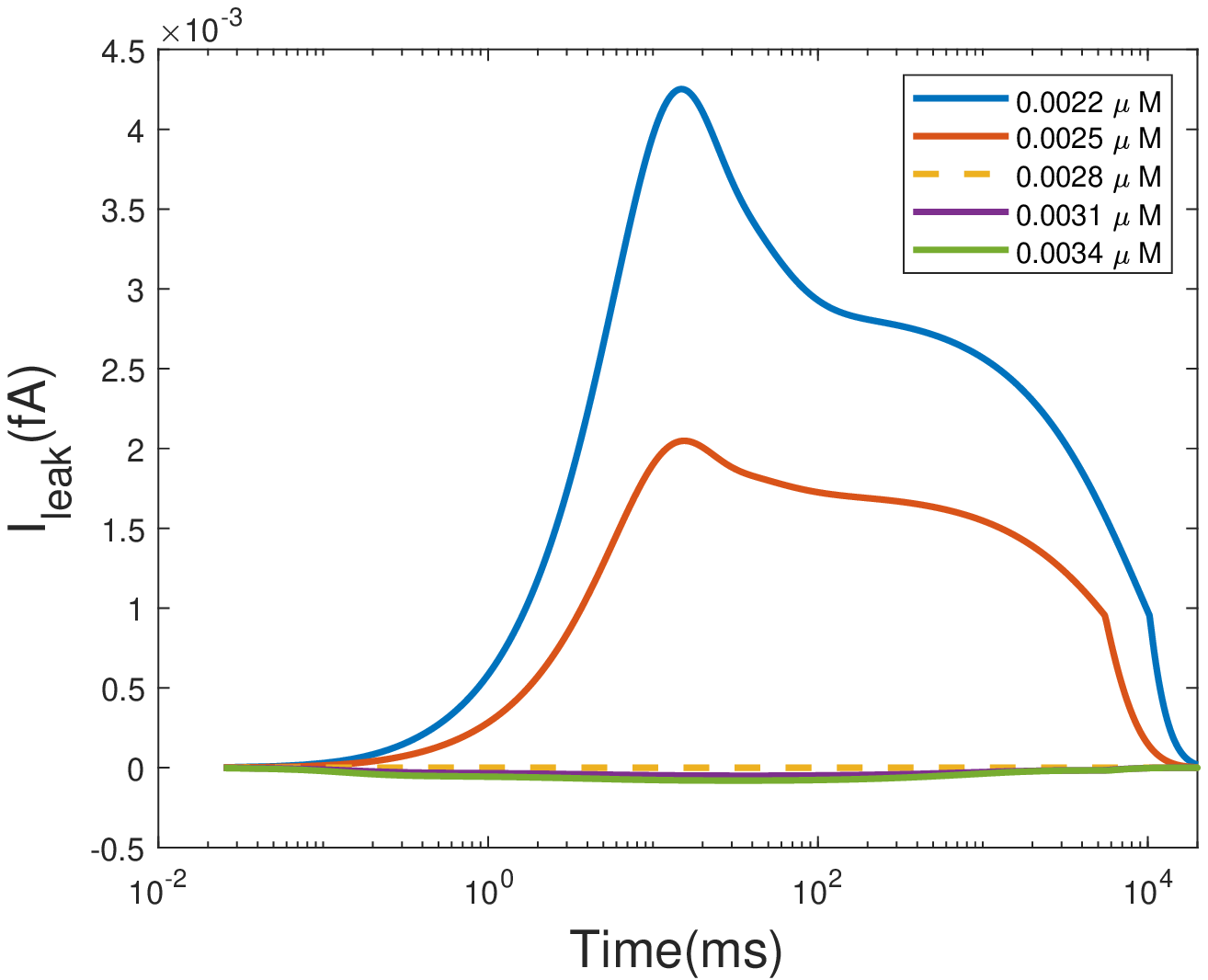}}
				\end{subfigure}
				
				\caption{Current with Different Oxygen concentration. (a) Pump  $I_{pump}$; (b) $I_{N2E}$; (c) $I_{N2B}$; (d) $I_{E2B}$; (e) $I_{E2X}$; (f) $I_{X2P}$; (g) $I_{leak}$. The dash lines are results with default parameters.}
				\label{fig:Current_ox}
			\end{figure}

			The effect of oxygen concentration in the reaction site is studied next. Fig.\ref{fig:Concentration_ox} shows the dynamics (i.e., time dependence) of the concentration of ions in different compartments at different oxygen concentrations. The dashed lines are computed with default parameters shown in Tables 1-2.  Panel  (a) shows that decreasing oxygen concentration at first decreases the reaction rate. The decrease in reaction rate produces the accumulation of electrons  (see panel  (b)) as they are supplied from the input source of constant electron current. The proton concentration in the  reaction site also increases: the reaction rate decreases as the accumulated electron attracts more protons from the E242 site (see Fig. \ref{fig:Current_ox} (c)). Since the pump strength depends on the reaction rate, the pump current $I_{Pump}$ also decreases (as shown in Fig. \ref{fig:Current_ox}(a)). The decrease in pump current  induces the increase at E242 and the decrease  at the PLS (Proton Loading) site (see  Fig. \ref{fig:Current_ox} (d)-(e)). Then the accumulated protons and electrons modify the reaction rate  $\mathcal{R}  $ in Eq. \eqref{reaction}. The reaction rate increases  until    $\mathcal{R} = -\frac{S_vI_e}{2F}$. That is in fact the same equilibrium as determined by the  default parameters. 
			
			At the same time, due to the accumulation of protons in E242, the chemical potential
			$\mu_E$ is larger than then N side $\mu_N$. The result is an activation of the rectifier making the current $I_{N2E}$  zero (see Fig. \ref{fig:Current_ox} (b)).  This action depends on the rectifier, There is no rectifier between the reaction site and N site, so the behavior is quite different.When the protons are accumulated in B, the current $I_{N2B}$ is negative. Similarly, fewer protons are pumped to the  Protein Loading Site PLS site. The chemical potential at that site $\mu_X$ is smaller than the P side $\mu_P$. Negative current is also observed in Fig. \ref{fig:Current_ox}  (f).
			
\begin{figure}[!ht]
				\centering
				\begin{subfigure}[]{
						\includegraphics[width=3.in]{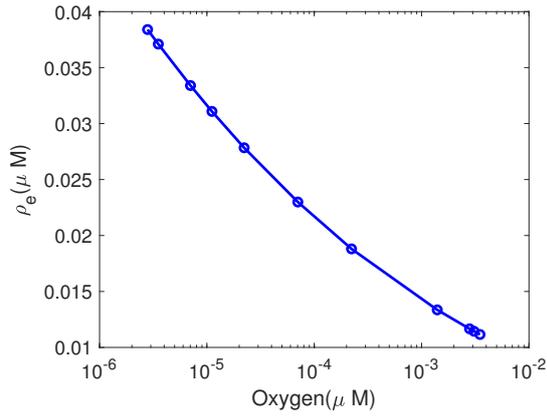}}
				\end{subfigure}
				\begin{subfigure}[]{
						\includegraphics[width=3.in]{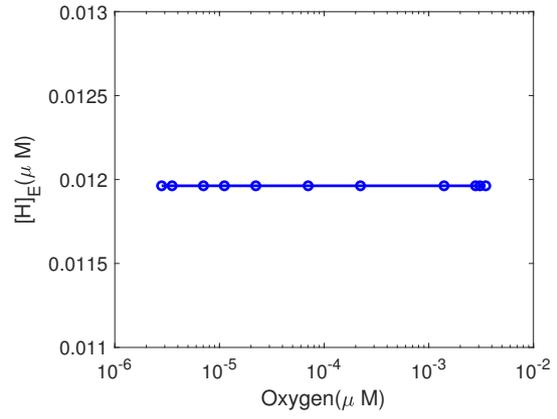}}
				\end{subfigure}
				\begin{subfigure}[]{
						\includegraphics[width=3.in]{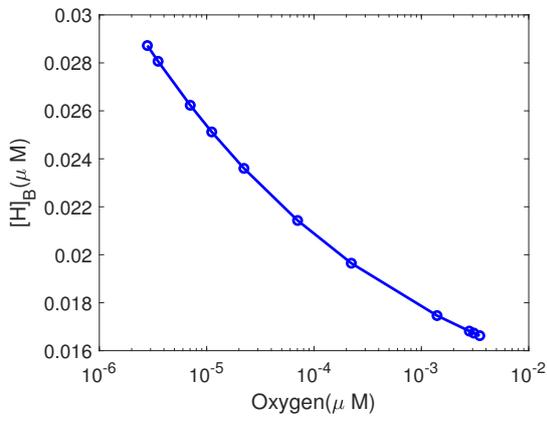}}
				\end{subfigure}
				\begin{subfigure}[]{
						\includegraphics[width=3.in]{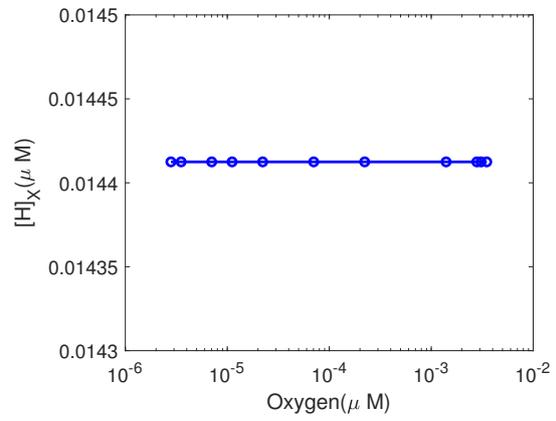}}
				\end{subfigure}
				\caption{Concentration at equilibrium states with different oxygen concentration. (a) Electron concentration $\rho_e$;  (b) $[H]_E$; (c) $[H]_B$;(d) $[H]_X$.  }
				\label{fig:Concentration_ox_eq}
			\end{figure}

\begin{figure}[!ht]
				\centering
				\begin{subfigure}[]{
						\includegraphics[width=3.in]{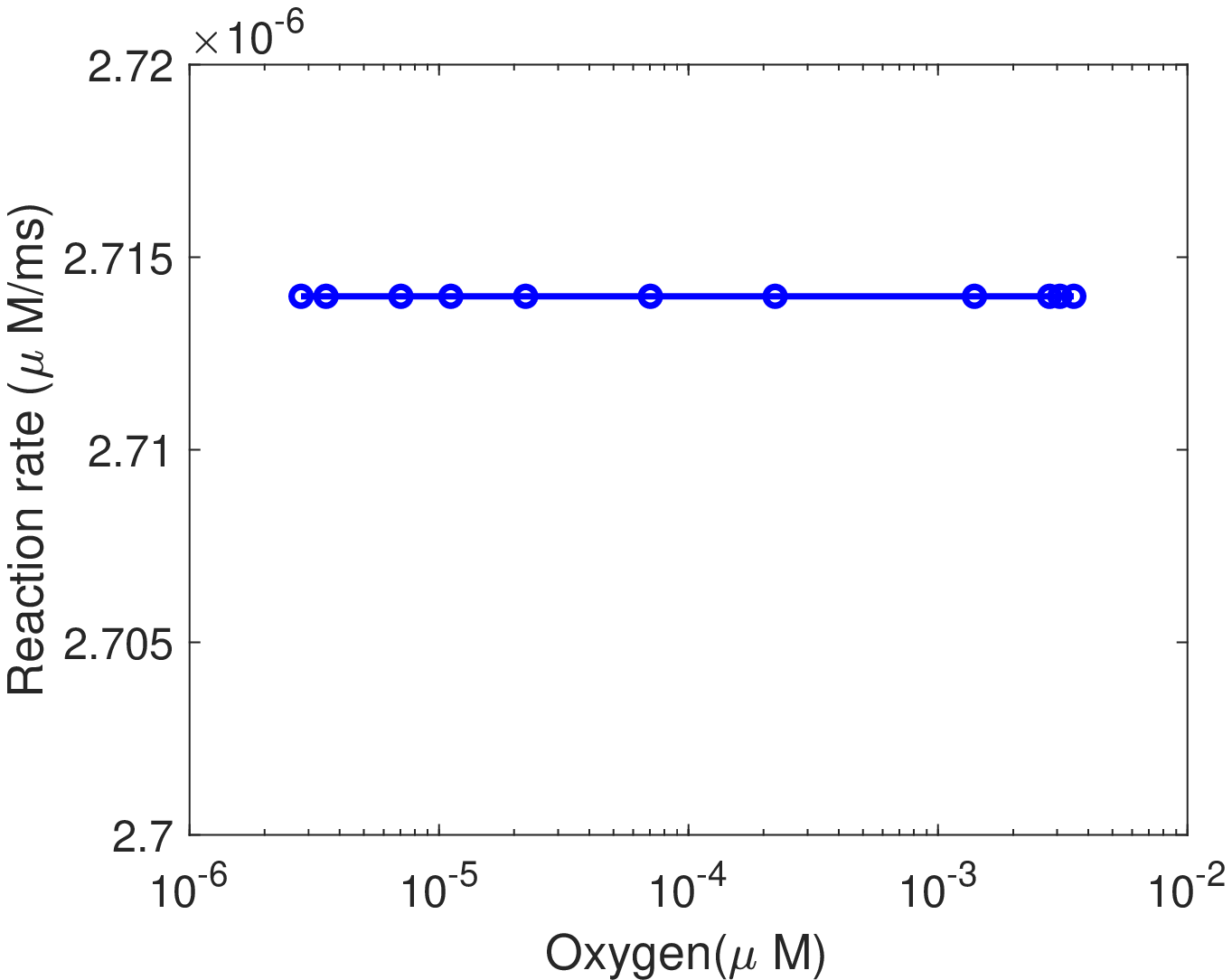}}
				\end{subfigure}
				\begin{subfigure}[]{
						\includegraphics[width=3.in]{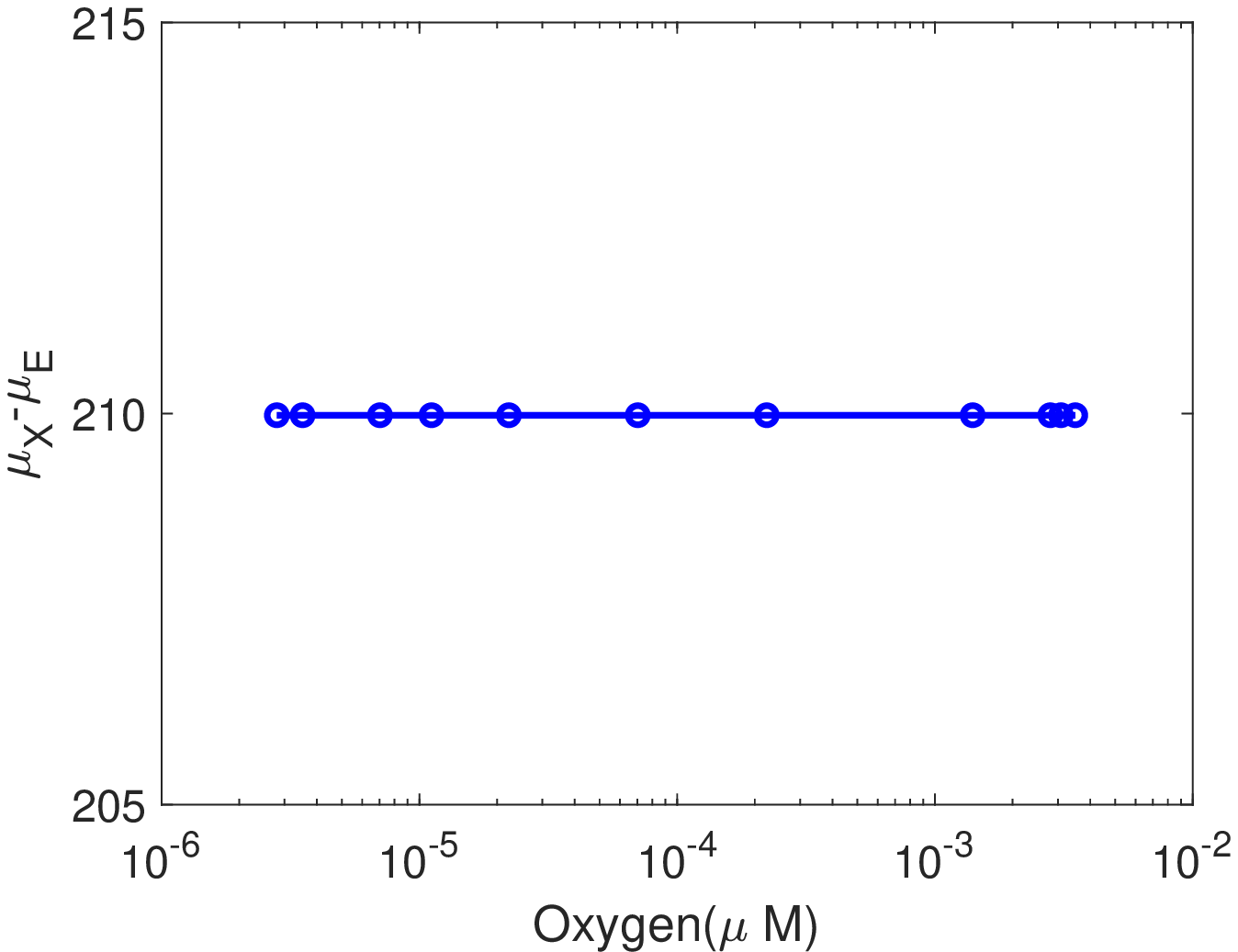}}
				\end{subfigure}
				\begin{subfigure}[]{
						\includegraphics[width=3.in]{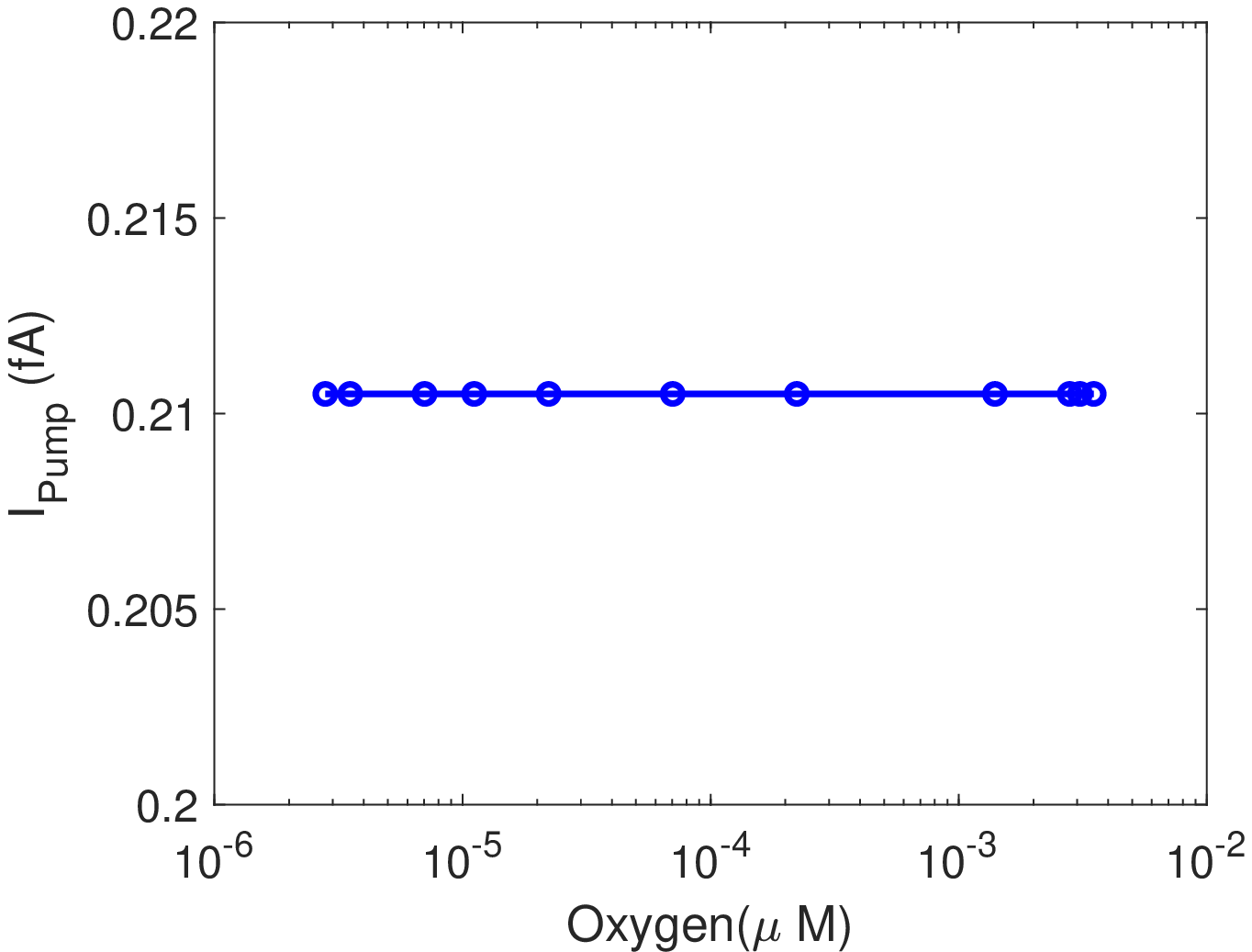}}
				\end{subfigure}
				\begin{subfigure}[]{
						\includegraphics[width=3.in]{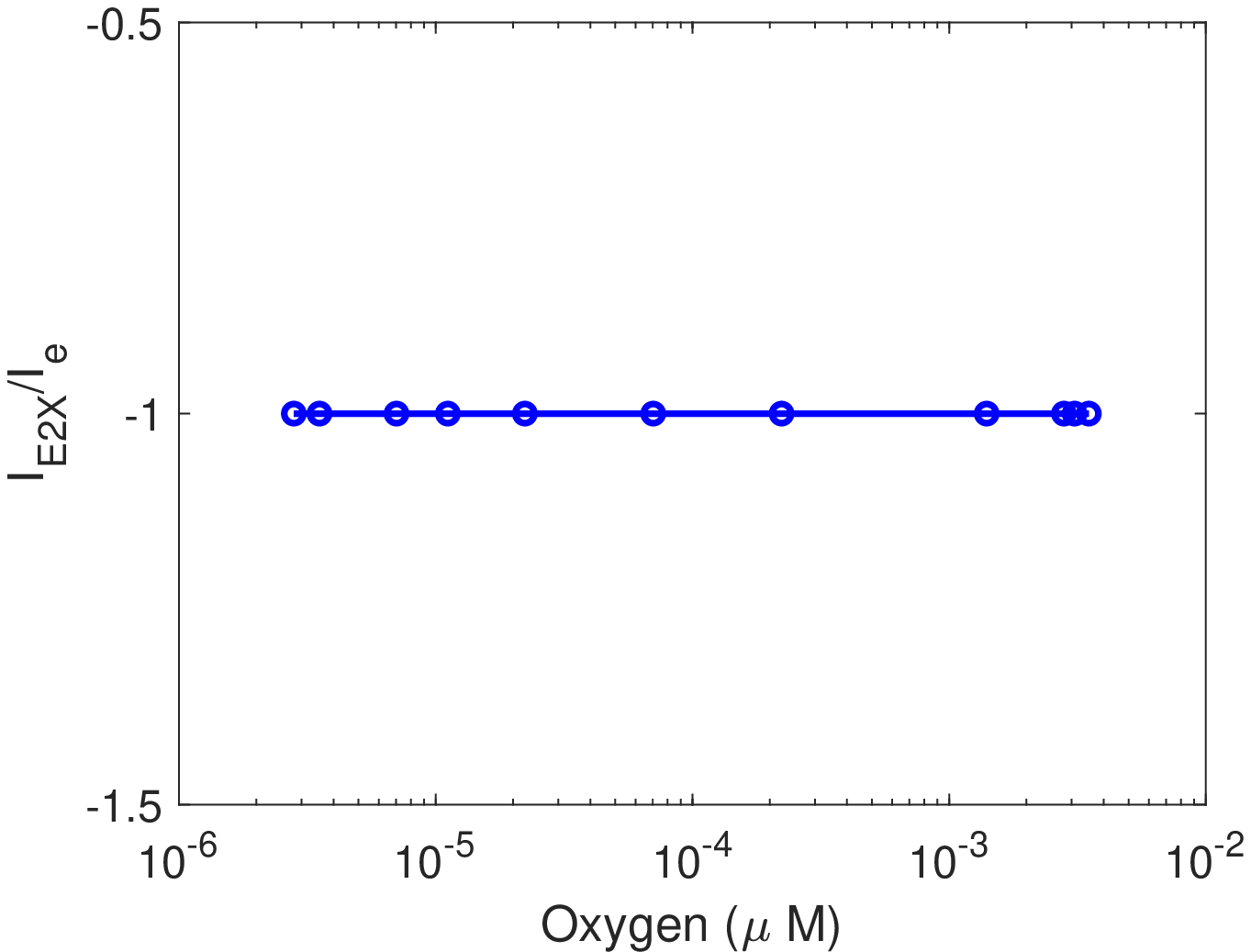}}
				\end{subfigure}
				\caption{Pump efficiency at equilibrium states with different oxygen concentration $[H]_N$ and different threshold. (a) Reaction rate;  (b) $\mu_X -\mu_E$; (c) $I_{Pump}$;(d) $I_{E2X}/I_e$. }
				\label{fig:Pump_ox_eq}
			\end{figure}

			We confirm and extend the above observations by changing the oxygen concentration from $10^{-6}\mu M$ to $10^{-2}\mu M$.  
			As shown in  Fig. \ref{fig:Concentration_ox_eq} and Fig. \ref{fig:Pump_ox_eq},  the decrease of oxygen concentration (at the equilibrium) changes just the concentrations of electron and proton at the BNC at the equilibrium state. 
			 The reaction rate at equilibrium keeps constant due to constant supplyment of electron flux $I_e$ and Eq.\eqref{rhoequation}.    
			In this case, the proton transportation follows the normal pattern as shown in Fig. \ref{fig:protonflow} a.  
			
			\subsection{Effect of the Switch}
			In this section,  the effect of the switch on the complex IV function is studied when the oxygen level in the reaction site is very low $2.8E-5\mu M$ which is  1\%  of the default value.    Especially, here we assume the switch is perfect $SW_0=0$ and defective ones with $SW_0= 1E-1,-2,-3 fA$.   
			
\begin{figure}[!ht]
				\centering
				\begin{subfigure}[]{
						\includegraphics[width=3.in]{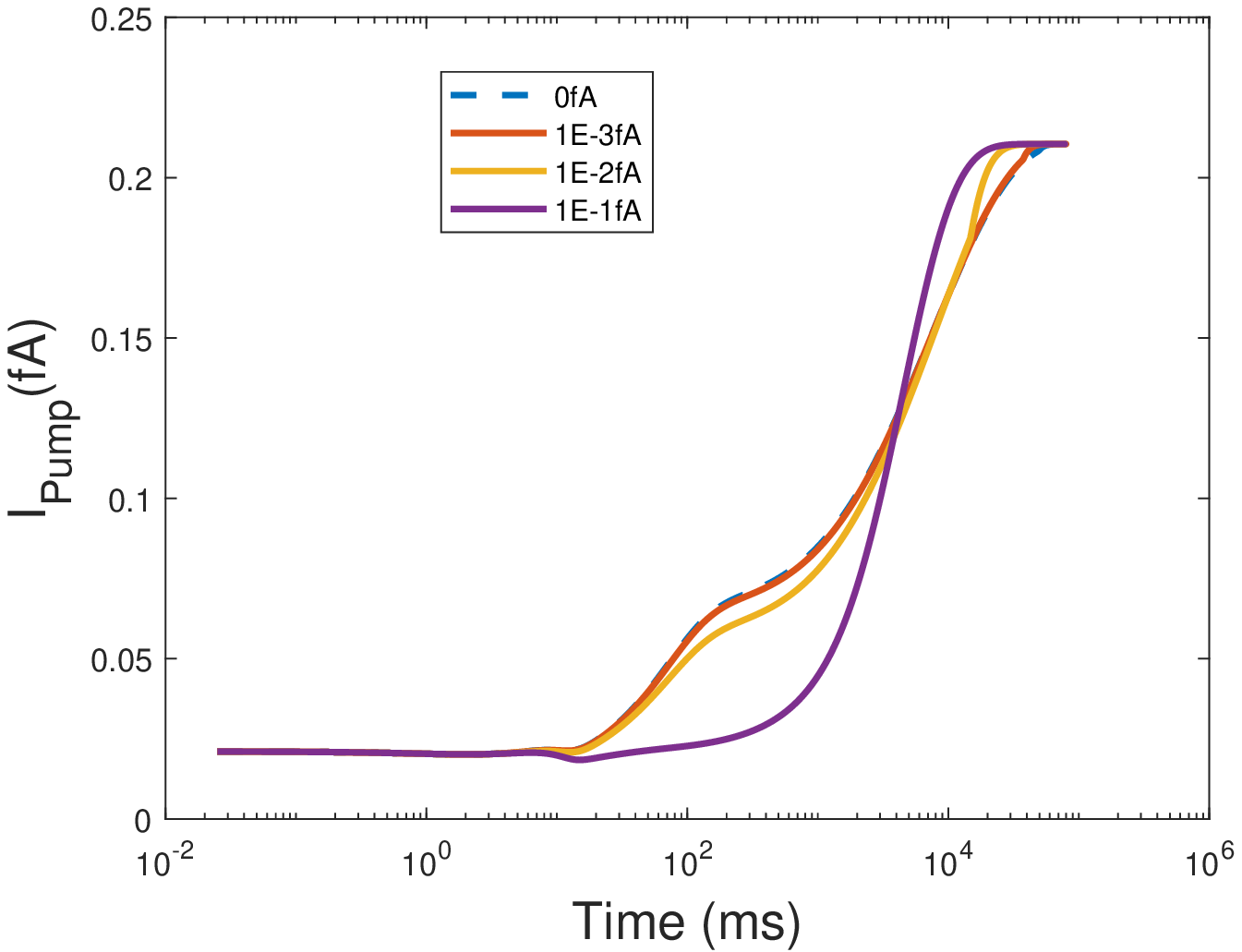}
					}
				\end{subfigure}
				\begin{subfigure}[]{
						\includegraphics[width=3.in]{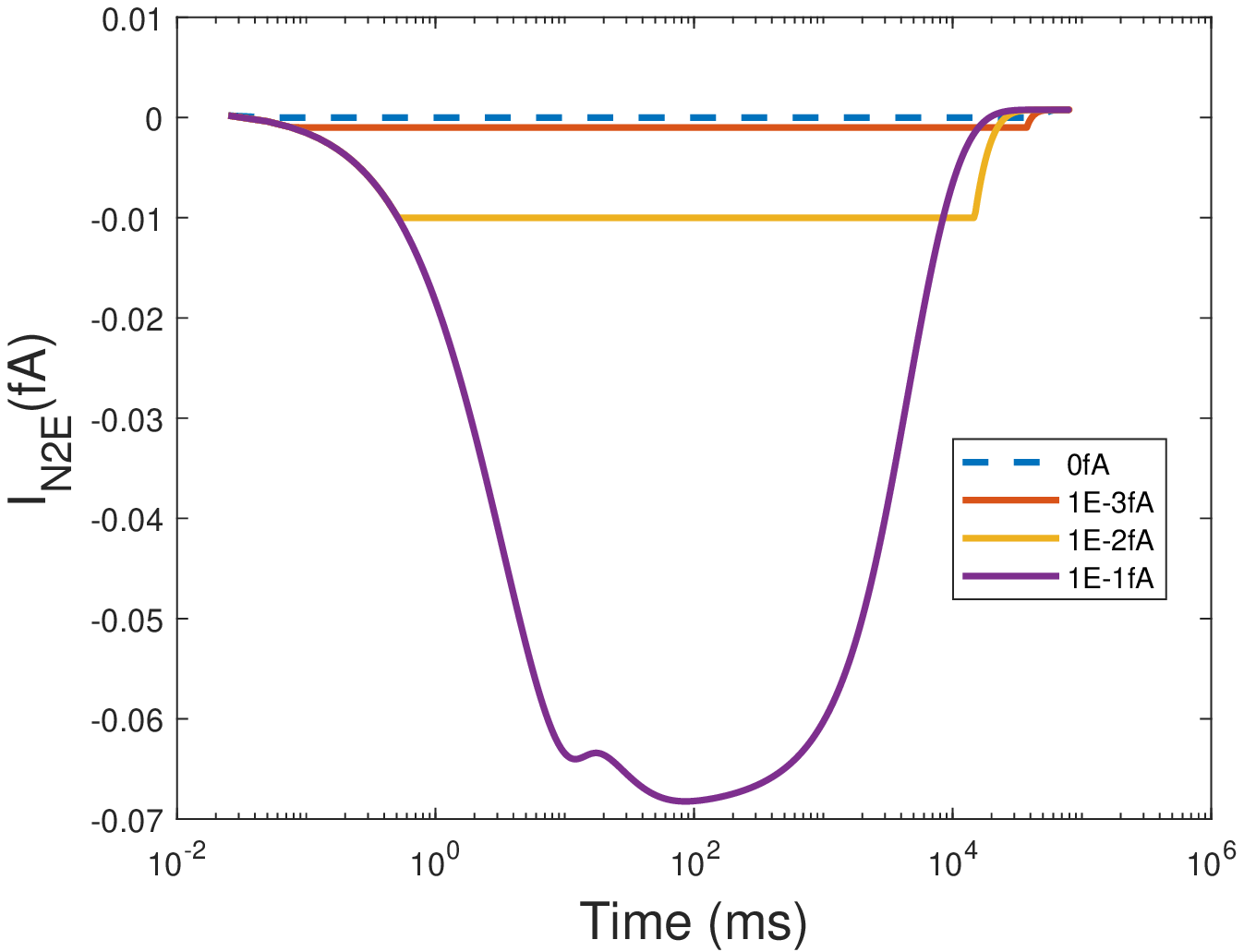}}
				\end{subfigure}
				\begin{subfigure}[]{
						\includegraphics[width=2.in]{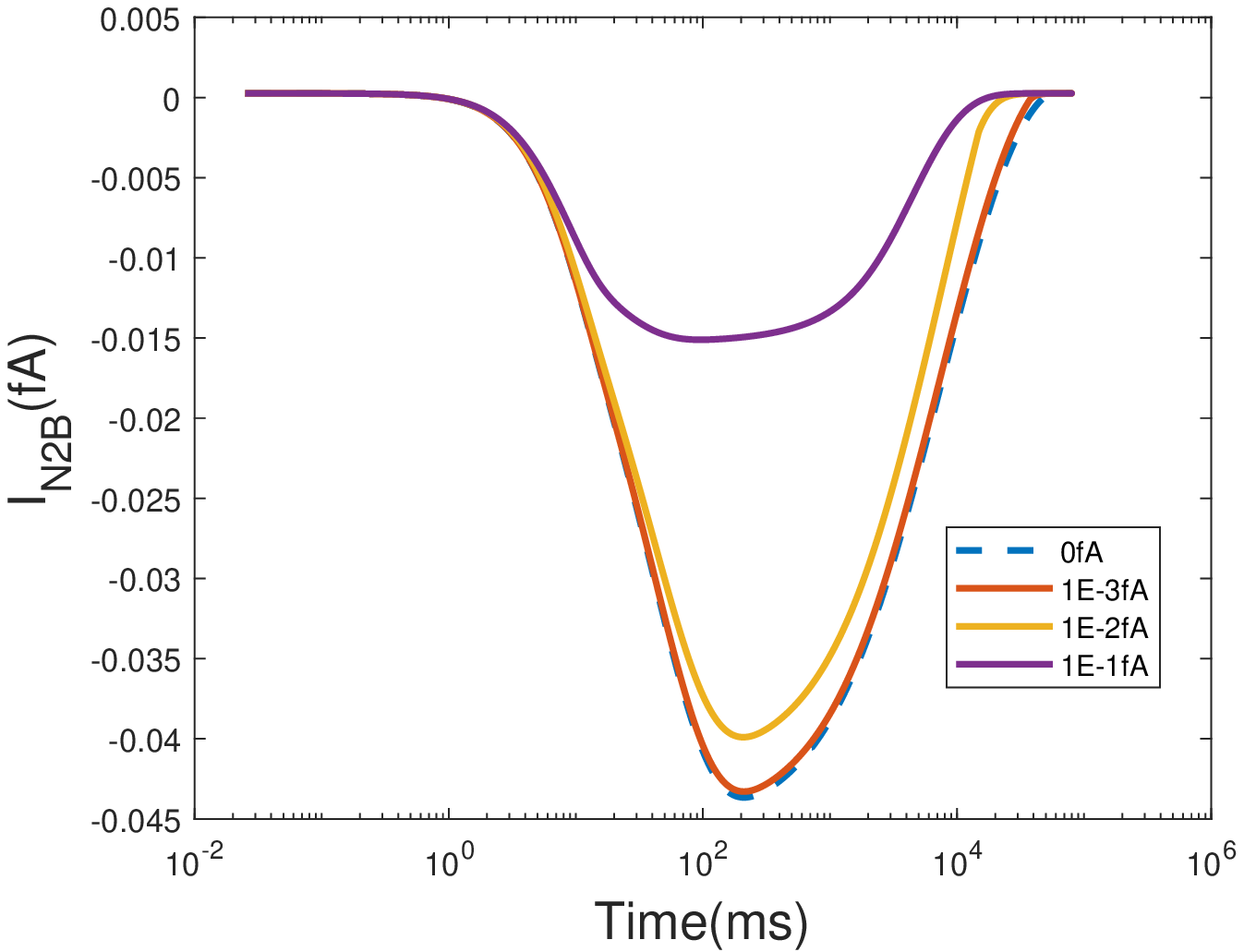}}
				\end{subfigure}
				\begin{subfigure}[]{
						\includegraphics[width=2.in]{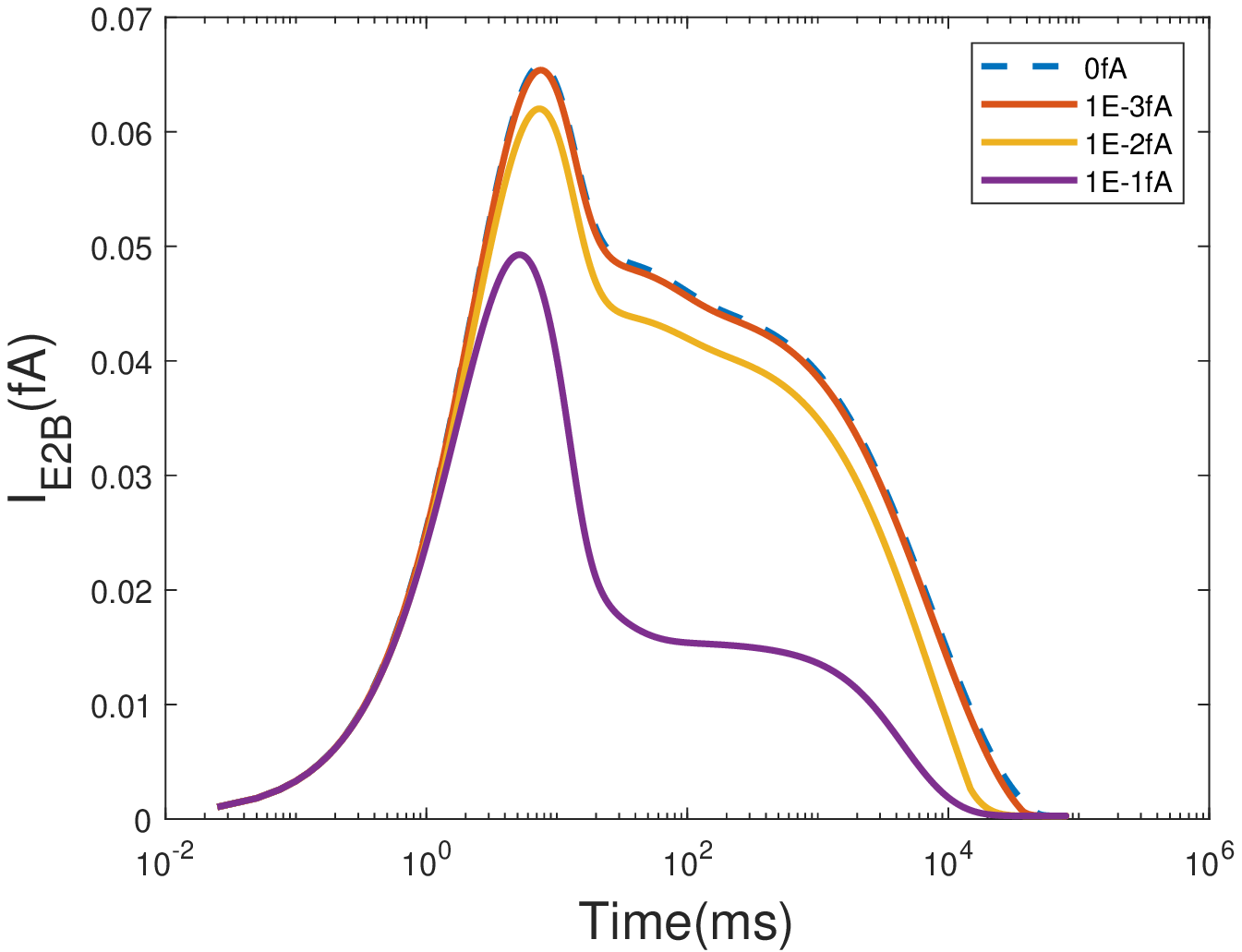}}
				\end{subfigure}
				\begin{subfigure}[]{
						\includegraphics[width=2.in]{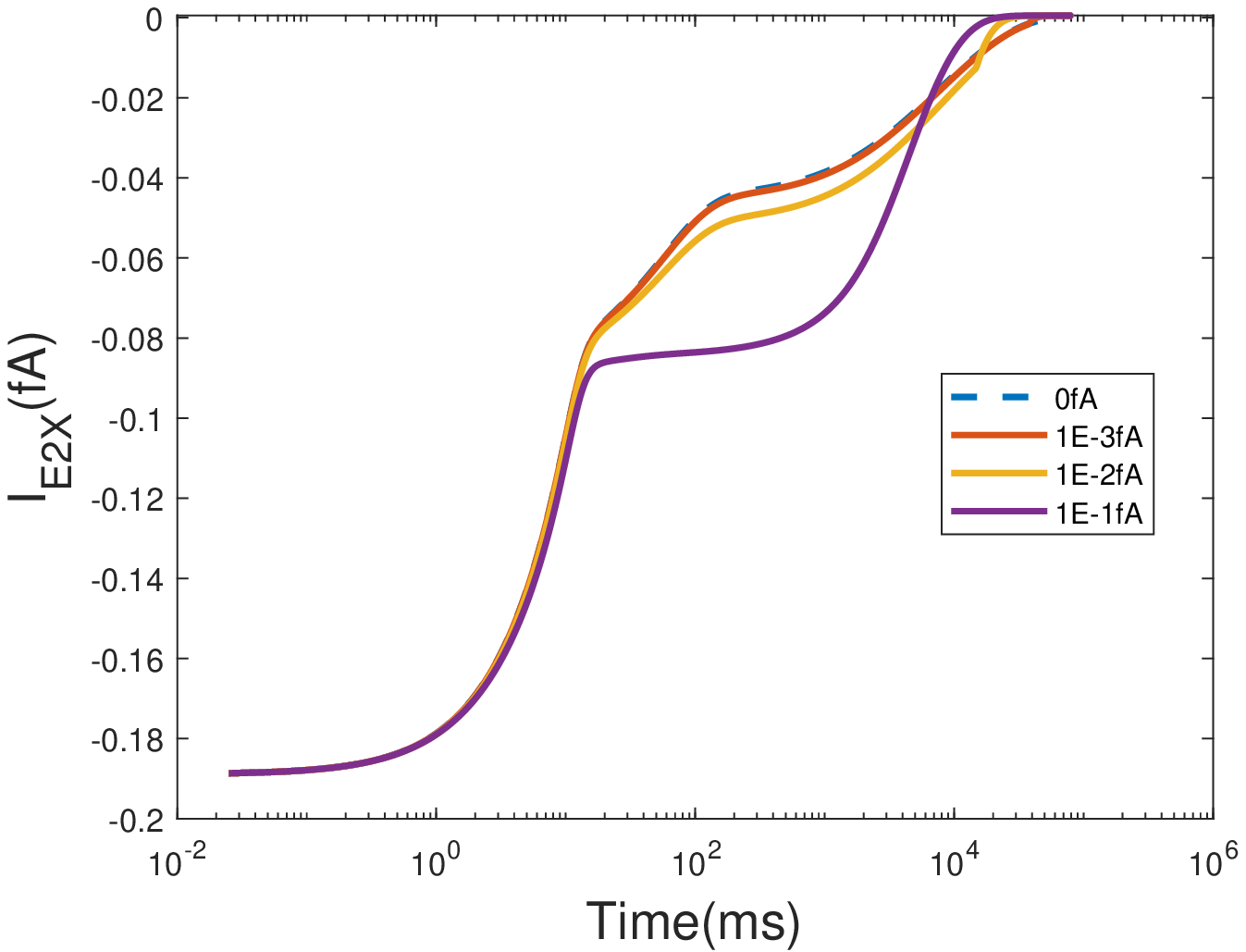}}
				\end{subfigure}
				\begin{subfigure}[]{
						\includegraphics[width=2.in]{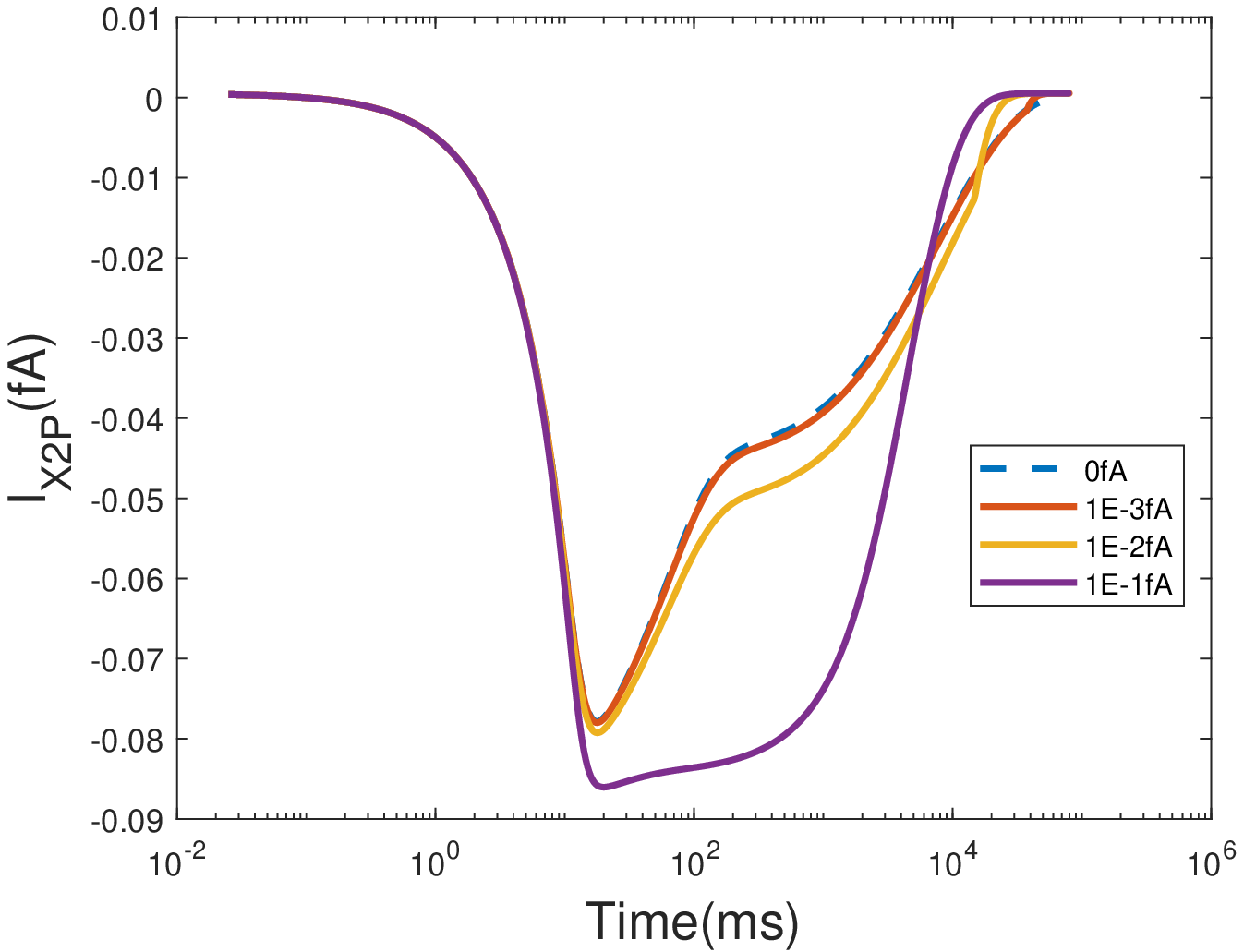}}
				\end{subfigure}
				\begin{subfigure}[]{
						\includegraphics[width=2.in]{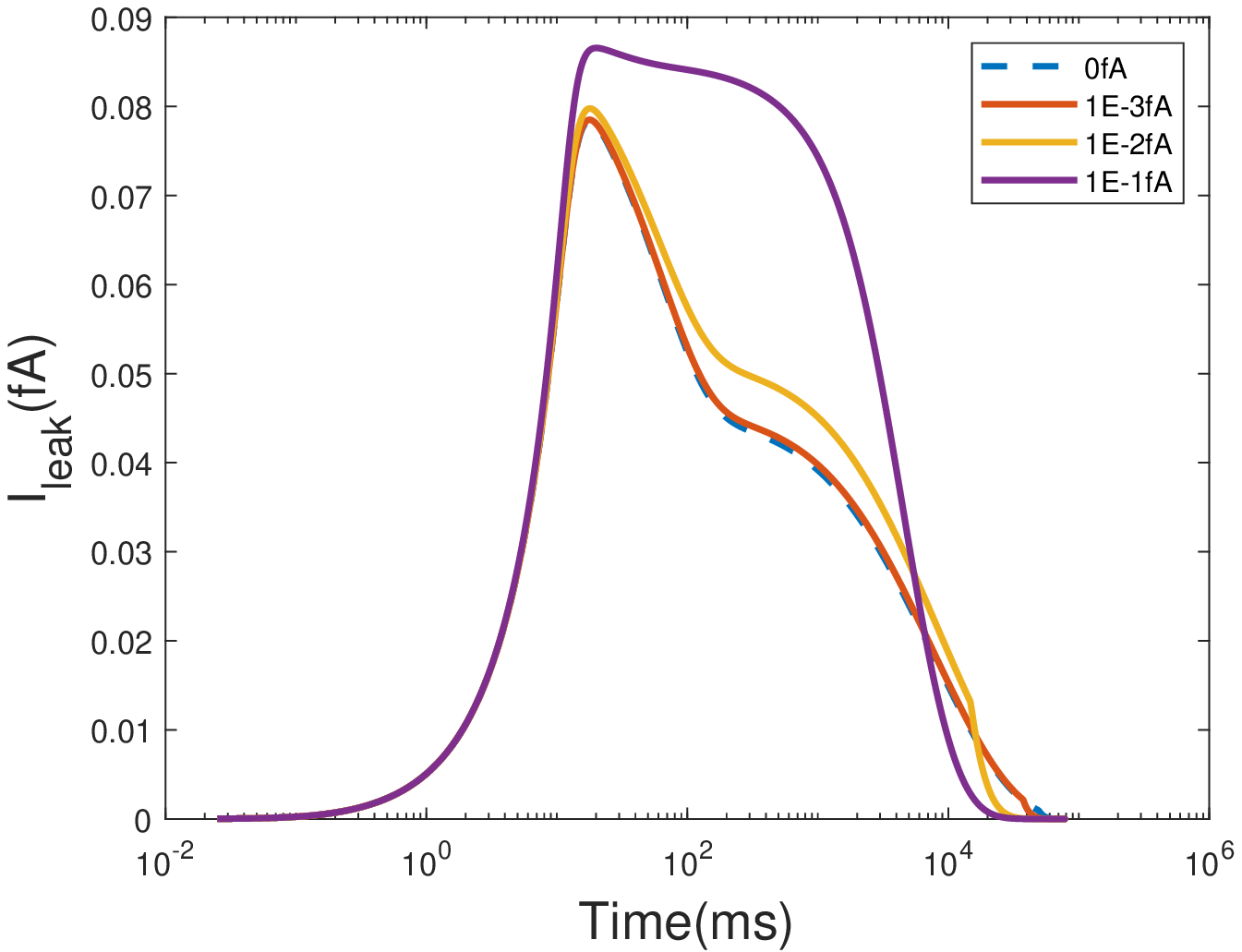}}
				\end{subfigure}
				
				\caption{Current with   Oxygen concentration  $2.8E-5\mu M$ and switch threshold. (a) Pump  $I_{pump}$; (b) $I_{N2E}$; (c) $I_{N2B}$; (d) $I_{E2B}$; (e) $I_{E2X}$; (f) $I_{X2P}$; (g) $I_{leak}$. The dash lines are results with default parameters.}
				\label{fig:Current_ox_sw}
			\end{figure}

\begin{figure}[!ht]
				\centering
				\begin{subfigure}[]{
						\includegraphics[width=3.in]{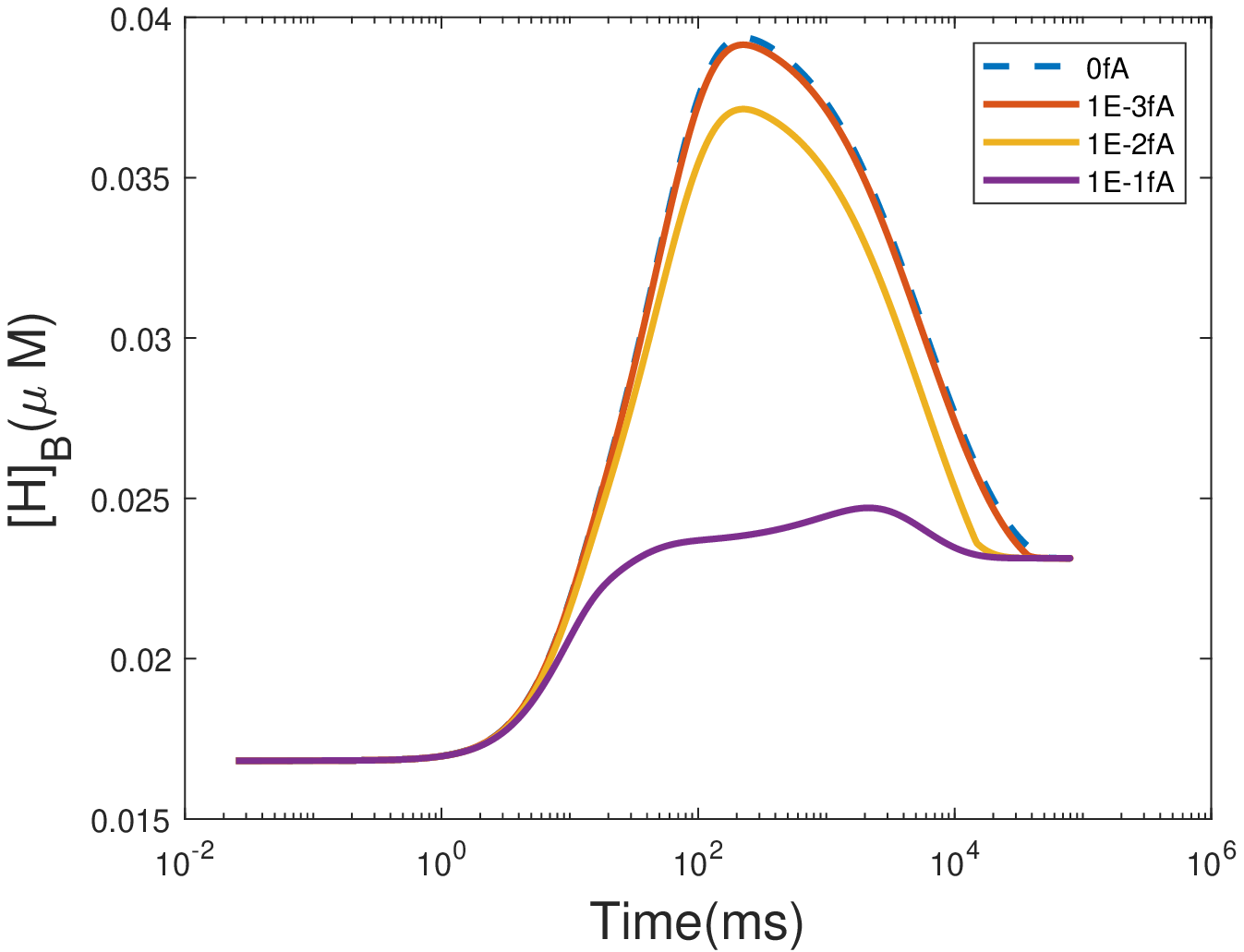}
					}
				\end{subfigure}
				\begin{subfigure}[]{
						\includegraphics[width=3.in]{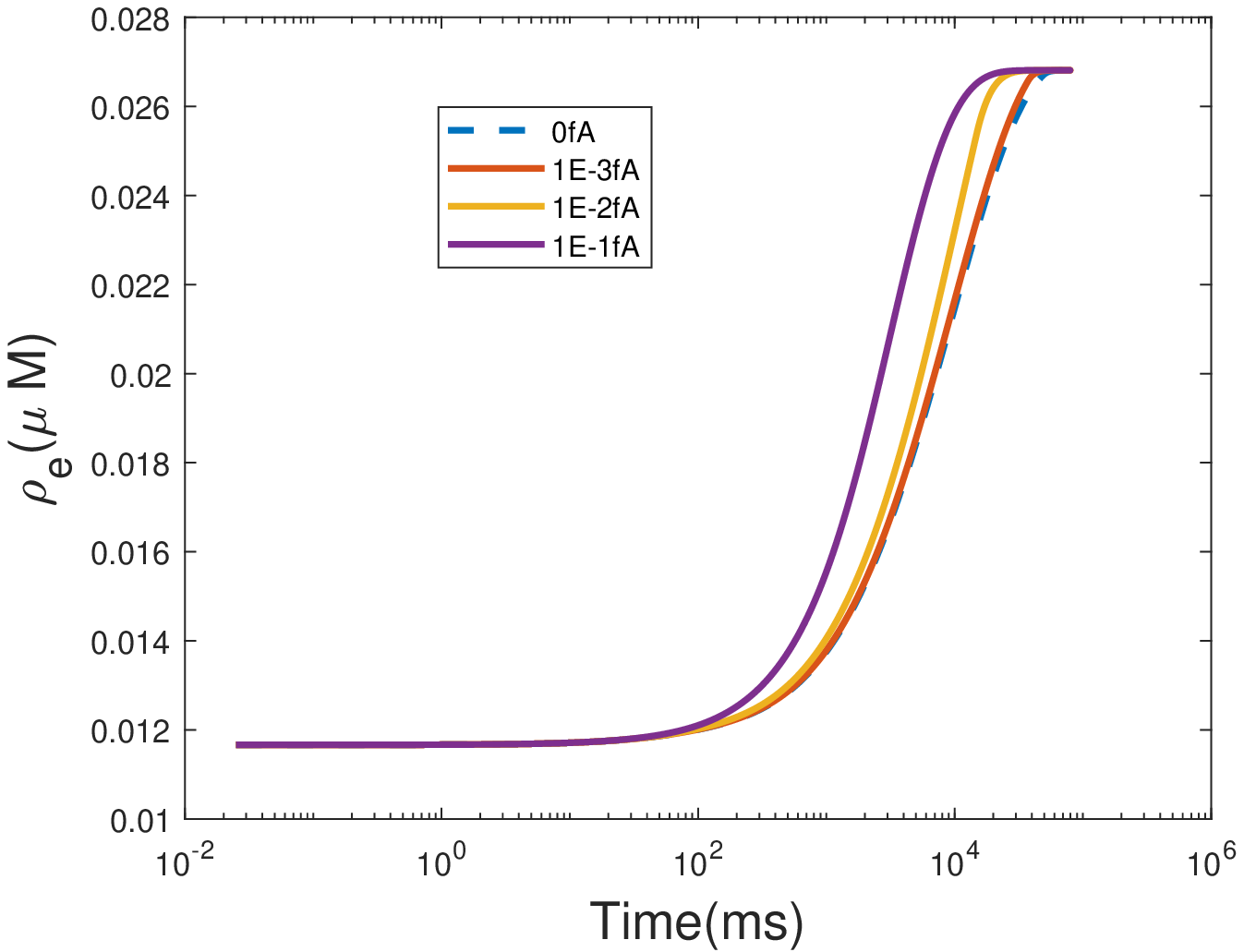}}
				\end{subfigure}
				\begin{subfigure}[]{
						\includegraphics[width=3.in]{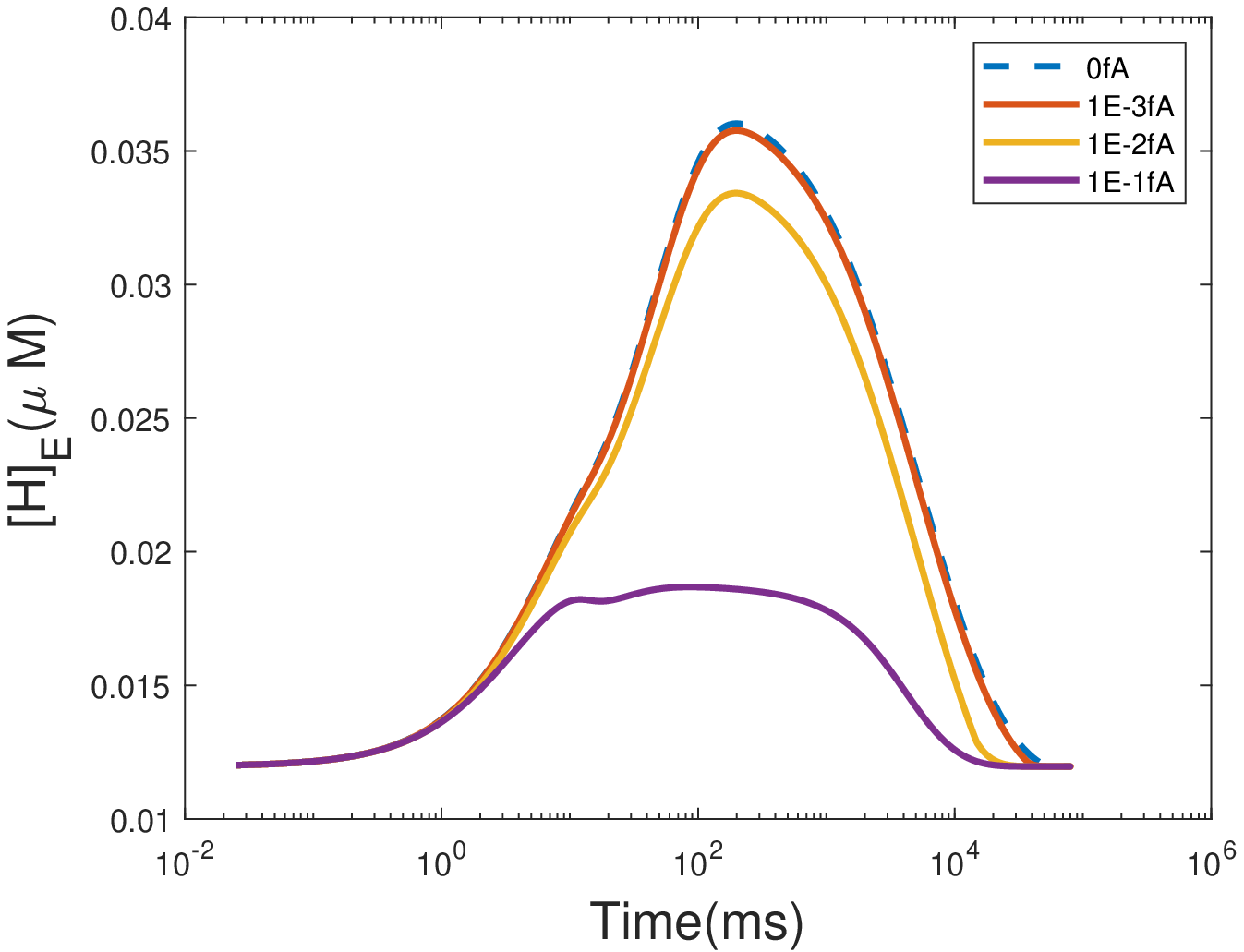}}
				\end{subfigure}
				\begin{subfigure}[]{
						\includegraphics[width=3.in]{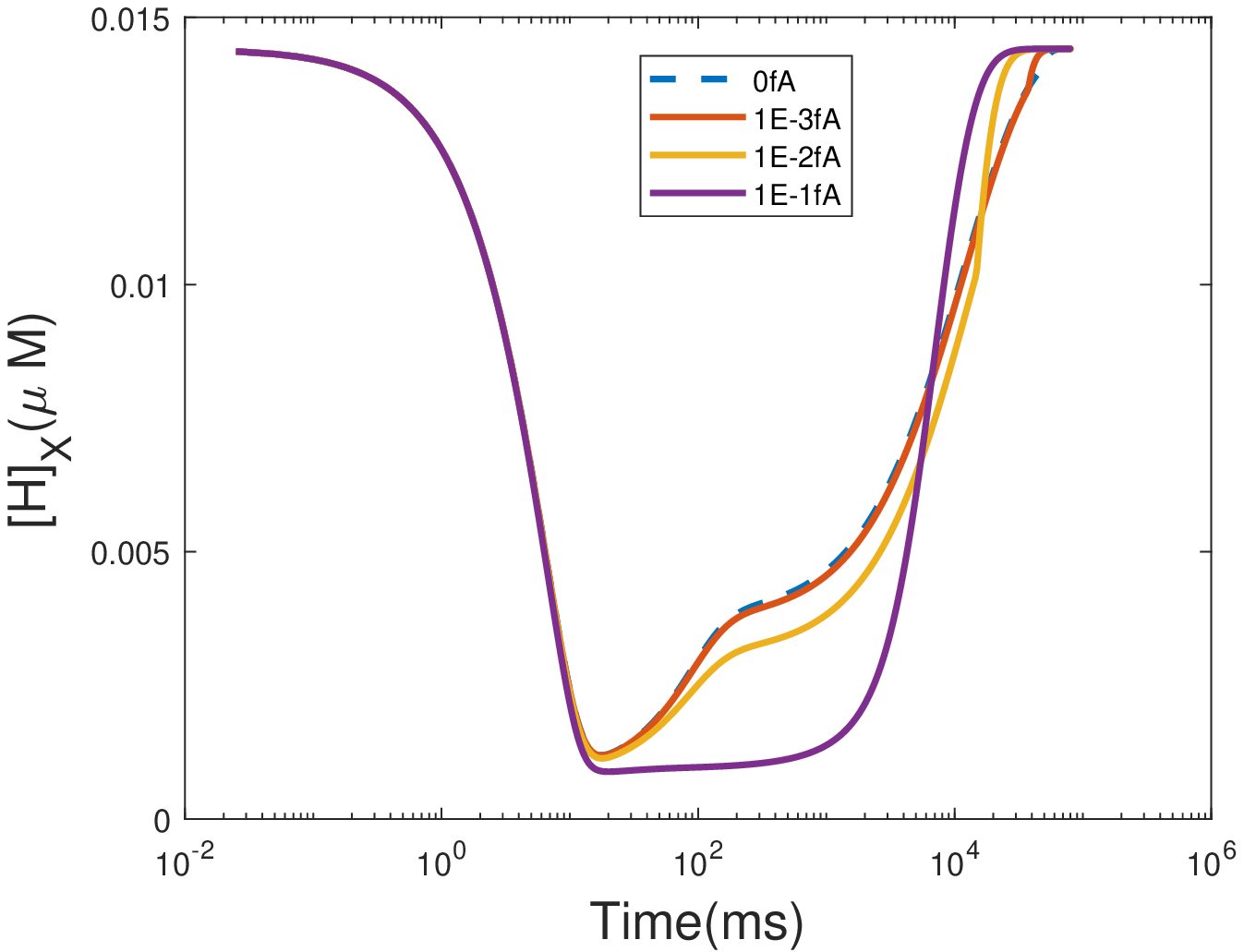}}
				\end{subfigure}
				\caption{Concentration with Different  $SW_0$. (a) Electron concentration $\rho_e$; (b) $[H]_B$; (c) $[H]_E$; (d) $[H]_X$. The dash lines are results with default parameters.}
				\label{fig:Concentration_ox_sw}
			\end{figure}

			In Fig. \ref{fig:Current_ox_sw}, Panel (a) is consistent with the results of the last section: the pump current decreases when the oxygen concentration is low. The pump current recovers when electrons and protons accumulate. When the switch allows larger counterflow, the pump current starts increases later but recovered faster to the default value.  The switch on the current $I_{N2E}$ is shown in Panel (b).  With small  $SW_0= 0 ,1E-2,-3 $,   the current is truncated if the counterflow current magnitude is larger than $SW_0$. When the switch allows larger counterflow, most of the accumulated  protons at E242 flow back from E242 are N side and  less flow from E242 to reaction site B  (Fig.\ref{fig:Current_ox_sw} (c)-(d)).   And due the decrease of chemical potential in E242, more protons are leaking back from  Protein Loading Site PLS site to the E242 (Fig.\ref{fig:Current_ox_sw} (e)) which induces more protons flows from P side to the  Protein Loading Site PLS site (Fig.\ref{fig:Current_ox_sw} (f)).   The corresponding dynamics of protons and electron are shown in Fig. \ref{fig:Concentration_ox_sw}.

		\end{document}